\documentclass[twocolumn,svgnames]{aastex63}
\usepackage{hyperref}
\usepackage{scrextend}
\deffootnote[0.45em]{0.2em}{0.2em}{\textsuperscript{\thefootnotemark}\,}
\usepackage{amsmath,url}
\usepackage{amssymb}
\usepackage[para,online,flushleft]{threeparttable}
\usepackage{algorithm}
\usepackage{amsfonts}
\usepackage{mathrsfs}
\usepackage{booktabs}

\usepackage{rotating}
\usepackage{graphicx}
\usepackage{subfigure}
\usepackage{lineno}
\usepackage[toc,page]{appendix}

\usepackage[]{xcolor}

\usepackage{natbib}
\defcitealias{nuria2023}{NTA23}
\defcitealias{zhao_2021}{XZ21}

\title{Hydrogen Column Density Variability in a Sample of Local Compton-Thin AGN II}

\newcommand{\chandra}{\textit{Chandra}}
\newcommand{\xmm}{XMM-\textit{Newton}}
\newcommand{\nustar}{\textit{NuSTAR}}
\newcommand{\suzaku}{\textit{Suzaku}}

\newcommand{\mytorus}{\texttt{MYTorus}}
\newcommand{\uxclumpy}{\texttt{UXCLUMPY}}
\newcommand{\borus}{\texttt{borus02}}
\newcommand{\xspec}{\texttt{XSPEC}}
\newcommand{\nh}{N$_{\rm H,los}$}
\newcommand{\nhav}{N$_{\rm H,av}$}
\newcommand{\nheq}{N$_{\rm H,eq}$}

\hypersetup{
    colorlinks=true,
    linkcolor=purple,
    filecolor=magenta,      
    urlcolor=SlateBlue,
    citecolor=BlueViolet,
    pdftitle={Overleaf Example},
    pdfpagemode=FullScreen,
    }

\shorttitle{Hydrogen Column Density Variability in a Sample of Local Compton-Thin AGN II}
\shortauthors{Pizzetti et al.}

\begin{document}

\title{Hydrogen Column Density Variability in a Sample of Local Compton-Thin AGN II}

\author{A. Pizzetti}
\affiliation{Department of Physics and Astronomy, Clemson University,  Kinard Lab of Physics, Clemson, SC 29634, USA}

\author{N. Torres-Alb\`{a}}
\affiliation{Department of Physics and Astronomy, Clemson University,  Kinard Lab of Physics, Clemson, SC 29634, USA}

\author{S. Marchesi}
\affiliation{Department of Physics and Astronomy, Clemson University,  Kinard Lab of Physics, Clemson, SC 29634, USA}
\affiliation{Dipartimento di Fisica e Astronomia (DIFA), Università di Bologna, via Gobetti 93/2, I-40129 Bologna, Italy}
\affiliation{INAF - Osservatorio di Astrofisica e Scienza dello Spazio di Bologna, Via Piero Gobetti, 93/3, 40129, Bologna, Italy}

\author{J. Buchner}
\affiliation{Max-Planck-Institut f\"{u}r extraterrestrische Physik, Giessenbachstra{\ss}e 1, D-85748 Garching, Germany}

\author{I. Cox}
\affiliation{Department of Physics and Astronomy, Clemson University,  Kinard Lab of Physics, Clemson, SC 29634, USA}

\author{X. Zhao}
\affiliation{Department of Astronomy, University of Illinois at Urbana-Champaign, Urbana, IL 61801, USA}

\author{S. Neal}
\affiliation{Department of Physics and Astronomy, Clemson University,  Kinard Lab of Physics, Clemson, SC 29634, USA}

\author{D. Sengupta}
\affiliation{Dipartimento di Fisica e Astronomia (DIFA), Università di Bologna, via Gobetti 93/2, I-40129 Bologna, Italy}
\affiliation{INAF - Osservatorio di Astrofisica e Scienza dello Spazio di Bologna, Via Piero Gobetti, 93/3, 40129, Bologna, Italy}

\author{R. Silver}
\affiliation{NASA Goddard Space Flight Center, Greenbelt, MD, 20771, USA}

\author{M. Ajello}
\affiliation{Department of Physics and Astronomy, Clemson University,  Kinard Lab of Physics, Clemson, SC 29634, USA}

\begin{abstract}
\noindent We present the multi-epoch analysis of 13 variable, nearby ($\rm z\lesssim0.1$), Compton-thin ($10^{22}<\rm{N_H}<1.5\times 10^{24}\mathrm{\,cm^{-2}}$) active galactic nuclei (AGN) selected from the 105-month BAT catalog. Analyzing all available archival soft and hard X-ray observations, we investigate the line-of-sight hydrogen column density (\nh) variability on timescales ranging from a few days to approximately 20 years.
Each source is analyzed by simultaneously modeling the data with three physical torus models, providing tight constraints on torus properties, including the covering factor, the cloud dispersion, and the torus average hydrogen column density (\nhav). For each epoch, we measure the \nh\ and categorize the source as `\nh\ Variable', `Non-variable in \nh', or `Undetermined' based on the degree of variability.
Our final sample includes 27 variable, Compton-thin AGN after implementing another 14 AGN analyzed in our previous work.
We find that all sources require either flux or \nh\ variability. We classify $37\%$ of them as `\nh\ Variable', $44\%$ as `Non-variable in \nh', and $19\%$ as `Undetermined'. Noticeably, there is no discernible difference between geometrical and intrinsic properties among the three variability classes, suggesting no intrinsic differences between the \nh-variable and non-variable sources. We measure the median variation in \nh\ between any observation pair of the same source to be $25\%$ with respect to the lowest \nh\ measure in the pair. Furthermore, $48\%$ of the analyzed sources require the inclusion of a Compton-thick reflector in the spectral fitting. Among these, the $30\%$ exhibits recorded 22 GHz water megamaser emission, suggesting a potential shared nature between the two structures.

\end{abstract}

\section{Introduction} 

Active Galactic Nuclei (AGN) are among the brightest and most energetic objects in the Universe. They are powered by a supermassive black hole (SMBH) at the center of the host galaxy, accreting matter from an accretion disk surrounded by a toroidal structure of gas and dust, the obscuring torus \citep{antonucci_1993,urry_padovani}. Ultraviolet photons arising from the accretion disk get up-scattered to X-rays via inverse Compton scattering by hot electrons close to the accretion disk \citep{haardt1994,ramosailmeida_ricci_2017,gianolli2023}. Being produced in the very center of the AGN, X-ray photons are powerful messengers that probe the physics of the accretion system and the matter in its surroundings. Although recent studies \citep[see, e.g.,][]{buchner2017,andoni3e2022} have shown a minor part of the obscuration to be generated on galactic scales, submillimeter, infrared, optical and X-ray studies \citep[see, e.g.,][and references therein]{Risaliti2002,risaliti_2005, simpson2005,Murphy2009, furui2016,Combes2019, honig2019, garcia_burillo2021} infer the presence of a torus-like structure of molecular and atomic gas and dust on parsec scales as the origin of the obscuration and reprocessing of the incoming AGN radiation. 

The atomic and molecular gas density in the torus is typically parameterized in X-rays as the line-of-sight neutral hydrogen column density \nh. When the X-ray-obscuring matter has a column density which is equal to or larger than the inverse of the Thomson cross-section $\sigma_T$, \nh $\geq \sigma_T^{-1}\simeq 1.5 \times 10^{24}$\,cm$^{-2}$, the source is defined as Compton-thick (CT) AGN \citep[e.g.,][]{ross1979, comastri2004}, and we define it as obscured Compton-thin when $10^{22}$\,cm$^{-2}<\rm N_{\rm H,los}< 1.5 \times 10^{24}$\,cm$^{-2}$.
 Despite the presence of other spectral components, such as thermal emission from hot plasma in the host galaxy \citep[e.g.,][]{comastri2004, torres-alba_2018}, the 2-10\,keV X-ray spectrum of obscured AGN is characterized by the photoabsorption of the main power-law photons by the gas in the line-of-sight (the so-called \textit{transmitted component} of the X-ray spectrum). Infrared and X-ray analyses \citep[e.g.,][]{Risaliti2002,risaliti_2005,bianchi_2005,Schartmann2005,risaliti2007,nenkova2008a, nenkova2008b,Sanfrutos2013,Markowitz_2014,Laha2020, buchner2021} have ruled out the initial idea of a homogeneous torus, favoring a clumpy scenario in which the torus is a dynamic structure made of clouds of different densities. Indeed, recent studies have shown how the X-ray obscuration is likely to be produced by multiple absorbers on various spatial scales \citep{ramosailmeida_ricci_2017, Buchner_2019,marchesi2022, pizzetti2022, nuria2023}, matching the description of an obscurer with a complex structure made of clouds with different column densities. In addition, \nh\ variability studies on large source samples \citep[e.g.,][]{Markowitz_2014,hernandez2015,Laha2020}, detect only a few sources to be variable, suggesting a scenario in which a more intricate system of clouds of different sizes, densities and radial velocity is preferred over a simple patchy torus.
The modeling of the AGN broadband X-ray spectra and its variability between multiple epochs is thus crucial in revealing precious information about the density and the location of the absorbing media surrounding the central engine and how it connects with the large-scale structures of the host galaxies \citep[see, e.g.,][]{risaliti_2005,Balokovic2018,marchesi2022, pizzetti2022,nuria2023}. 

In this work, we present the multi-epoch X-ray analysis of a sample of 13 potentially \nh\ variable obscured AGN, each of which has multiple X-ray observations covering both the hard and the soft regions of the X-ray spectrum. In most cases, observations are taken over a period of $\sim20$ years. The spectral analysis is performed by simultaneously fitting all the X-ray spectra available for each source; this approach is proven to be a powerful tool to constrain overall torus parameters that are not expected to change in the considered timescales (such as the covering factor, the torus inclination angle and the torus average column density), while allowing the estimation of \nh variability, as shown by \citet{Balokovic2018}and \citet{marchesi2022}. In order to have a larger statistical sample, we include in the analysis the sample studied by \cite{nuria2023}, as well as NGC 6300 (Sengupta et al.,  in prep.) and NGC 7479 \citep{pizzetti2022}. The overall properties of the whole sample of galaxies are thus presented.

The paper is structured as follows: in Sections \ref{sec:Observ} and \ref{sec:analysis}, we describe the sample selection, the data reduction and the X-ray spectral analysis processes. We outline the methods used to classify the sources as `Variable', `Non-variable' and `Undetermined' in Section \ref{section:var_estimates} and in Section \ref{sec:results}, we present the obtained results. The computation of the equatorial column densities for \uxclumpy\ is presented in Section \ref{sec:nheq_uxclumpy}. We discuss the results obtained from the variability analysis performed on the full sample of galaxies in Section \ref{sec:discussion}, and we report our conclusions in Section \ref{sec:conclusion}. Information about the best-fit parameters, spectra, \nh-variability plots, and comments on each individual source are presented in Appendices \ref{appendix:xray_fitting_allsources}, \ref{sec:allspectra}, \ref{sec:variability_plots}, and \ref{appendix:individual_sources}, respectively.

The cosmological parameters assumed in this work are compatible with a flat $\Lambda$CDM cosmology with H$_0$=69.6\,km\,s$^{-1}$\,Mpc$^{-1}$, $\Omega_m$=0.29 and $\Omega_\Lambda$=0.71 \citep{bennett2014}. All reported uncertainties are at a 90$\%$ confidence level unless otherwise stated.

%
%
\section{Sample selection}\label{sec:Observ}

The sample analyzed in this work is complementary to the one analyzed in \cite{nuria2023} -  \citepalias[][hereafter]{nuria2023}. Both samples have been selected from the work of \cite{zhao_2021} - \citepalias[hereafter][]{zhao_2021}, which analyzed 93 Compton-thin AGN with \nh\ between $10^{23}-10^{24}$\,cm$^{-2}$, selected from the 105-month Burst Alert Telescope \citep[BAT,][]{Barthelmy2005}-catalog \citep{Oh2018}. All the sources analyzed in \citetalias{zhao_2021} were fit with at least one \nustar\ observation and one soft X-ray observation, either from \xmm, \chandra, or \textit{Swift}-XRT. This guarantees low and high-energy X-ray coverage, which is necessary to disentangle the degeneracy between reflection and line-of-sight components.

Of the 93 sources, 22 had simultaneous \nustar\ and soft X-ray observations, and 13 were analyzed using \textit{Swift}-XRT, which typically exhibits a very low signal-to-noise ratio. In 31 of the remaining 58 sources, \citetalias{zhao_2021} detected either intrinsic flux or \nh\ variability, determined via simultaneous X-ray spectral analysis of one \nustar\ and one soft X-ray observation for each source.
We note that in this preliminary analysis by \citetalias{zhao_2021}, for each source, both \nh\ and intrinsic flux variability were considered in order to obtain a good fit. Considering the \nh\ and flux variability can be highly degenerate, performing a simultaneous multi-epoch analysis of all the observations available for each of the 31 sources is then pivotal to better understand the origin of variability, as shown by \citet{pizzetti2022, marchesi2022} and \citetalias{nuria2023}.
Out of the 31 variable galaxies, NGC 7479 has been analyzed in a pilot project by \cite{pizzetti2022}, 12 have been analyzed in \citetalias{nuria2023}, and two are being excluded due to the lack of new available data with respect to the one analyzed by \citetalias{zhao_2021} (ESO 201-IG004 and 4C+73.08); NGC 6300, MRK 477 and NGC 7582 are being analyzed in single-source projects (Sengupta et al.,  in prep., Torres-Alb\`{a} et al., in prep.), due to a large number of observations available. This leaves us with 13 sources that are studied in this work. The list of the total 52 observations analyzed in this paper is presented in Table \ref{tab:all_observ}.

\begin{table*}[h!]
\centering
 \caption{Observation details and properties of the sources analyzed in this work. }
 \label{tab:all_observ}
\vspace{.1cm}
\resizebox{\textwidth}{!}{%
  \begin{tabular}{cccccccc}
 
       \hline
       \hline
    \textbf{Source Name}& \textbf{R.A.} & \textbf{Dec}& \textbf{Redshift} & \textbf{Instrument}  & \textbf{ObsID}     &    \textbf{Exposure Time  } &   \textbf{Obs Date}  \\ 
    & J2000 & J2000 & &&&[ks] &\\
    (1) & (2) & (3) & (4) & (5) & (6) & (7) & (8)\\
    \hline
    
 NGC 454E & 01:14:24.93 & -55:23:49.31 &  0.012 & \suzaku\     &  704009010    &   	128.95        & 	2009-04-29  \\
    
 && & & \xmm\        &  0605090301   &      29.9      &  2009-11-05  \\
    
 & & && \chandra\ - 1   &   13891   &     	5.1        &   2012-02-19 \\
    
  && & &\nustar\ - 1     &  60061009002    &   	24.2         & 2016-02-14  \\
    
 & && & \nustar\ - 2    &  60760003002    &   25.4          &  2021-02-14  \\
    
 & & && \chandra\ - 2    &   23812   &     	15.1    &   2021-02-14 \\
    \hline

    MRK 348 & 00:48:47.14 & +31:57:25.08 & 0.015 &  \xmm\ - 1        &  0067540201   &      49.5       &  	2002-07-18  \\
    
     & & &&\suzaku\     &  703029010    &   	87.5        & 	2008-06-28  \\
     
   & & && \chandra\    &   12809   &     	95.0        &   2010-10-13 \\
    
  & & &&  \xmm\ - 2    &   0701180101   &     	13.1    &   2013-01-04   \\
    
   & & && \nustar\ - 1     &  60160026002    &   	21.5        & 2015-10-28\\
    
  & & &&  \nustar\ - 2    &  60701017002    &   104.8          &  2022-02-06   \\
  
     \hline
 NGC 4992 & 13:09:05.59 & +11:38:02.82 & 0.025  &\chandra\ - 1   &   6277   &     	3.7        &   2005-07-25 \\
    
  & & & & \suzaku\     &  701080010    &   	38.8       & 	2006-06-18 \\
        
   & & & &\xmm\        &  0312192101   &      	16.4       &  2006-06-27  \\
 
   & & & &\nustar\- 1     &  60061239002    &   	23.5         & 2015-01-27  \\

   & & & &\chandra\ - 2    & 	23707   &     	6.3    &   2021-04-04  \\
   
     \hline 
        ESO 383-18 & 13:33:26.10 & -34:00:53.31 & 0.012 &\xmm\        &  0307000901   &      	16.1       &  2006-01-10 \\
    
    & & & &\nustar\ - 1     &  60061243002    &   	17.3         & 2014-09-11  \\
    
    & & & &\nustar\ - 2     &  60261002002    &   		106.5         & 	2016-01-20  \\
    
    & & & &\chandra\   &   23687   &     		2.1        &   2021-04-03 \\
   
  \hline   
   
  MRK 417 & 10:49:30.92 & +22:57:52.37 & 0.032 & \xmm\        &  0312191501   &      	14.3       &  2006-06-15 \\
    
   & && &\suzaku\     &  702078010    &   	41.5          & 	2007-05-18  \\
    
  && &  & \nustar\     &  60061206002    &   	20.7      & 	2017-02-20  \\

   && && \chandra\   &   23724   &     		5.7         &   2022-07-10 \\
    
      \hline

 MCG-01-05-047  & 01:52:49.004 & -0.3:26:48.56 & 0.017 &\suzaku\     &  704043010   &   	46.9        & 		2009-06-25  \\
    
   & &&   &  \xmm\        &  0602560101   &      	14.9       &  2009-07-24  \\
       
   & && &\nustar\     &  60061016002    &   	13.4         & 		2012-11-30  \\
   
   && & &\chandra\   &   26146   &     		10.1        &   	2021-10-10 \\
  
       \hline

 ESO 103-35 & 18:38:20.32  & -65:25:39.14 & 0.013 &\xmm\        &  0109130601   &      12.9  &  2002-03-15  \\
    
  & & &  &\suzaku\     & 703031010  &  91.4   & 2008-10-22           \\
    
  && &  &\nustar\ - 1     &  60061288002    &   27.3         & 2013-02-24   \\
    
   & && &\nustar\ - 2     &  60301004002    &   43.8     & 	2017-10-15  \\
    
  \hline
      
   NGC 1142 & 02:55:12.23 & -00:11:00.80 &  0.028  & \xmm\        & 	0312190401   &     11.9    &  	2006-01-28  \\
      
   &&&& \suzaku\ - 2     & 701013010    &   	101.6        &    2007-01-23  \\
    
   &&& &\suzaku\ - 1     & 702079010    &   	40.6        & 	2007-07-21 \\

  &&& & \nustar\     &  60368001002   &   	20.7         & 	2017-10-14  \\
       \hline
 
IRAS 16288+3929  & 16:30:32.65 & 39:23:3.15 & 0.030 &  \suzaku\     & 806150010    &   	18.7        & 	2011-08-18 \\
    
  &&& &  \xmm\        &  0671900101   &     31.0     &  2011-09-08\\
    
  &&& &  \nustar\     &  60061271002   &   	16.1         & 	2012-09-19 \\

       \hline 
 
ESO 263-13 &10:09:48.21 & -42:48:40.4 & 0.032 &   \suzaku\     & 702120010  &   	45.3        & 		2007-06-01  \\
 
  &&&&  \xmm\        & 0501210101   &      26.3       &  		2007-06-14         \\
    
  &&&&  \nustar\     &  60061098002	    &   22.8      & 	2015-10-13  \\

       \hline

 Fairall 272 & 08:23:01.10 & -4:56:5.50 & 0.022 & \xmm\        & 0501210501   &      	11.914       &  		2007-10-14  \\
    
   & & &  &\nustar\     &  60061080002    &   	24.3        & 		2014-01-10 \\
    
  & & & &  \chandra\   &   	23703   &     		2.9        &   		2021-03-28  \\
      
       \hline

  LEDA 2816387 & 3:56:19.97 &  -62:51:28.73 & 0.107 &  \nustar\- 1     &  	60201034002    &   	26.5         & 	2016-05-06   \\
    
  & & & & \xmm\        &  0802450201   &      21.0       &  2017-11-20                 \\
  
  & & &  &\nustar\- 2    &  60160172002    &   	24.6          &  2021-06-04  \\
    
   \hline

  2MASX J06411806+3249313 & 6:41:18.04 & +32:49:31.6 & 0.047 &  \xmm\        &  0312190901   &      	13.9       &  2006-03-11  \\
    
  & & &  & \chandra\    &   22040   &     10.1       &   2018-12-25 \\
    
 & & &  & \nustar\     &  60061071002    &   	18.3         & 2018-12-31   \\

    
  \hline  
           
\end{tabular}}
\begin{tablenotes}
    \textbf{Notes:} (1) Source name; (2) and (3) Source R.A. and Dec (J2000 Epoch); (4) Source redshift from NED (NASA/IPAC extragalactic database \url{https://ned.ipac.caltech.edu/}); (5) Telescope used in the analysis; (6) Observation ID; (7) Nominal exposure time in ks; (8) Observation starting date.\\ The sources are arranged in descending order based on the number of observations.
\end{tablenotes} 
  
\vspace{.2cm}
\end{table*}

\section{Data reduction}
\label{sec:datareduction}
In this Section, we present the data reduction processes applied to the \chandra, \xmm, \suzaku, and \nustar data used in this work. 
\subsection{Chandra}

The data were reduced with the CIAO \citep[v4.14;][]{fruscione2006} software and the \chandra\ Calibration Data Base \texttt{caldb} 4.9.7 adopting standard procedures. The source and the background spectra were extracted using the CIAO \texttt{specextract} tool. For the source, we selected a circle of $5"$; the background was extracted either from an annulus centered in the source or a circle positioned a few arcseconds away after a visual inspection to avoid the presence of any other source in the field. The spectra were grouped between 3 - 15 counts per bin (see Appendix \ref{appendix:individual_sources} for details), using the \texttt{grppha} tool. 

\subsection{\xmm}
The observations were reduced using the Science Analysis System \citep[SAS;][]{jansen2001} version 20.0.0. For each observation, we inspected the lightcurve at energies E$>$10\,keV in order to remove high particle background. 
We extracted the spectra from a region of 30" for the source and 40" for the background for all the detectors after visually inspecting each image to avoid contamination from nearby X-ray sources. 
Finally, we binned the spectra to have at least 15 counts per bin. 

\subsection{\suzaku}
The \suzaku\ data were extracted following the ABC guide\footnote{\url{https://heasarc.gsfc.nasa.gov/docs/suzaku/analysis/abc/}} from HEASARC. For each observation, we extracted the spectra from both the frontside (XI0, XI3) and back-side (XI1) illuminated chips unit of the X-ray Imaging Spectrometers (XIS) running the \texttt{aepipeline}. To extract the spectra, we used a region of 60"-260" depending on the source\footnote{Following the recommendation on the region selection in the ABC guide.}. The response, ancillary and background files were generated running the tasks \texttt{xisrmfgen}, \texttt{xissimarfgen} and \texttt{xisnxbgen}, respectively. We then grouped the data to a minimum of 25 counts per bin.

\subsection{\nustar}
The \nustar\ data were retrieved from both focal plane modules, FPMA and FPMB. The event data files were calibrated and cleaned using the \nustar\ \texttt{nupipeline} script version 0.4.8. and the Calibration Database (CALDB) v.20210427 as response file. Then, we used the \texttt{nuproducts} script to generate the  ARF, RMF and light-curve files. The source spectrum was extracted from a 75" circular region centered on the source's optical position. 
After a visual inspection, we used a region of the same size to extract the background spectrum from each module to avoid contamination from other sources. Lastly, the \nustar\ spectra were grouped with the \texttt{grppha} task with a minimum of 15 counts per bin.

%
%

\section{X-ray spectral analysis}\label{sec:analysis}
The X-ray spectra were analyzed in \xspec\ \cite{Arnaud1996} version 12.12.1 in HEASOFT version 6.30.1.

We analyzed the torus properties of each source in the sample by simultaneously fitting the X-ray spectra using three models based on Monte-Carlo simulations that can self-consistently describe the primary AGN emission with the Compton-thick gas (N$_{\rm H,los}>1.5\times 10^{24}$\,cm$^{-2}$) in the surrounding torus. The three torus models used in this work are \borus\ \citep{Balokovic2018}, \mytorus\ \citep{Murphy2009,Yaqoob2012} and \uxclumpy\ \citep{Buchner2019}.

To all models, we add a collisionally-ionized diffuse gas emission component \citep[\texttt{apec},][]{smith2001} in order to account for the soft X-ray emission from the central region of the host galaxy. The galactic absorptions measured by \cite{Kalberla2005} are included to consider the absorption due to the gas in the Milky Way. Finally, we model the fraction of the intrinsic AGN power law that is scattered without being reprocessed by the obscuring material by multiplying the intrinsic power law (\texttt{cutoffpl} in the model) by the fraction of the scattered emission, $F_s$. To account for intrinsic flux variability between different epochs, we apply a multiplicative constant, $C_{\rm AGN}$, to the intrinsic power-law model. 


In \texttt{XSPEC} notation, the source model is defined as:
\begin{eqnarray}
 Model =  phabs * \{apec +C_{\rm AGN}*(TorusModel+\nonumber \\ +F_s*cutoffpl)\}.  
\end{eqnarray}

Here, the \textit{TorusModel} is one of the aforementioned torus models.

All the models are fitted in the 0.6\footnote{The \mytorus\ model is not accurate below 0.6\,keV.}-30/70\,keV range, depending on the energy at which the \nustar\ background counts overtook the source emission for each source. 
To be consistent with \cite{balokovic2020} and their study on the high-energy cutoff of local obscured AGN, we froze the high-energy cutoff for all three models to 300\,keV. Furthermore, using only \nustar\ data (up to 70\,keV in a best-case scenario), we would be unable to constrain the cut-off energy. 
 



\subsection{borus02}\label{subsec:borus02}

The \borus\ model assumes a uniform-density sphere with two conical cutouts filled with a cold, neutral, static medium with solar abundance. In this model, the line-of-sight inclination angle $\theta_i$ and the covering factor $C_F$ (the fraction of the sky covered by the torus as seen from the central engine) are variable parameters.

Since \borus\  models only the reflection component of the AGN emission, which accounts for the Compton-hump and the fluorescent emission lines, the absorption of the torus clouds and the scattered component were added manually.

In \xspec, the model has the following configuration:
\begin{eqnarray}
 \label{eq:borus02}
    TorusModel = \text{borus02\_v170323a.fits}+ \nonumber\\ 
     +zphabs*cabs*cutoffpl.
\end{eqnarray}
where \texttt{zpbabs} and \texttt{cabs} account for the photoelectric absorption due to the cold medium and the Compton scattering losses along the line of sight, respectively.

\subsection{MYTorus}\label{subsec:myt}
The \mytorus\ model considers a cylindrical, azimuthally symmetric torus with a fixed half-opening angle of $60^{^{\circ}}$, filled with a uniform neutral cold reprocessing material. In this model, the main components of an obscured AGN X-ray spectrum (the line of sight, the reflection and the fluorescent emission line component) are treated self-consistently by the use of three different tables, as can be seen in Eq.~\eqref{eq:myt}. The Compton-scattered and the emission lines components are weighted differently by adding multiplicative constants $A_S$ and $A_L$, respectively. 

In this work, we use the model in the decoupled configuration \citep{Yaqoob2012, yaqoob2015}, which allows the disentanglement of the line-of-sight column density from the average one (\nhav). 
In the decoupled configuration, the zeroth-order continuum (the continuum photons that escaped the torus without being scattered) is independent of the inclination angle, which is fixed to be $\theta_i=90^{\circ}$. In this way, the zeroth-order continuum is independent of geometry and becomes a line-of-sight quantity. 
In order to take into account the possible patchiness and configurations of the torus and of the consequent Compton-scattering and lines features, these two components are considered both in an edge-on and face-on configuration. In the first case, the inclination angle, set to be $\theta_{i,S,L}=90^{^{\circ}}$, mimics the forward scattering and it is weighted by $A_{S,L90}$; this means that we are accounting for a more uniform torus, as the photons are primarily reprocessed by the obscuring material that is lying between the AGN and the observer.
In the second case, $\theta_{i,S,L}=0^{^{\circ}}$ accounts for a backward scattering and $A_{S,L0}$ is the weighting constant. This second scenario is more likely to happen when the torus presents a more patchy structure, in which the photons scattered by the back side of the torus have less chance to interact again with the material before reaching the observer. When $A_{S,L90}$ and $A_{S,L0}$ are left free to vary, we refer to the configuration as ``decoupled free", and a ratio between the two constants can give a qualitative idea of which emission is more prominent, thus giving us an indication of the inclination angle of the torus. In addition, a ratio between N$_{\rm H,los}/$N$_{\rm H,av}$ can approximate the clumpiness of the torus.

In \xspec, the model is as follows:
\begin{eqnarray}
 \label{eq:myt}
  TorusModel =  \text{mytorus\_Ezero\_v00.fits} * zpowerlw + \nonumber \\ 
  +A_{\rm S,0} * \text{mytorus\_scatteredH300\_v00.fits} + \nonumber \\ 
  +A_{\rm L,0} * \text{mytl\_V000010nEp000H300\_v00.fits} +\nonumber \\ 
  +A_{\rm S,90} * \text{mytorus\_scatteredH300\_v00.fits} + \nonumber \\ 
  +A_{\rm L,90} * \text{mytl\_V000010nEp000H300\_v00.fits}.
\quad
\end{eqnarray}


\subsection{UXCLUMPY}\label{subsec:uxclumpy}

\uxclumpy\ \citep[]{Buchner2019} is a model constructed to reproduce and model the column density and the cloud eclipsing events in AGN tori in terms of their angular sizes and frequency. 

\uxclumpy\ mainly differs from the two aforementioned models in that the torus clumpiness and the cloud dispersion are included in the model. Indeed, the model is constructed to reproduce a cloud distribution with different hydrogen column densities based on observed eclipse event rates \citep[]{Markowitz_2014, Buchner_2019} assuming the clouds to be on circular Keplerian orbits on random planes, for simplicity. The dispersion of the distribution is modulated by TOR$\sigma$ ($\sigma \in [0-84]$), where a large value stands for a large dispersion of the clouds. To model strong reflection features \citep[see, e.g.,][]{pizzetti2022}, \uxclumpy\ allows for an additional \textit{inner ring} of CT material, whose covering factor is measured by the CTKcover parameter (CTKcover $\in [0-0.6]$). Both TOR$\sigma$ and CTKcover provide a powerful, consistent way to probe the torus geometry by modeling the clouds and the extension of the reflector. TOR$\sigma$, CTKcover, and the line-of-sight inclination angle $\theta_{i}$ are free parameters during the fit. 

In the previously discussed model geometries, \nhav\ and \nh\ are closely related. In contrast, \uxclumpy\ is a unified obscurer model, where one clumpy torus geometry defined by TOR$\sigma$ and CTKcover can be observed under a wide range of \nh\ ($10^{20} - 10^{26}\mathrm{\,cm}^{-2}$). The corresponding equatorial hydrogen column density \nheq\ has so far not been published. For comparison with the other models, we compute for each TOR$\sigma$ and CTKcover combination the \nheq\ in Section \ref{sec:nheq_uxclumpy}.

In \texttt{XSPEC}, the model is as follows:
\begin{eqnarray}
 \label{eq:uxclumpy}
    TorusModel = \text{uxclumpy-cutoff.fits}+ \nonumber\\ +F_s*\text{uxclumpy-cutoff-omni.fits}.
\end{eqnarray}
The first table includes the torus transmitted and reflected component. In contrast, the second one, multiplied by the scattering fraction, takes into account the presence of a \textit{warm mirror emitter}, a volume-filling gas between the clumps that is, in part, responsible for the scattering of the intrinsic AGN powerlaw.

\section{Variability estimates}\label{section:var_estimates}
In this section, we briefly summarized the method we implemented to measure the \nh\ variability in the studied sample. For a complete explanation, we invite the reader to check Section 5 of \citetalias{nuria2023}.
 The method proposed by \citetalias{nuria2023} and reported below uses two different estimators of source variability, the \textit{reduced $\chi^2$} ($\chi^2_{red}$) and the \textit{p-value}. 
\subsection{Reduced $\chi^2$/reduced stat comparison}\label{subsection:chicomparison}
To test for variability, we compare the $\chi^2_{red}$ of the best fit to the X-ray data performed with \texttt{xspec} (which assumes both intrinsic flux and \nh\ variability), to the 
best possible fits performed when fitting the data under three assumptions:
1) no variability at any epoch is taken into account, either in intrinsic flux or \nh, ($\chi^2_{red}$ No Var); 2) only \nh\ variability is considered ($\chi^2_{red}$ No C$_{AGN}$ Var); 3) only intrinsic flux variability at any epoch is considered ($\chi^2_{red}$ No \nh Var). We then define a tension between the data and the model as $T=|1-\chi^2_{red}|/\sigma$, where $\sigma=1/\sqrt{N}$ is the standard deviation of the Gaussian that best represents the $\chi^2$ distribution, and N the number of degrees of freedom of the spectral fit. We consider model A to fit a source significantly better than model B when $\rm T_A<3$ and $\rm T_B>5$ \citep[see][for more details]{andrae2010}. 

This method classifies sources as \nh-variable by comparing the best-fit T with the non-\nh-variable T. If both models yield a $\rm T<3 $, we classify the source as non-variable in \nh, as \nh\ variability is not required to fit the data. Similarly, in the case both models (variable and non-variable) return $\rm T>5 $, we consider the difference in the significance between the two models and discuss whether the inclusion of \nh-variability improves the fit or not (see, e.g., NGC 4388 in \citetalias{nuria2023}, ESO 383-18 and ESO 263-18 in Appendix \ref{sec:eso383-18}, \ref{sec:eso263-13}, respectively). We classify the source as `Undetermined' when the three torus models yield different classifications. We note that in the cases in which the reduced statistics is used, the distribution does not necessarily follow a Gaussian, as the $\chi^2$ does for a large number of degrees of freedom. Thus, the interpretation of T in such cases is not straightforward. We still provide the value as a reference. 

We note that, for some sources, a combination of $\chi^2$ and C-statistics was applied to fit the data (see individual source comments in Appendix \ref{appendix:individual_sources}). In such cases, we use \textit{reduced stat} instead of $\chi^2_{red}$. The reduced $\chi^2$/reduced stat and p-values for each source and all models are listed in Table \ref{tab:2MASXJ06411806+3249313_fitting} and Table \ref{tab:NGC454e_fitting} through \ref{table:leda_fitting}.

\subsection{p-value}\label{subsection:pvalue}  We estimate the probability that the source is non-variable in \nh\ (null hypothesis) by computing a new $\chi^2$, defined as follows:
\begin{equation}
\chi^2_{N_H} = \sum_{i} \frac{(\mathrm{N_{H,los,i}}-\langle\mathrm{N_{H,los}}\rangle)^2}{\delta(\mathrm{N_{H,los,i})^2}}.\\
\end{equation}
which derives from fitting the \nh\ for all epochs, for each source ($\rm{N_{H,los,i}}$) to the average \nh\ for that source ($\langle\mathrm{N_{H,los}}\rangle)$.
The $\chi^2$ is then converted into a p-value (null hypothesis: the source is not \nh-Variable) by following the method described in \citetalias{nuria2023}. We classify the source as `\nh\ variable' if p-value $<0.01$ for all the three models used, as `non-variable in \nh' if p-value $>0.01$. We classify the source as `undetermined' if the three models are in disagreement, i.e., the p-value is above the threshold for one model and below for the others. 

The final source classification is reported in Table \ref{tab:var_results}, in which we compare the $\chi^2_{red}$ and the p-value results for each model. We classify the source as `undetermined' when the two methods or the different fitting models disagree in the classification. 


\section{Results}\label{sec:results}
\begin{figure}[ht]
    \centering
    \includegraphics[trim={0cm 1cm 0cm 0cm},clip, width=0.5\textwidth]{ 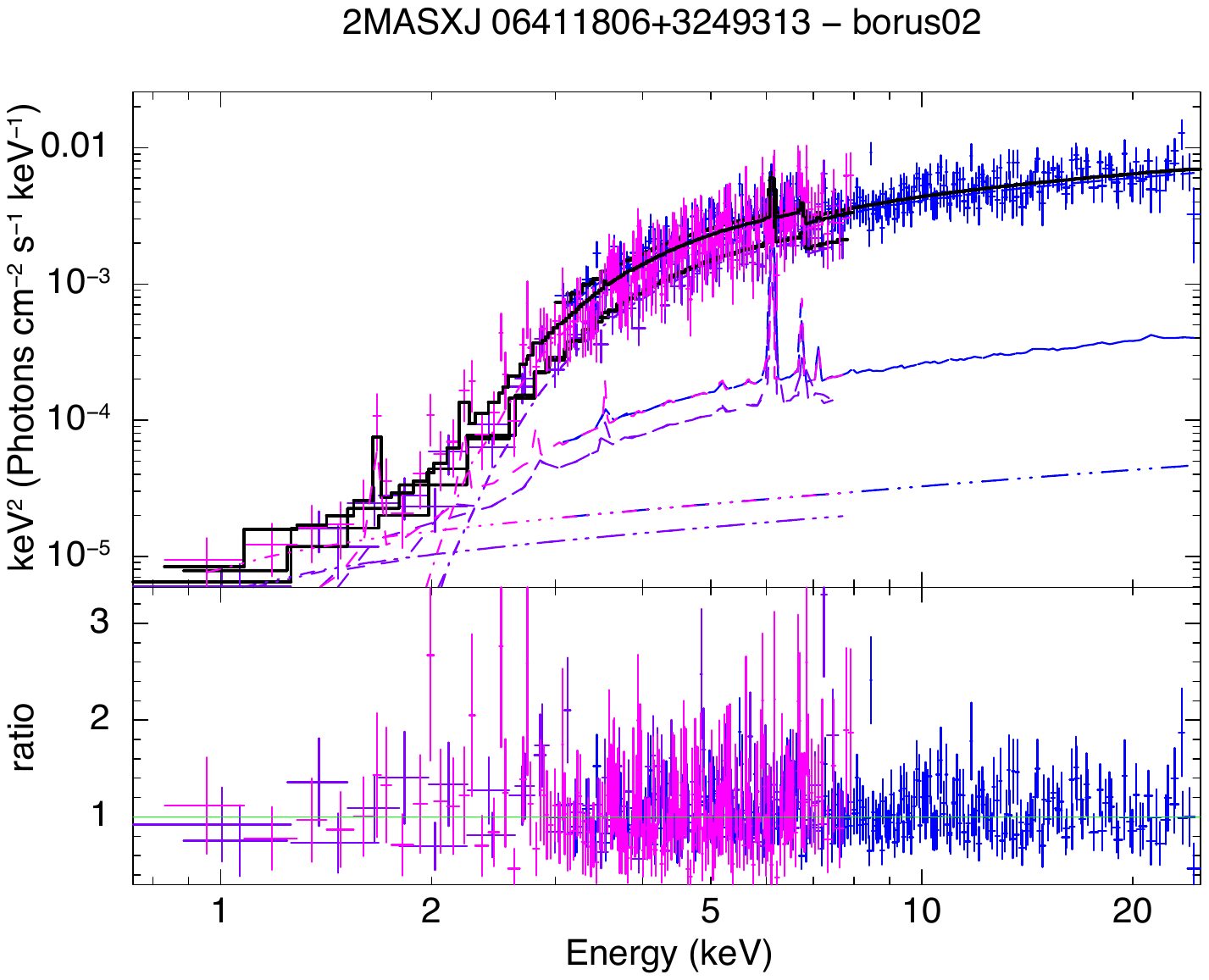}
    \caption{Unfolded \chandra\ (magenta), \xmm\ (purple) and \nustar\ (blue) 0.6-25\,keV combined spectrum of 2MASX J06411806+3249313 modeled with \borus. 
    The best-fit model is plotted with a solid black line. Reflection: dashed, Line-of-sight: dash-dotted, Scattering: dash-dot-dot-dot.}
    \label{fig:borus_spectrum}
\end{figure}

In this Section, we present the result of the X-ray spectral analysis performed on the 13 galaxies of the sample. We perform a simultaneous multi-epoch fit for each galaxy using the three models described above. 
We display the best-fit results (Table \ref{tab:2MASXJ06411806+3249313_fitting}), the \borus\ best-fit spectrum (Figure \ref{fig:borus_spectrum}) and the \nh-variability plot (Figure \ref{fig:nhvar_plots3}) for 2MASX J06411806+3249313 (2MASXJ 06411806 hereafter). The best-fit values, the \borus\ best-fit spectra and \nh-variability plots for the remaining 12 sources are reported in Appendix \ref{appendix:xray_fitting_allsources}, \ref{sec:allspectra}, and \ref{sec:variability_plots}, respectively. The comments on each individual source are reported in Appendix \ref{appendix:individual_sources}.

\begin{table*}
\centering
\begin{threeparttable}
\caption{X-ray fitting results for 2MASX J06411806+3249313}
\label{tab:2MASXJ06411806+3249313_fitting}
\renewcommand*{\arraystretch}{1.2}
\begin{tabular}{llccc}
  
  \textbf{Parameter} && \textbf{\borus}  & \textbf{\mytorus\ dec} & \textbf{\uxclumpy}\\  
  
 \hline\hline
 
     stat/d.o.f$^a$         &         & 500.44/542 &  499.72/543 &  509.73/541  \\ 
    
    red stat$^b$       &         &   0.92    &   0.92      & 0.94\\

   T$^c$       &  &   $1.78 $   &    $1.85 $   &  $1.34 $ \\

      p-value$^d$  && 0.60& 0.55& 0.37 \\
    
    \hline
\hline

  $kT^e$     &    &  --       &  --   & --\\

  apec norm$(\times 10^{-4})^f$      &  & --       &  --   & --\\

    \hline

    $\Gamma^g$    &     & $1.55_{-0.07}^{+0.09}$       &  $1.73_{-0.13}^{+0.12}$   & $1.77_{-0.06}^{+0.05}$\\  
 
    N$_{\rm H,av}^h\times$10$^{24}$\,cm$^{-2}$  && $0.09_{-0.05}^{+0.19}$       &  $0.35_{-0.16}^{+0.16}$   & $\dots$\\

    C$_F^i$   &       & $0.79_{-0.45}^{+0.14}$      & $\dots$       & $\dots$  \\

    cos$(\theta_i)^j$      &    & $0.95_{-u}^{+u}$    & $\dots$  & $\dots$        \\
    
    $\theta_i^k$                &    & $\dots$  & $\dots$     & $<11.01$       \\
    
    CTKcover$^l$ &  & $\dots$ &  $\dots$ &  $0.36_{-0.03}^{+0.02}$\\

    TOR$\sigma^m$&  & $\dots$ & $\dots$ & $<2.36$\\

    $A_{S,90}^n$         &          &  $\dots$      & $0.08_{-0.06}^{+0.41}$   & $\dots$   \\

    $A_{S,0}^o$       &  &  $\dots$   & $11.63_{-5.86}^{+10.57}$    & $\dots$  \\

     $F_s (\times 10^{-3})^p$       & & $6.47_{-2.79}^{+2.79}$       &  $6.45_{-4.03}^{+4.07}$   & $6.72_{-1.47}^{+7.59}$\\

     Norm $(10^{-3})^q$     && $1.85_{-0.33}^{+0.48}$       &  $1.61_{-0.29}^{+0.37}$   & $6.05_{-1.13}^{+1.86}$\\
    
    \hline
     
    &\xmm\             & $0.68_{-0.07}^{+0.08}$       &  $0.67_{-0.06}^{+0.07}$   & $0.76_{-0.11}^{+0.23}$\\
     
   $C_{\rm AGN}^r$ & \chandra         & $0.99_{-0.12}^{+0.14}$       &  $1.00_{-0.11}^{+0.12}$   & $1.13_{-0.18}^{+0.34}$\\  
   
     & \nustar\ & 1       & 1   & 1\\

     \hline
         &  \xmm\        & $15.53_{-1.73}^{+1.90}$       &  $14.69_{-2.09}^{+4.27}$   & $14.02_{-0.72}^{+1.77}$\\

    N$_{\rm H,los}\times$10$^{22}$$^s$ & \chandra\ &$14.29_{-1.74}^{+1.93}$       &  $13.35_{-2.03}^{+3.94}$   & $13.65_{-0.98}^{+1.59}$\\ 
 
   & \nustar\          & $11.51_{-2.10}^{+2.36}$       &  $9.03_{-2.49}^{+3.44}$   & $11.68_{-1.41}^{+0.94}$\\
 
   \hline
       \hline
    
   log L$_{NuSTAR, 2-10\,keV}^t$ & & $43.68_{-0.05}^{+0.05}$       &  $43.47_{-0.08}^{+0.08}$   & $44.08$\\
   log L$_{NuSTAR, 10-40\,keV}^u$ & & $43.56_{-0.04}^{+0.04}$       &  $43.04_{-0.23}^{+0.03}$   & $44.11$\\

   \hline
    \hline
    
   Red stat - No Var.$^v$ &&1.50 & 1.49 & 1.57 \\
   T       &  &   $11.72 $   &    $11.67 $   &  $13.54 $ \\

\hline

   Red stat - No C$_{AGN}^w$Var. && 0.98 & 0.96 &  0.95\\
     T       &  &   $0.47 $   &    $0.87 $   &  $1.07 $ \\
   \hline

   Red stat - No N$_{\rm H,los}^x$ Var. && 0.93 & 0.93 &  0.96\\
     T       &  &   $1.57 $   &    $1.54 $   &  $0.91 $ \\
   \hline

\hline
  \hline
\end{tabular}

\begin{tablenotes}
      \footnotesize
     \textbf{Notes:} a) Statistic (or $\chi^2$) over degree of freedom. We refer to `Stat' as a combination of C-stat and $\chi^2$ statistics. b) Reduced statistic (or $\chi^2$). c) Statistical Tension between the data and the model. d) p-value where the null hypothesis is that no \nh variability is found among different observations of the source. e) \texttt{apec} model temperature in units of keV. f) \texttt{apec} model normalization. g) Powerlaw photon index. h) Torus' average hydrogen column density in units of $10^{24}$\,cm$^{-2}$. i) Covering factor of the torus, as computed using \borus. j) Cosine of the inclination angle, as computed by \borus. cos$(\theta_i)$=0 represents an edge-on scenario. k) Inclination angle, as computed using \uxclumpy. $(\theta_i)$=90 represents an edge-on scenario. l) Covering factor of the inner ring of clouds, as computed using \uxclumpy. m) Cloud dispersion factor, as computed by \uxclumpy. n) Reflection component constant associated with an edge-on scenario, as computed by \mytorus. o) Reflection component constant associated with a face-on scenario, as computed by \mytorus. p) Scattering fraction. q) Normalization of the intrinsic AGN powerlaw. r) Cross normalization constant between observations with respect to the first \nustar\ observation, for which $C_{\rm AGN}$ is fixed to 1. s) Line of sight hydrogen column density of the torus in units of $10^{22}$\,cm$^{-2}$. t) Logarithm of the intrinsic luminosity of the first \nustar\ observation in the 2-10\,keV regime.  u) Logarithm of the intrinsic luminosity of the first \nustar\ observation in the 10-40\,keV regime. v) Reduced statistic (or $\chi^2$) and Tension when no variability is considered. w) Reduced statistic (or $\chi^2$) and Tension when no flux variability is considered. x) Reduced statistic (or $\chi^2$) and Tension when no \nh variability is considered. 
    \end{tablenotes}
    \end{threeparttable}
    
\end{table*}

The best-fit values for each model for 2MASXJ 06411806 are reported in Table \ref{tab:2MASXJ06411806+3249313_fitting} and are tabulated as follows: the first section reports the best-fit statistics (stat/degree of freedom and reduced statistics), as well as the p-value and tension T between the data and the best-fit, derived as described in Section \ref{section:var_estimates}. The second block reports the best-fit soft-emission properties. In the third section, we report the best-fit torus parameters, such as the covering factor, the inclination angle and the cloud dispersion. The fourth and fifth panels report the variability measurements in intrinsic flux ($C_{AGN}$) and \nh\ measured in the fitting process. The sixth section reports the logarithm of the intrinsic luminosity for the first \nustar\ observation in two energy ranges, 2-10\,keV and 10-40\,keV. The luminosity at any other epoch can be obtained by multiplying this value to the cross-normalization constant. The last three panels report the statistical analysis (reduced statistics and tension) derived by considering: 1) no variability (either intrinsic flux or \nh) at any epochs; 2) no intrinsic flux variability between the observations (i.e., the fit is performed by allowing only \nh\ variability); 3) no line-of-sight hydrogen column density variability is allowed between the observations (i.e., only intrinsic flux variability is permitted). By comparing the best-fit tension to the non-variable one, we can affirm all the sources require variability, either in intrinsic flux or \nh.

\subsection{Equatorial hydrogen column density - UXCLUMPY}\label{sec:nheq_uxclumpy}

This section presents the computation of the equatorial column densities for \uxclumpy{}. The \uxclumpy\ X-ray spectral model library consists of two parameters that define the clumpy torus geometry: TOR$\sigma$ and CTKcover. A grid of possible geometries was created by \cite{Buchner2019}, by generating a population of spherical blobs. \cite{Buchner2019} presented the angle-averaged column density distribution. This is achieved by sampling many directions from the center (the location of the X-ray corona) and computing the column density. Based on the same information, \cite{Boorman2023HEXPobscurer} presented the covering factor for various column densities.
Here, we analyze the column density distribution in the equatorial plane, averaged over azimuthal angles. This is achieved with the same sampling approach, but only within the equatorial plane.

The results are presented in Table \ref{tab:uxclumpy_nheq}.
When the covering factor of the inner ring of Compton-thick clouds (CTKcover) is above 0.2, the equatorial column densities are Compton-thick. At lower values, e.g., CTKcover=0, the equatorial plane is dominated by the cloud population. At TOR$\sigma$=0, there is no obscurer at all; therefore, the column density is equal to $10^{20}\rm{\,cm}^{-2}$, which is the typical obscuration given by the host galaxy. At TOR$\sigma>0$, the column density increases with decreasing TOR$\sigma$ because more clouds lie along the equatorial sightlines. In contrast, in the uniform density models \borus\ and \mytorus, the \nhav\ is not related to the opening angle in the same way, and rather sets the column density of the entire Compton reflecting material directly. This may explain the differences in the inferred \nheq\ and \nhav.
The code to compute \nheq\ starting from best-fit values of TOR$\sigma$ and CTKcover can be found at: \url{https://github.com/andrealunapizzetti/NH_eq-UXCLUMPY/}.

\begin{table}[h!]
\centering
\begin{tabular}{l c c c}

\textbf{Source}             & \textbf{TOR$\sigma$} & \textbf{CTKcover} & \textbf{\nheq} \\ 
&[$^\circ$]&&$[\times 10^{24}$\,cm$^{-2}]$\\ \hline\hline
MRK 348              & 27.99   & 0       & 3.46    \\ 
NGC 4992             & 28.18   & 0.30    &  41.67  \\ 
ESO 383-18          & 7.90    & 0       &  11.65     \\ 
 MRK 417              & 9.39    & 0       &   10.65   \\ 
MCG-01-05-047       & 27.23   & 0.21    &   40.36     \\ 
ESO 103-35          & 81.83   & 0.34    &  39.95 \\ 
NGC 1142            & 25.71   & 0.26    &  42.07 \\ 
IRAS 16288-3929     & 70.00   & 0.57    &  37.58   \\ 
ESO 263-13          & 7.03    & 0.25    &  52.46  \\ 
Fairall 272         & 21.27   & 0.16    &  28.27  \\
LEDA 2816387        & 7.03    & 0.27    &  52.94 \\ 
2MASXJ 06411806     & 2.36    & 0.36    &  47.15   \\ 
\hline
NGC 612   & 0.36  & 0.37   & 43.75    \\ 
NGC 788   & 12.6  & 0      &   8.77 \\ 
NGC 833   & 3.8   & 0      &  0.06 \\ 
NGC 835   & 6.8   & 0      & 8.80 \\ 
3C 105    & 15.9  & 0      & 7.19\\ 
4C+29    & 17.5  & 0     &  6.53    \\ 
NGC 3281  & 28.0    & 0     &  3.46   \\ 
NGC 4388  & 66.7  & 0      &   1.01  \\ 
IC 4518A  & 84.0    & 0.29    &  38.54  \\ 
3C 445    & 84.0    & 0      &   0.58  \\ 
NGC 7319  & 77.9  & 0     &  0.71    \\
3C 452    & 84.0    & 0      &   0.58  \\ \hline
NGC 7479  & 24.6  & 0.6      &  43.43  \\ \hline
NGC 6300  & 24.1 & 0.6       & 43.69 \\ \hline\hline
\end{tabular}
\caption{\uxclumpy\ equatorial hydrogen column density (\nheq) values for all the sources analyzed in this study, in \citetalias{nuria2023}, in \citet{pizzetti2022} and Sengupta et al., in prep.}
\label{tab:uxclumpy_nheq}
\end{table}

\section{Discussion}\label{sec:discussion}
In this section, we discuss the results obtained from the variability analysis performed on the full sample of 27 galaxies, 13 of which are analyzed in this paper, 12 in \citetalias{nuria2023}, one in \citet{pizzetti2022} and one in Sengupta et al., in prep. All the sources were selected from a pool of potentially \nh\ variable sources identified by \citetalias{zhao_2021}; our sample is thus intentionally biased toward variable sources (either in \nh\ or intrinsic flux).

Applying the method proposed by \citetalias{nuria2023} and explained in Section \ref{section:var_estimates}, we are able to classify the sources as shown in Table \ref{tab:var_results} and Table 3 of \citetalias{nuria2023}. Although the sample is biased toward variable sources, as mentioned above, only $37\%$ of the sources (10 out of 27) require \nh\ variability between the observations, while for $44\%$ of them (12 out of 27), we can confidently say no \nh\ variability is needed to reach a proper fit in the timeframe analyzed. For five sources, either intrinsic flux or \nh\ variability is required, but the data is insufficient to confidently distinguish between the two scenarios; these sources are thus classified as `Undetermined'. As proposed by \citetalias{nuria2023}, joint soft-hard X-ray observations could disentangle the two scenarios.

\subsection{Line-of-sight hydrogen column density variability over time}\label{subsec:nh_var}
\begin{figure}[ht]
    \centering
    \includegraphics[width=0.48\textwidth]{ 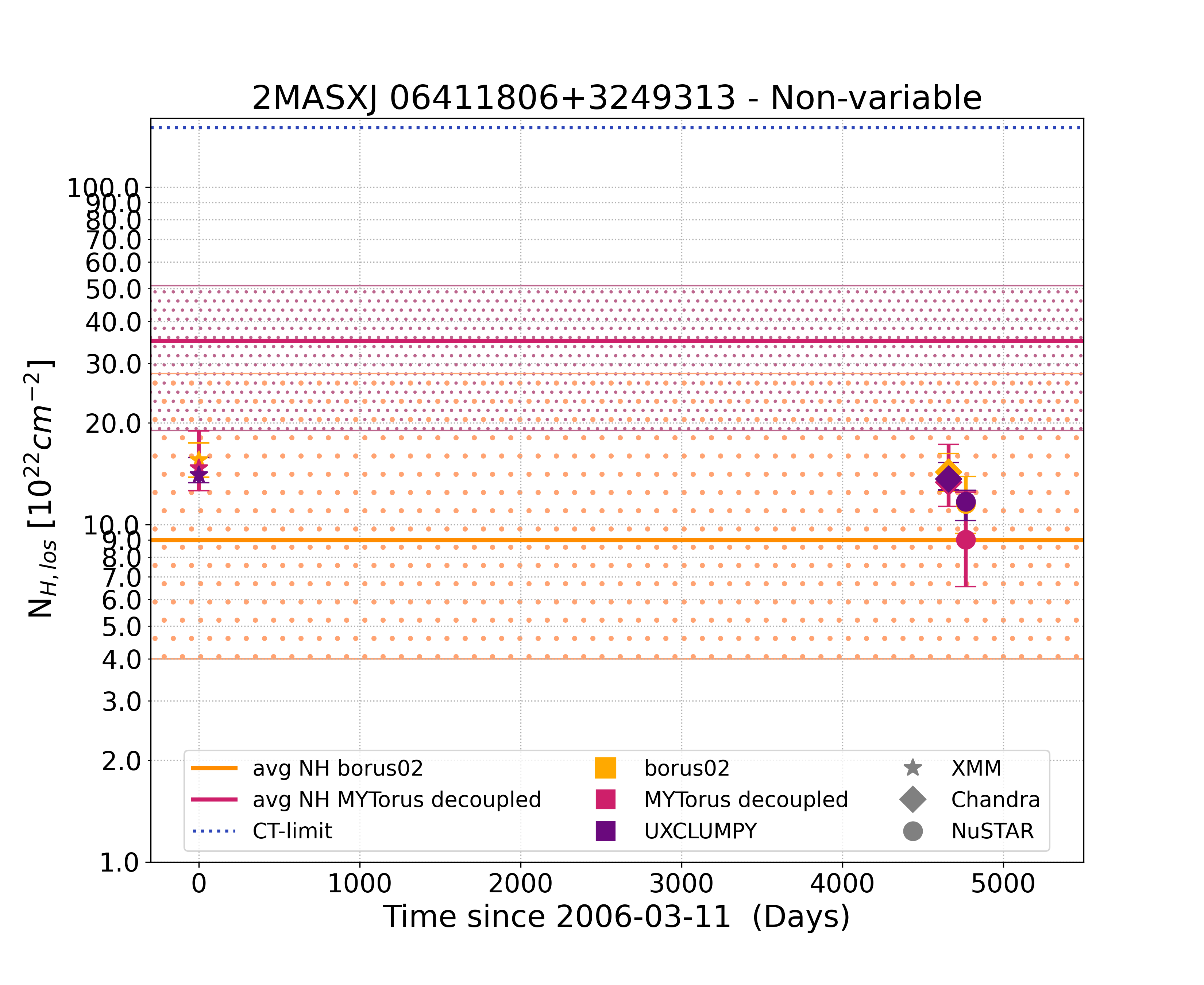}
    \caption{\nh\ ($\times 10^{22}\mathrm{\,cm}^{-2}$) of 2MASX J06411806+3249313  as function of time for \borus, \mytorus\ and \uxclumpy. Solid lines and shaded areas represent \nhav\ and associated uncertainties for \mytorus\ and \borus. The blue dotted line corresponds to the Compton-thick limit ($150\times 10^{22}\mathrm{\,cm}^{-2}$).}
    \label{fig:nhvar_plots3}
\end{figure}

Figure \ref{fig:nhvar_plots3} shows the l.o.s. hydrogen column density as a function of time for 2MASXJ 06411806, classified as `Non-variable in \nh'. 2MASXJ 06411806 is the only galaxy of the sample for which we observe the \nh\ values to lie within the range of average hydrogen column density, resembling the scenario observed by  \citetalias{nuria2023} for NGC 612, NGC 833 and 3C 105. The other sources, whose \nh\ variability plots can be seen in Appendix \ref{sec:variability_plots} (Figures \ref{fig:nhvar_plots}, \ref{fig:nhvar_plots2}), present two main behaviors: for 5 out of 13, the clouds (\nh) are denser than the average density of the torus, while for 6 out of 13 we observe the torus to be denser than the clouds; a similar denser torus has also been observed in NGC 7479 \citep{pizzetti2022}. Similar to NGC 3281 in \citetalias{nuria2023},  MRK 417 exhibits characteristics aligning with both scenarios, depending on the considered model (see Section \ref{sec:mrk417} for more details). 

The first scenario can be interpreted by considering the medium responsible for X-ray obscuration (the obscurer) as consisting of overdense clouds dispersed within a less dense inter-cloud medium, acting as a \textit{thin reflector}. This scenario was previously suggested by \citetalias{nuria2023}, and it hypotheses that, while the denser clouds also reflect, their reflection component (or thick reflector) is missed by the modeling due to its spectral shape being more similar to that of the line of sight component \citepalias[see, Figure 3 in][for more details]{nuria2023}. This idea is consistent with the two-phase medium invoked by \citet{stakevski2011} and \citet{Siebenmorgen2015} to explain infrared observations. 

The second scenario, where the overall density of the torus exceeds the column density along the line of sight, can be explained by considering the reflector as an area of higher density positioned near the X-ray emitting region, while the obscurer represents an underdense cloud distribution. The absence of observed absorption by the reflector (otherwise resulting in solely higher line-of-sight column densities) can be explained if we consider it to be geometrically thin \citep[i.e., a warped disk - see][and Subsection \ref{subsec:inner_ring} for more details]{lawrence2010}, capable of intercepting a sufficient amount of incoming X-ray radiation to produce pronounced reflection features, yet permitting a portion of the radiation to be absorbed by the clouds \citep[see, e.g., NGC 7479 in ][]{pizzetti2022}. Both cases favor a scenario where the media mainly responsible for absorption and reflection may be dissociated within the torus, as proposed in \citet{pizzetti2022} and \citet{nuria2023}. Interestingly, we do not find any correlation between the two scenarios and the variability class, as well as between \nhav\ and the intrinsic luminosity (Pearson coefficient$=-0.21$) and between \nhav\ and the Eddington ratio (Pearson coefficient$=-0.18$).

\subsection {Torus properties as a function of variability class}
Figure \ref{fig:histograms} shows bar plots of the averaged best-fit properties for all sources, grouped by variability class.
Panel a) displays the time-averaged value of the line-of-sight column density ($\langle$\nh $\rangle$), while b) shows the average hydrogen column density of the torus; the difference between the two can be interpreted as an indicator of the torus clumpiness/inhomogeneity (i.e., the smaller the value, the more homogeneous the torus - panel c). Torus geometrical properties, such as the torus inclination angle, the covering factor and the cloud dispersion, are shown in panels d), e), and f). Intrinsic AGN properties, such as the 2-10\,keV luminosity and the Eddington ratio are shown in panels g) and h). To obtain the Eddington ratio $\lambda_{\rm Edd}=L_{\rm Bol}/L_{\rm Edd}$, we first computed the bolometric luminosity for each source, by applying the bolometric correction reported in \citet{vasudevan2010} to L$_{2-10\rm{\,keV}}$. We then divided this value with the Eddington luminosity $L_{\rm Edd}=\frac{4\pi G M_{\rm BH}m_p c}{\sigma_T}$, where $m_p$ is the mass of the proton, $M_{\rm BH}$ is the mass of the black hole as reported in \citet{koss2022}, and $\sigma_T$ is the Thompson cross-section. As seen in the bar plot, there does not seem to be any substantial difference between the three classes of variability. 

Upon initial inspection of the plots, it becomes evident that, within errors, there appears to be no discernible distinction among the three populations. 
Applying the Anderson-Darling test\footnote{The Anderson-Darling-test probes whether two samples originate from the same parent population.} \citep{Stephens1974_andersondarlingtest} to the three datasets for each best-fit property, the three samples are compatible with originating from the same parent population, requiring no intrinsic difference between the sources classified as variable, non-variable or undetermined. This is in contrast to the tentative trends suggested by \citetalias{nuria2023} when analyzing the first 12 sources, which suggested there may be a trend for variable sources to have higher average obscuration and broader cloud distributions. 

This lack of difference between the variability classes may point to two different scenarios. One scenario could involve distinguishing between sources with variable hydrogen column density (\nh\ variable) and those without (non-variable in \nh), yet their average geometric characteristics show no significant distinctions. This leads us to inquire about the underlying causes of the observed variability or its absence. Another possibility would be that the absence of detected variability in \nh\ does not necessarily imply that the \nh\ of the source has remained constant over time. Instead, it indicates that within the timeframe and the spectral quality examined in this study, variations in \nh\ were not observed. However, it remains uncertain whether it is likely to detect sources exhibiting no variability across multiple observations spanning several years (see, e.g., NGC 7479 in \citealp{pizzetti2022} or 3C 445 in \citetalias{nuria2023}), while others demonstrate variability between each observation, occurring at various possible timescales \citepalias[see, e.g., NGC 4388 in][]{nuria2023}.
Future targeted monitoring campaigns and a larger sample will help discern between the two proposed scenarios.

\begin{figure*}
  \centering
\includegraphics[width=0.47\textwidth]{ 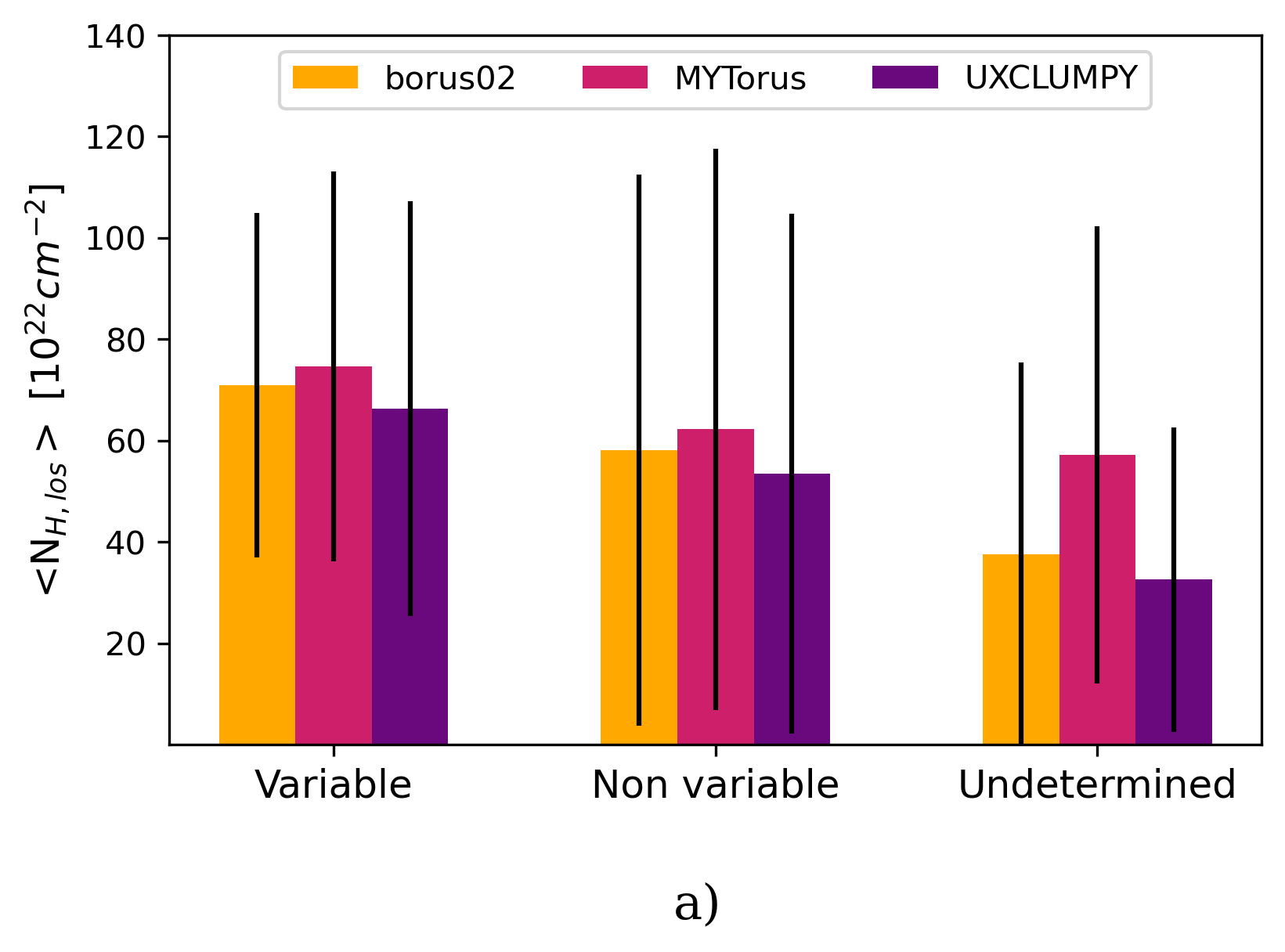}
\includegraphics[width=0.47\textwidth]{ 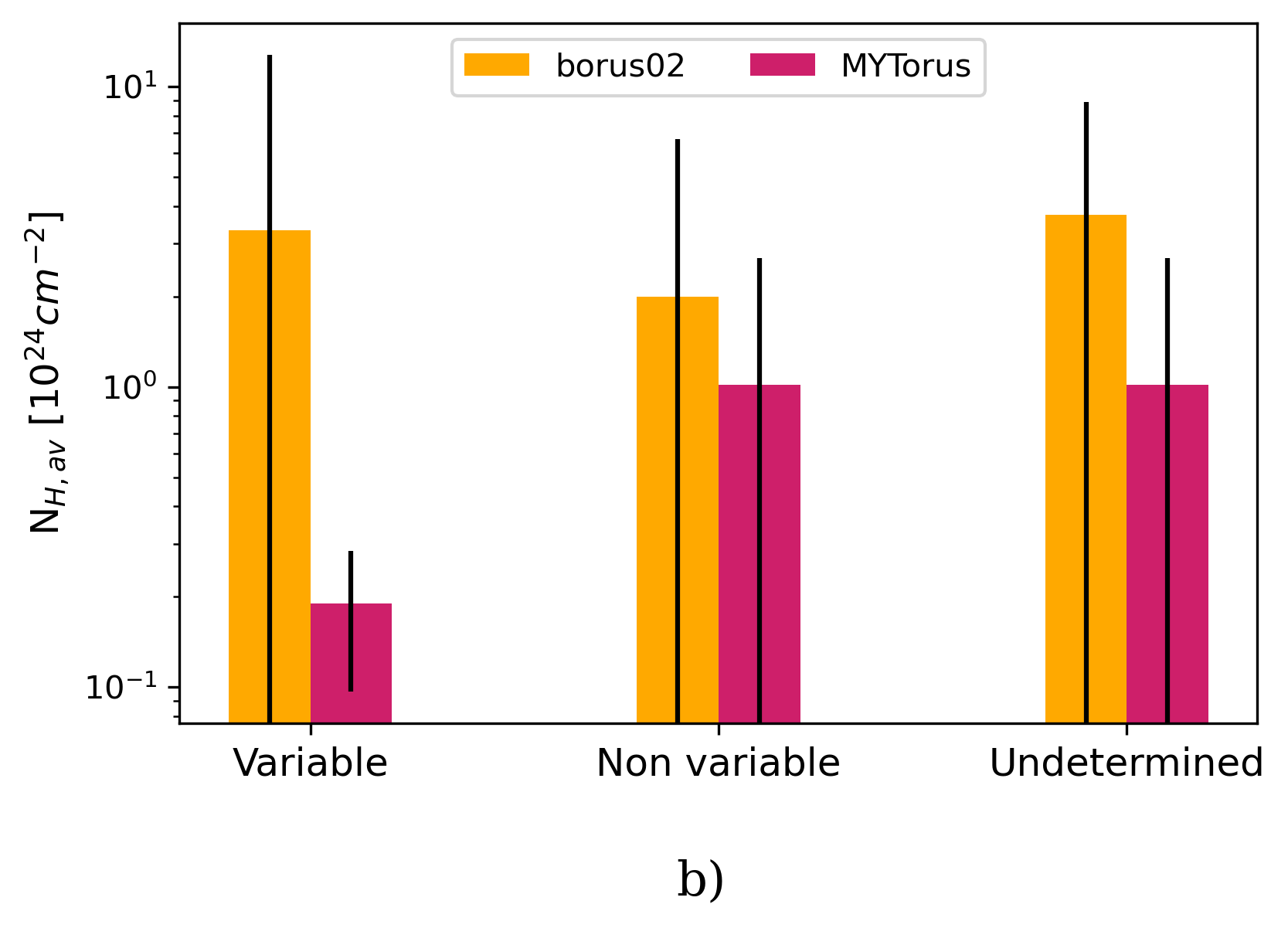}
\includegraphics[width=0.47\textwidth]{ 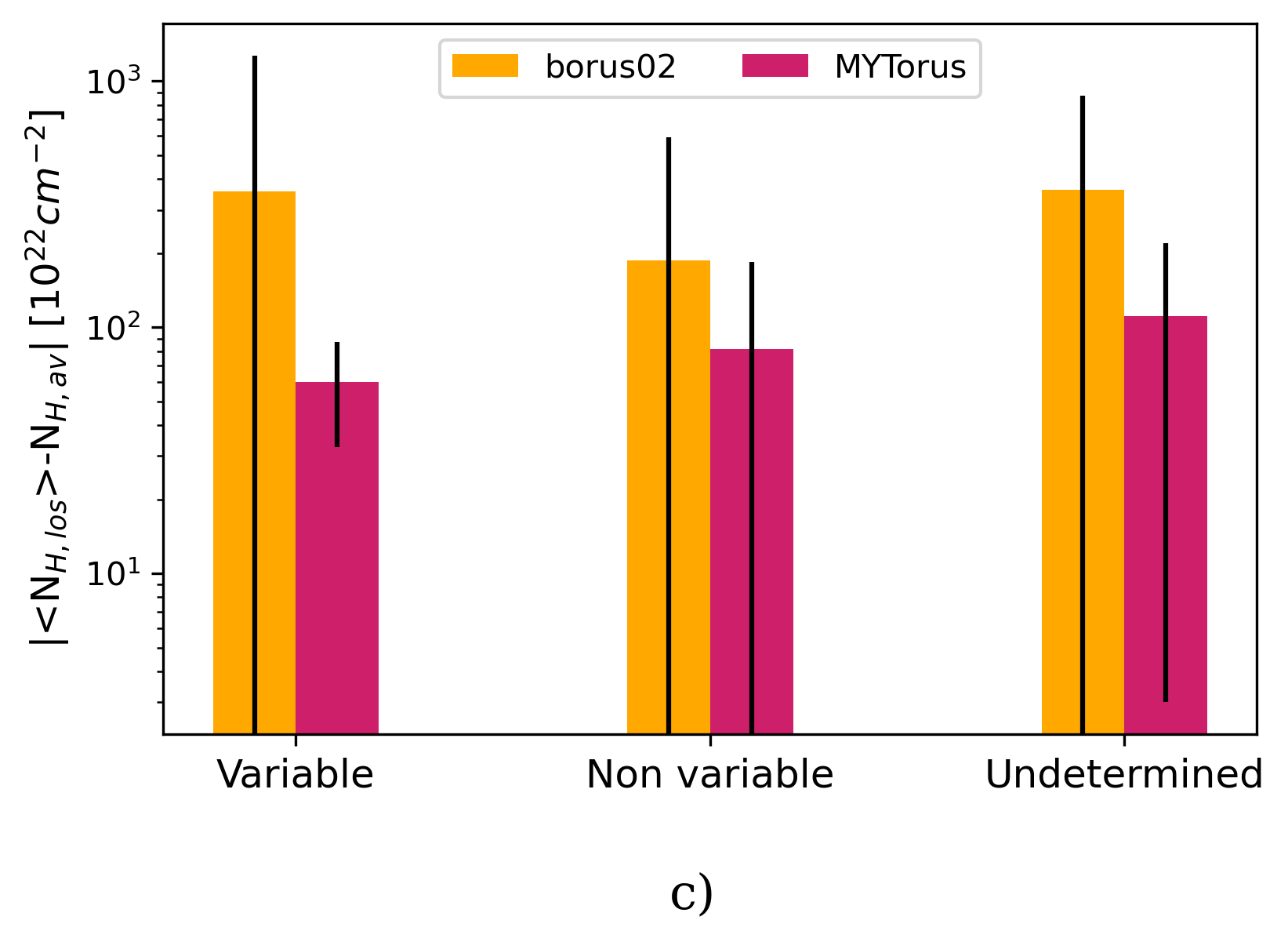}
\includegraphics[width=0.47\textwidth]{ 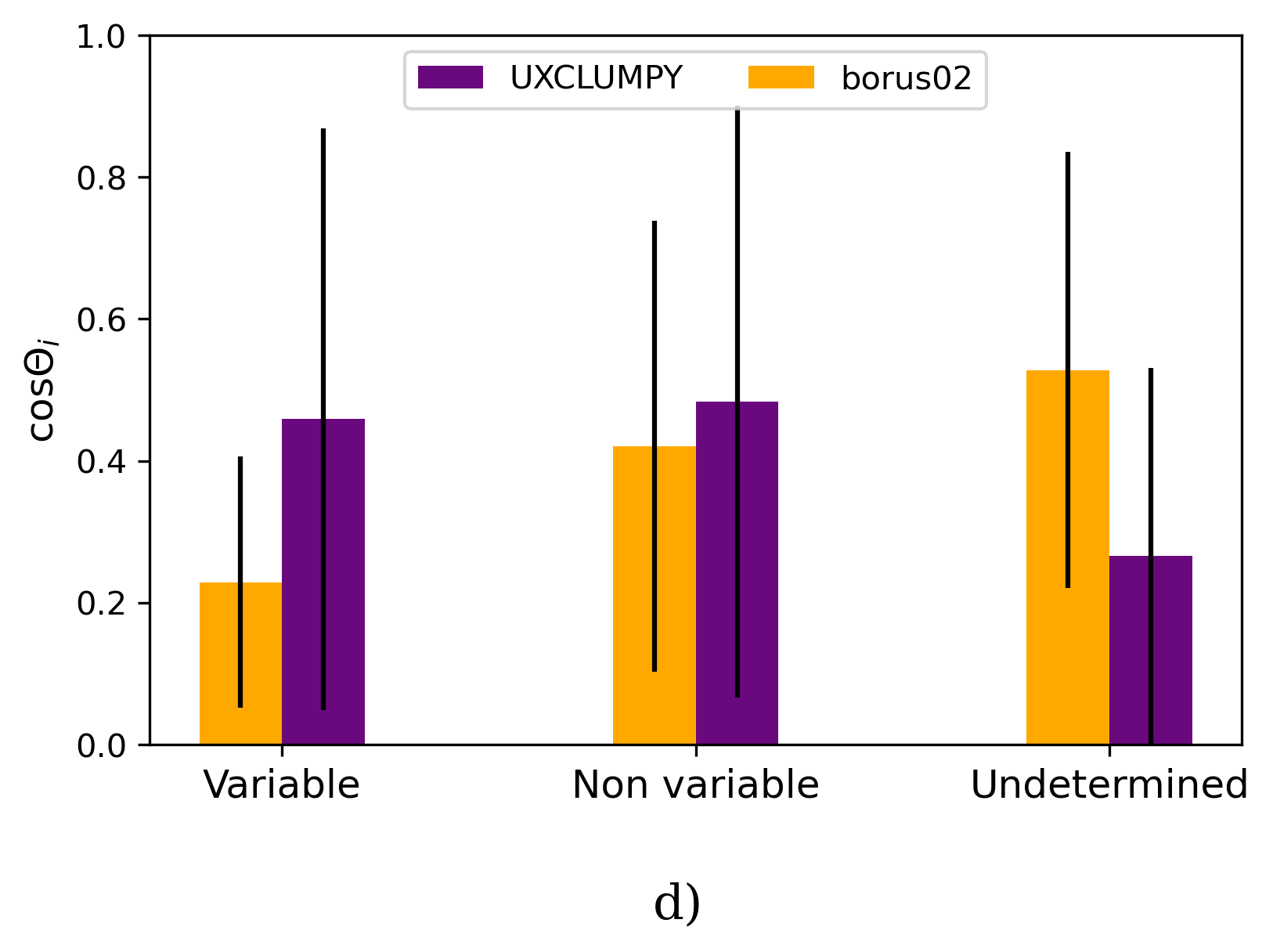}
\includegraphics[width=0.47\textwidth]{ 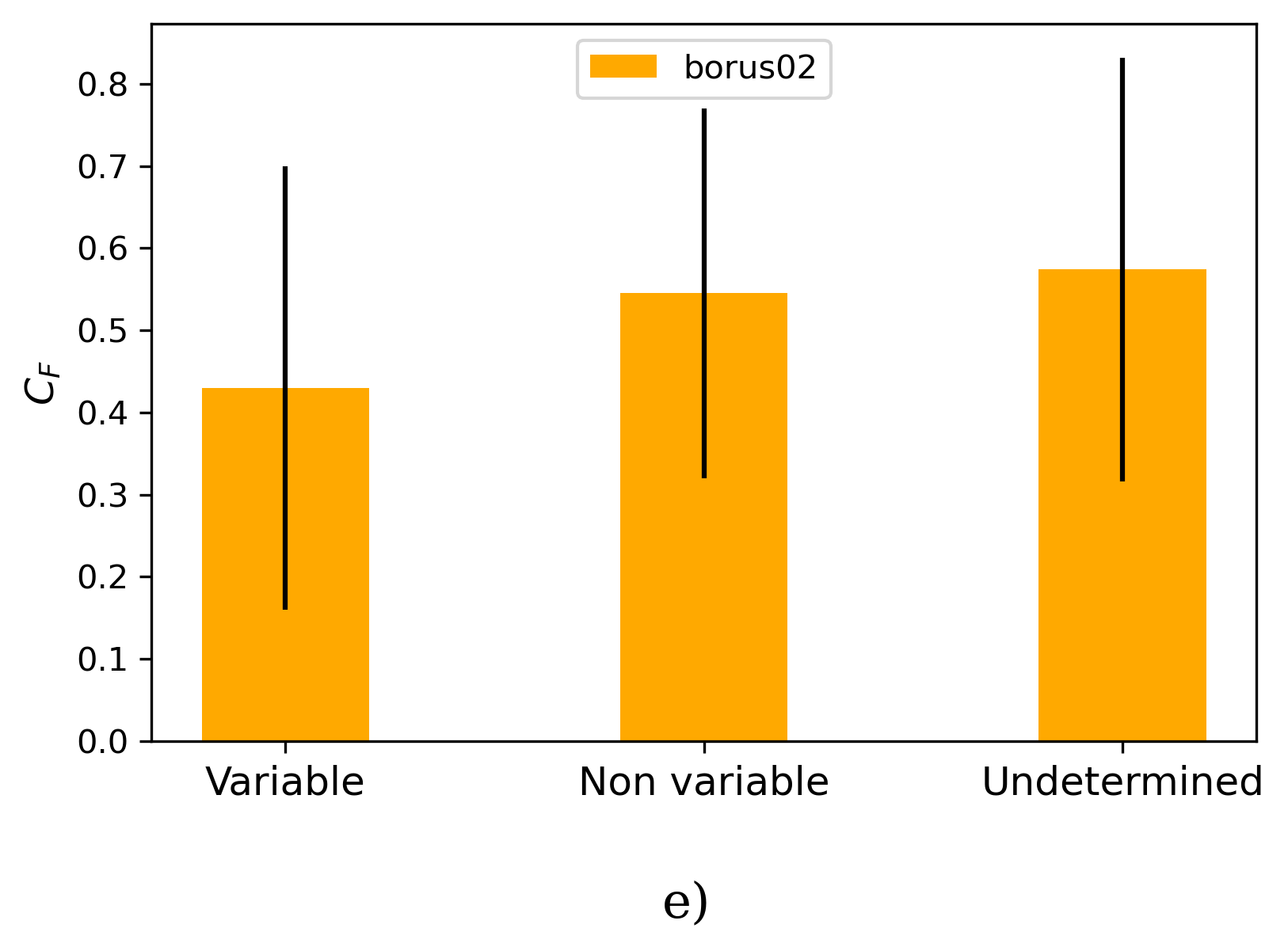}
\includegraphics[width=0.47\textwidth]{ 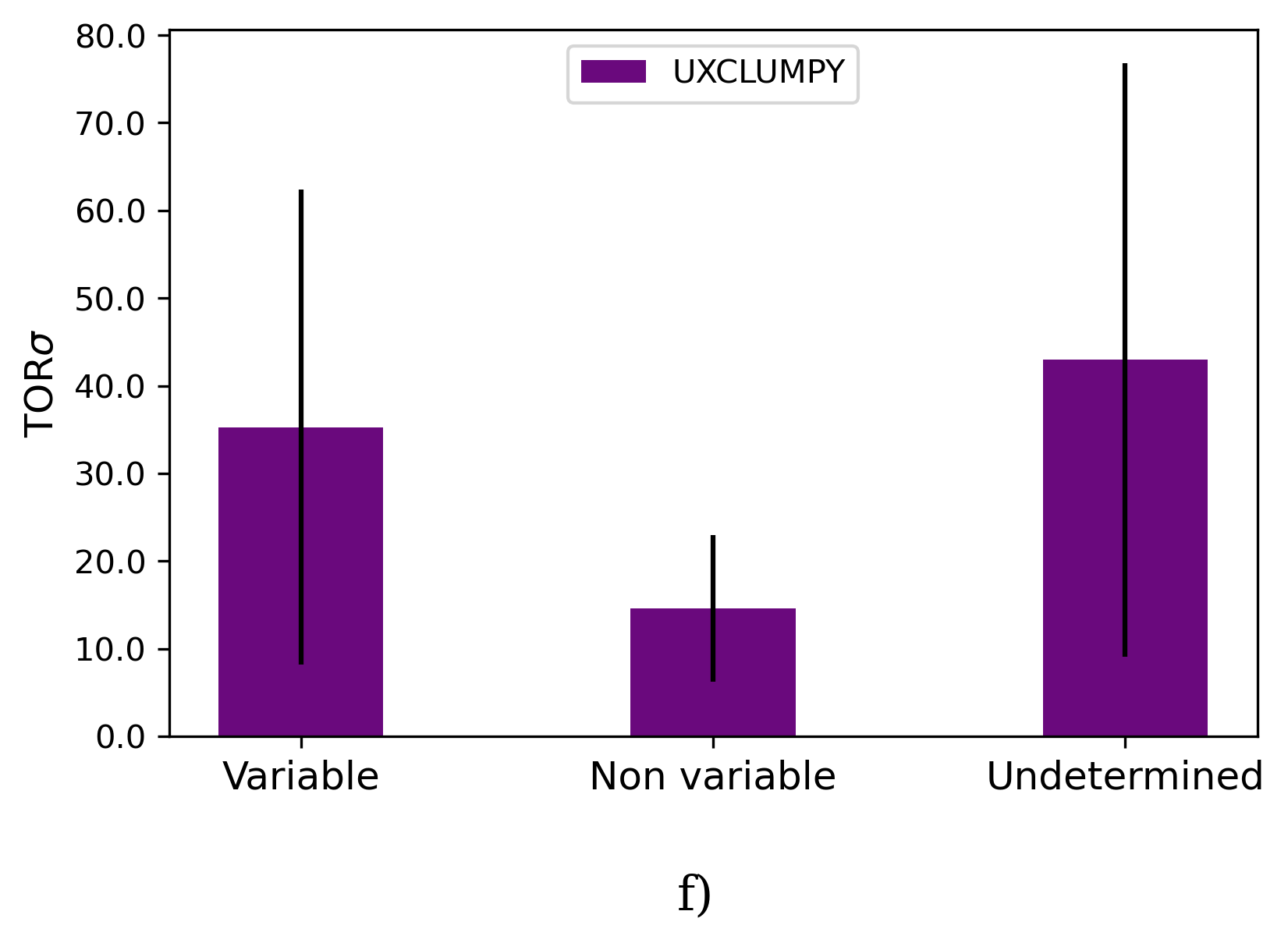}
\end{figure*}

\begin{figure*}[!ht]
  \centering
\includegraphics[width=0.47\textwidth]{ 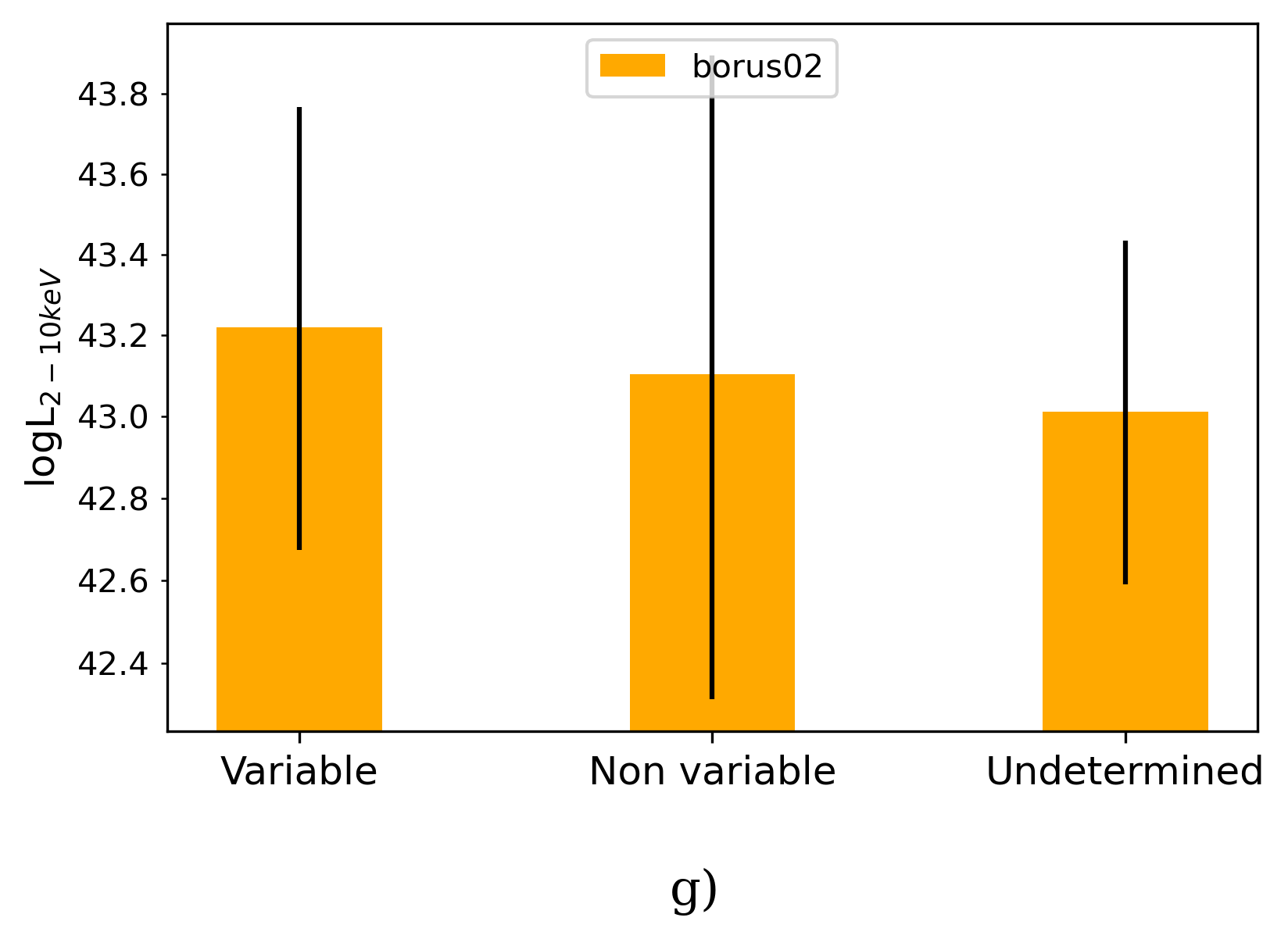}
\includegraphics[width=0.47\textwidth]{ 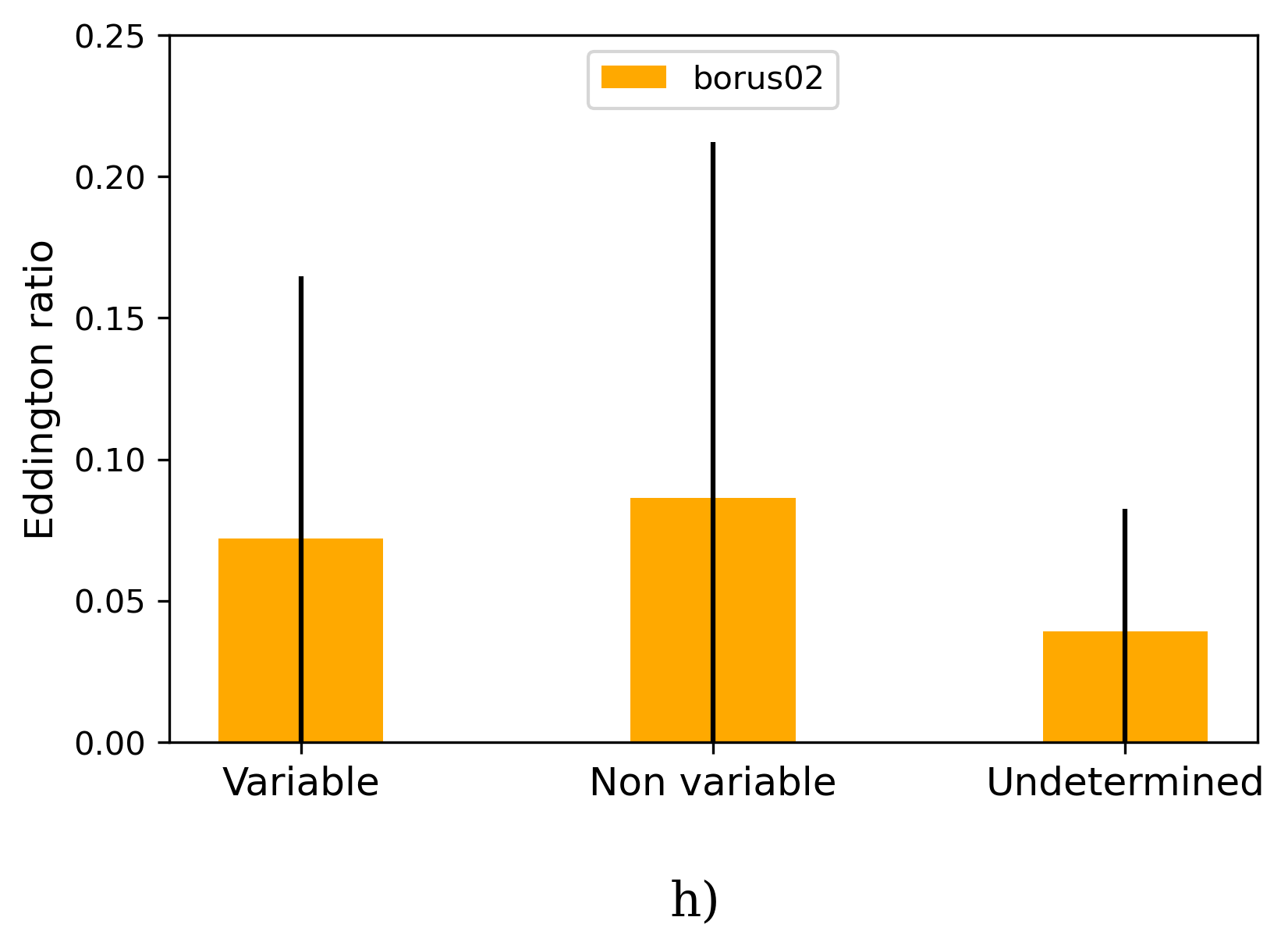}

\caption{Bar plots displaying the averaged best-fit values for all sources, binned by variability class. The average of each parameter is computed across the sources within each variability class, with the error bar representing the standard deviation of each distribution. The models are color-coded as follows: yellow for \borus, magenta for \mytorus, and violet for \uxclumpy. \\ \textbf{a)} Time average of the \nh\ for each single source. \textbf{b)} Average hydrogen column density (to be considered as the column density of the reflector). \textbf{c)} Absolute value of the difference between the average \nh and \nhav. \textbf{d)} Cosine of the torus inclination angle as measured by \borus\ and \uxclumpy. \textbf{e)} Covering factor as computed by \borus. \textbf{f)} Cloud dispersion (TOR$\sigma$) as computed by \uxclumpy. \textbf{g)} Logarithm of the intrinsic luminosity of the first \nustar\ observation in the 2-10\,keV regime. We report only \borus\ measurements for clarity. \textbf{h)} Eddington ratio calculated by applying the bolometric correction reported in \citet{vasudevan2010} to the L$_{2-10\rm{\,keV}}$ measured by \borus, and using the black hole masses as reported in \citet{koss2022}. }
\label{fig:histograms}
\end{figure*}

\subsection{Inner reflector and water megamaser disks}\label{subsec:inner_ring}
As proposed in \citet{Buchner2019}, and shown in \citet{pizzetti2022, nuria2023}, a sole clumpy model is an inadequate description of the broad X-ray spectra of certain nearby obscured AGN. Indeed, an additional component consisting of dense ($\rm{N_{H,los}}>10^{25}$\,cm$^{-2}$) reflecting material is required to reproduce the Compton hump (peaking around $\sim 20-50$\,keV) observed in some local obscured AGN.  
This dense \textit{inner ring} is predicted to be on the inner side of the torus-accretion disk transition, although neither observational nor spectral data have yet confirmed its presence, position or size (see Pizzetti et al., in prep. for insights into an ALMA campaign aimed at locating the putative inner ring in NGC 7479). Among the sources discussed in this work, 13 out of 27 ($\sim 48\%$) require the inclusion of the aforementioned \textit{inner ring} in the fitting process. Notably, only one (NGC 7479) is Compton-thick, with the remaining 25 sources being Compton-thin. This scenario, featuring a Compton-thick reflector and Compton-thin absorber, could be explained by considering the distinction between the torus medium responsible for the reflection (inner ring) and the absorption (clouds) and assuming a \textit{warped} geometry of the inner ring, as proposed by \citet{Buchner2019,buchner2021}. Warped disk geometries naturally arise from non-planar accretion \citep[for a review see, ][]{petterson1997}, and have been proposed to be the source of Compton-thick obscuration and reflection \citep[i.e.,][]{lawrence2010, buchner2021}. Notoriously, the 22 GHz water megamaser\footnote {Extragalactic water masers are generally called megamasers because of their high luminosity with respect to Galactic masers \citep{Masini2016}.} disks are known to be warped \citep[see, e.g., ][]{greenhill2003,greenhill2003b, Herrnstein2005,kuo2011}. The 22 GHz maser line emitted by water molecules originates in high-density regions in the innermost part of the nuclear structure. \citet{Masini2016} localized the maser disk to be outside the accretion disk, i.e., within the torus. This is supported by considering that the presence of water molecules requires a dense \citep[n$_{\rm H_2}\gtrsim 10^7$\,cm{$^{-3}$},][]{bennert2004} dusty environment, positioning the maser disk outside the dust sublimation radius. 
Such findings imply a potential shared nature, or at least common geometrical and spatial properties, between the thick inner reflector and the water megamaser disks, as proposed by \citet{buchner2021}, particularly given that the majority of water megamasers are detected in Seyfert 2 and Compton-thick AGN \citep{Greenhill2008}. 

Among the galaxies requiring the inner ring in their fit, approximately $30\%$ (4\footnote{We do not include MRK 348 as it hosts a `jet' megamaser, rather than a disk maser \citep{peck2003}. } out of 13) exhibit recorded water megamaser emission (ESO 103-35 \cite{braatz1996}, NGC 6300 \cite{greenhill2003}, NGC 7479 \cite{braatz2008}, and IRAS 16288+3929 \cite{greenhill2009}; Megamaser Cosmology Project\footnote{\url{https://safe.nrao.edu/wiki/bin/view/Main/PublicWaterMaserList}.}). On the other hand, NGC 4388 hosts a 22 GHz maser disk \citep{braatz2004} but does not require the inner reflector to fit the spectra\footnote{We note the fit does not firmly rule out the presence of the component.}.
The partial discrepancy in the correlation between the presence of the inner reflector and the water megamaser disk could be attributed to several factors: firstly, the need for an almost perfectly edge-on geometry to produce maser amplification, which may not always align with the AGN studied here, where characterizing the torus inclination angle proves highly degenerate. Secondly, as shown in \citet{panessa2013}, there exists an observational bias in terms of distance, with the detection fraction of water megamasers being approximately $26\%$ in a sample of Seyfert galaxies located within 20 Mpc, implying that some of the more distant AGN requiring an inner reflector may host a megamaser disk too faint to be detected. Moreover, the inner reflector may be located in a dust-free region of the torus, precluding the presence of masing water molecules. Lastly, not all the sources analyzed in this work have been observed in maser campaigns; thus, a targeted observing campaign would benefit the understanding of the possible connection between the inner ring and the water megamaser disk.

\subsection{\nh\ variations as function of time}\label{subsec:deltanh_vs_time}
\begin{figure}[ht]

  \centering
\begin{minipage}{\linewidth}
    \centering
    
      {\includegraphics[width=\linewidth]{ 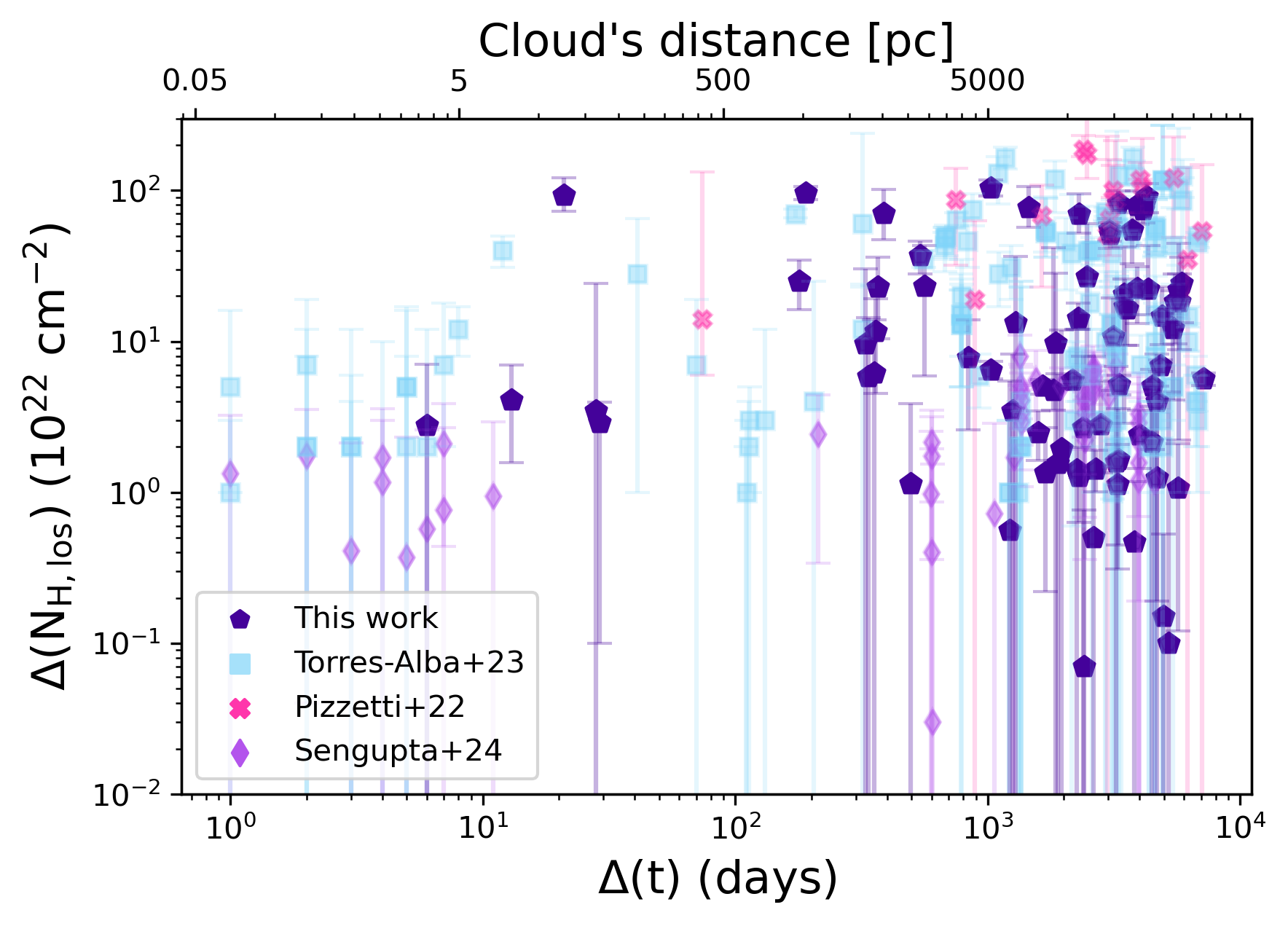}}

      {\includegraphics[width=\linewidth]{ 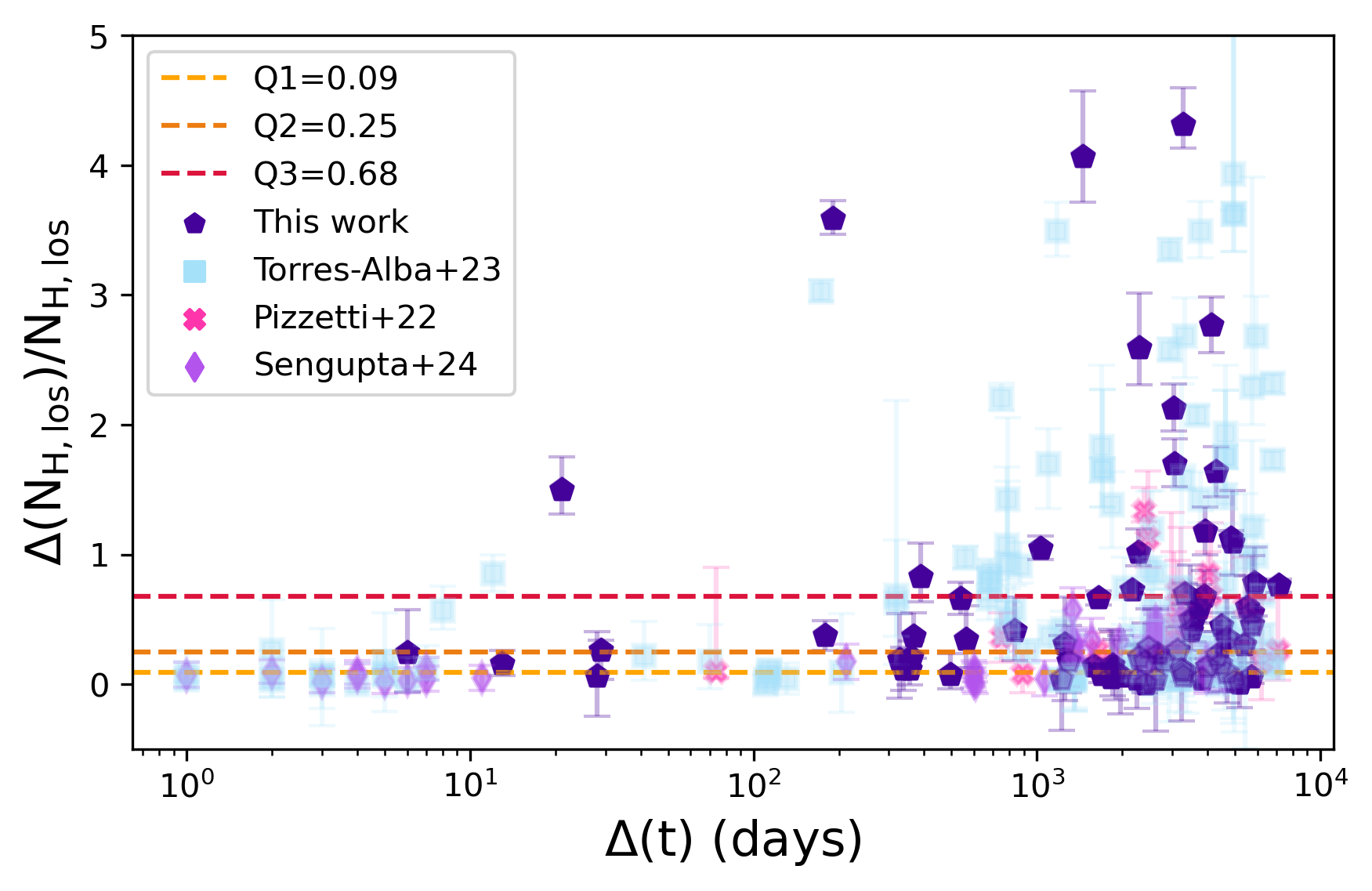}}
   
  \end{minipage}\quad

  \caption{\textit{Top:} Variation in \nh\ between all pairs of observations for each source relative to the time span between the two observations. The upper x-axis serves as a reference to indicate the position where the obscuring material would be located to induce an eclipse within that timeframe, following the approach proposed by \citet{risaliti_2005}. In this approach, we note that the cloud's distance is computed by assuming the observed $\Delta$\nh\ is generated by a single cloud eclipsing event, which is not necessarily the case.
  \textit{Bottom:} Fractional fluctuation of \nh\ between two observations of the same source, normalized to the lowest \nh\ measure in each pair. The $25\%$, $50\%$, $75\%$ quartiles of the distribution, (respectively, Q1, Q2 and Q3) are shown as dashed horizontal lined.}
  \label{fig:nhlos_vs_time}
\end{figure}

In the top panel of Figure \ref{fig:nhlos_vs_time}, we present the variations in \nh\ observed between any pair of available observations for each source, plotted against the time interval between the two observations. As all the three models are compatible with each other, we show the results obtained with \borus. For reference, we provide the position where the obscuring material would be located to induce an eclipse within that timeframe, following the approach proposed by \citet{risaliti_2005}.
As shown in the figure, minor variations in \nh\ are evident across all time intervals between observations. However, significant fluctuations in \nh\ ($\Delta\rm{N_{H,los}}> 50\times10^{22}$\,cm$^{-2}$) are more probable with larger time intervals ($>$100 days). This could be interpreted as a consequence of the heterogeneity and dimensions of the single clouds, as proposed by \citetalias{nuria2023}. However, it is important to note that this study is biased towards longer timescales. The observations analyzed in this work were not part of monitoring campaigns but were conducted randomly over the years (e.g., to monitor supernova explosions, ultra-luminous X-ray sources, and star-forming regions within the galaxies). Therefore, only dedicated campaigns designed to monitor \nh\ variability can accurately populate the plot \citep[see, e.g., the dedicated monitoring campaign for NGC 1358 in][]{marchesi2022}. 

The lower panel of  Figure \ref{fig:nhlos_vs_time} presents the \nh\ variation between two observations (of each source) normalized to the lowest \nh\ value in each pair. The $25\%$, $50\%$ and $75\%$ quartiles of the distribution are also provided, demonstrating that when randomly selecting two observations of the same source, the median variability between the two is $25\%$ of the lower \nh\ value. For a quarter of the observation pairs in the sample, the increase is above $70\%$. The results differ from those reported by \citetalias{nuria2023}, who obtained higher quantiles, suggesting that, on average, the combined sample exhibits smaller variations in \nh.

\begin{table*} 
\centering
\begin{threeparttable}
\caption{$N_{\rm H,los}$ variability results.}
\label{tab:var_results}
\renewcommand*{\arraystretch}{1.4}
\begin{tabular}{lcc|cc|cc|c}
\textbf{Source}  & \multicolumn{2}{c}{\borus} &\multicolumn{2}{c}{\mytorus}  & \multicolumn{2}{c}{\uxclumpy} & \textbf{Classification}\\
{} & $\chi^2_{\rm red}$& p-val. & $\chi^2_{\rm red}$& p-val. & $\chi^2_{\rm red}$& p-val. \vspace{0.1cm}\\ \hline\hline
NGC 454E & Y* & Y & Y& Y & -- & -- & Variable\\
MRK 348 & Y* & Y &  Y* & Y  & Y* & Y & Variable\\
NGC 4992 & N & N & N & N & N & N & Non-variable\\
ESO 383-18 & N & N & N & N & N & N & Non-variable\\
MRK 417 & N & N & N & N & N & N & Non-variable\\
MCG-01-05-047 & N & N & Y & Y  & N & N & Undetermined\\
ESO 103-35 & Y & N &N  & N & Y & N & Undetermined\\
NGC 1142 & Y & Y & Y & Y & Y & Y & Variable\\
IRAS 16288+3929 & Y & Y & Y & Y & Y & Y & Variable\\
ESO 263-13 & Y* & Y & Y* & Y & Y* & Y & Variable\\ 
Fairall 272 & Y & Y & Y & Y & N & N & Undetermined \\
LEDA 2816387 & N & N & N & N & N & N & Non-variable \\
2MASX J06411806+3249313  & N & N & N & N & N & N & Non-variable \\
\hline
NGC 7479 & N & N & N & N & N & N & Non-variable\\
\hline
NGC 6300 & N & N & N & N & N & N & Non-variable\\
\hline\hline
\end{tabular}
\vspace{0.5cm}
\begin{tablenotes}
{\textbf{Notes:} $N_{\rm H,los}$-variability determinations using the $\chi^2$$_{\rm red}$ and the p-value methods described in Section \ref{section:var_estimates}. \newline N: Non-variable. Y: Variable. Y*: Variable, see Appendix \ref{appendix:individual_sources} for details. See \citet{pizzetti2022} and Sengupta et al., in prep. for details about NGC 7479 and NGC 6300, respectively. No reliable fit was obtained for NGC 454E using \uxclumpy\ (see Appendix \ref{sec:ngc454e} for more details.) }
\end{tablenotes}
\end{threeparttable}
\end{table*}

\section{Conclusions}\label{sec:conclusion}
This work presents the simultaneous multi-epoch analysis of 13 likely-variable Compton-thin galaxies selected from a sample of heavily obscured AGN previously analyzed by \citetalias{zhao_2021}. Each source has been analyzed using three torus models, \borus, \mytorus, and \uxclumpy, yielding insights into geometrical properties such as the covering factor and cloud dispersion, as well as the photon index and average hydrogen column density, among other parameters. The line-of-sight hydrogen column density was determined for each source at each epoch, allowing for the categorization of each galaxy into `\nh-Variable', `Non-variable in \nh' or `Undetermined', depending on the estimated degree of variability. We then discussed the \nh\ variability properties of the full sample of 27 galaxies (13 analyzed in this work, 12 in \citetalias{nuria2023}, one in \citet{pizzetti2022} and one in Sengupta et al., in prep.). We summarize here our conclusions:
\begin{enumerate}
    \item $37\%$ of the sources (10 out of 27) exhibit \nh\ variability, indicating that variations in \nh\ are necessary to achieve a proper fit across the available observations. For $44\%$ of the sources (12 out of 27), we can confidently say no \nh\ variability is required over the analyzed timeframe. However, all sources necessitate either flux or \nh\, (or both), variability, aligning with the anticipated outcomes based on the sample selection criteria.
    \item By comparing the torus geometrical and intrinsic properties among the three classes of variability, it becomes evident that, within errors, there appears to be no distinction between the `Variable', `Non-variable', and `Undetermined' sources, suggesting no intrinsic differences between the variable and non variable sources. 
    \item For $85\%$ of the total sample (23 out of 27), we measured a difference between the \nh\ and \nhav\, in support of the clumpy torus scenario; this implies the potential disconnection of the material responsible for reflection and absorption, as suggested by \citetalias{nuria2023}. For 12 out of 27 sources, the clouds (\nh) are denser than the average density of the torus (\nhav), suggesting a scenario in which overdense clouds are dispersed in a less dense inter-cloud medium, acting as a thin-reflector. For 9 out of 27 sources, the average density of the torus is found to be denser than the clouds, indicating the presence of a geometrically thin (i.e., warped disk), dense reflector. Two sources exhibit characteristics aligning with both scenarios, depending on the considered model.
    \item $48\%$ (13 out of 27) of the analyzed sources require the inclusion of a Compton-thick reflector (\textit{inner ring}) in the fitting process. Among these, four exhibit recorded 22 GHz water megamaser emission, suggesting a potential shared nature and spatial properties between the inner reflector and the megamaser disk. The partial discrepancy in the correlation between the presence of the inner reflector and the water megamaser disk is also discussed.
    \item The median variation in \nh\ between any two observations of the same source is of $25\%$ with respect to the lowest \nh\ value in the pair. This value is less than what was previously reported in \citetalias{nuria2023}, where the median variation was $36\%$, indicating that, on average, the entire sample displays smaller variations in \nh. In a quarter of the observation pairs in the sample, the fractional difference in \nh\ between two observations exceeds $70\%$.
    
    \item We present the equatorial hydrogen column density as computed for the \uxclumpy\ model and provide an open-source tool to compute the \nheq\ from CTKcover and TOR$\sigma$ obtained from spectral fitting.
\end{enumerate}
\clearpage

\section*{acknowledgements}

A.P., N.T., I.C. and M.A. acknowledge funding from NASA  (contracts 80NSSC23K1611, 80NSSC23K0484, 80NSSC23K0146). 
The scientific results reported in this article are based on observations
made by the X-ray observatories NuSTAR, XMM-Newton and Chandra. This research has made use of the NuSTAR Data Analysis Software (NuSTARDAS) jointly developed by the ASI Space Science Data Center (SSDC, Italy) and the California Institute of Technology (Caltech, USA), and the NASA/IPAC Extragalactic Database (NED), operated by the Jet Propulsion Laboratory, California Institute of Technology, under contract with NASA. We acknowledge the use of the software packages XMM-SAS and HEASoft.
\software{CIAO \citep[v4.14,][]{fruscione2006}, XSPEC \citep{Arnaud1996}, HEAsoft (HEASARC 2014), SAS (Gabriel et al. 2004) \citep{gabriel2004_SAS}}.

\bibliographystyle{aasjournal}
\bibliography{bibliography}

\begin{thebibliography}{}
\expandafter\ifx\csname natexlab\endcsname\relax\def\natexlab#1{#1}\fi
\providecommand{\url}[1]{\href{#1}{#1}}
\providecommand{\dodoi}[1]{doi:~\href{http://doi.org/#1}{\nolinkurl{#1}}}
\providecommand{\doeprint}[1]{\href{http://ascl.net/#1}{\nolinkurl{http://ascl.net/#1}}}
\providecommand{\doarXiv}[1]{\href{https://arxiv.org/abs/#1}{\nolinkurl{https://arxiv.org/abs/#1}}}

\bibitem[{{Andonie} {et~al.}(2022){Andonie}, {Bauer}, {Carraro}, {Ar{\'e}valo}, {Alexander}, {Brandt}, {Buchner}, {He}, {Koss}, {Ricci}, {Salinas}, {Solimano}, {Tortosa}, \& {Treister}}]{andoni3e2022}
{Andonie}, C., {Bauer}, F.~E., {Carraro}, R., {et~al.} 2022, A\&A, 664, A46, \dodoi{10.1051/0004-6361/202142473}

\bibitem[{{Andrae} {et~al.}(2010){Andrae}, {Schulze-Hartung}, \& {Melchior}}]{andrae2010}
{Andrae}, R., {Schulze-Hartung}, T., \& {Melchior}, P. 2010, arXiv e-prints, arXiv:1012.3754, \dodoi{10.48550/arXiv.1012.3754}

\bibitem[{{Antonucci}(1993)}]{antonucci_1993}
{Antonucci}, R. 1993, \araa, 31, 473, \dodoi{10.1146/annurev.aa.31.090193.002353}

\bibitem[{{Arnaud}(1996)}]{Arnaud1996}
{Arnaud}, K.~A. 1996, in Astronomical Society of the Pacific Conference Series, Vol. 101, Astronomical Data Analysis Software and Systems V, ed. G.~H. {Jacoby} \& J.~{Barnes}, 17

\bibitem[{Balokovi{\'{c}} {et~al.}(2018)Balokovi{\'{c}}, Brightman, Harrison, Comastri, Ricci, Buchner, Gandhi, Farrah, \& Stern}]{Balokovic2018}
Balokovi{\'{c}}, M., Brightman, M., Harrison, F.~A., {et~al.} 2018, ApJ, 854, 42, \dodoi{10.3847/1538-4357/aaa7eb}

\bibitem[{{Balokovi{\'c}} {et~al.}(2020){Balokovi{\'c}}, {Harrison}, {Madejski}, {Comastri}, {Ricci}, {Annuar}, {Ballantyne}, {Boorman}, {Brandt}, {Brightman}, {Gandhi}, {Kamraj}, {Koss}, {Marchesi}, {Marinucci}, {Masini}, {Matt}, {Stern}, \& {Urry}}]{balokovic2020}
{Balokovi{\'c}}, M., {Harrison}, F.~A., {Madejski}, G., {et~al.} 2020, \apj, 905, 41, \dodoi{10.3847/1538-4357/abc342}

\bibitem[{{Barthelmy} {et~al.}(2005){Barthelmy}, {Barbier}, {Cummings}, {Fenimore}, {Gehrels}, {Hullinger}, {Krimm}, {Markwardt}, {Palmer}, {Parsons}, {Sato}, {Suzuki}, {Takahashi}, {Tashiro}, \& {Tueller}}]{Barthelmy2005}
{Barthelmy}, S.~D., {Barbier}, L.~M., {Cummings}, J.~R., {et~al.} 2005, \ssr, 120, 143, \dodoi{10.1007/s11214-005-5096-3}

\bibitem[{{Bennert} {et~al.}(2004){Bennert}, {Schulz}, \& {Henkel}}]{bennert2004}
{Bennert}, N., {Schulz}, H., \& {Henkel}, C. 2004, \aap, 419, 127, \dodoi{10.1051/0004-6361:20034497}

\bibitem[{{Bennett} {et~al.}(2014){Bennett}, {Larson}, {Weiland}, \& {Hinshaw}}]{bennett2014}
{Bennett}, C.~L., {Larson}, D., {Weiland}, J.~L., \& {Hinshaw}, G. 2014, ApJ, 794, 135, \dodoi{10.1088/0004-637X/794/2/135}

\bibitem[{{Bianchi} {et~al.}(2005){Bianchi}, {Guainazzi}, {Matt}, {Chiaberge}, {Iwasawa}, {Fiore}, \& {Maiolino}}]{bianchi_2005}
{Bianchi}, S., {Guainazzi}, M., {Matt}, G., {et~al.} 2005, \aap, 442, 185, \dodoi{10.1051/0004-6361:20053389}

\bibitem[{{Boorman} {et~al.}(2023){Boorman}, {Torres-Alb{\`a}}, {Annuar}, {Marchesi}, {Pfeifle}, {Stern}, {Civano}, {Balokovi{\'c}}, {Buchner}, {Ricci}, {Alexander}, {Brandt}, {Brightman}, {Chen}, {Creech}, {Gandhi}, {Garc{\'\i}a}, {Harrison}, {Hickox}, {Kammoun}, {LaMassa}, {Lanzuisi}, {Marcotulli}, {Madsen}, {Matt}, {Matzeu}, {Nardini}, {Piotrowska}, {Pizzetti}, {Puccetti}, {Sicilian}, {Silver}, {Walton}, {Wilkins}, \& {Zhao}}]{Boorman2023HEXPobscurer}
{Boorman}, P.~G., {Torres-Alb{\`a}}, N., {Annuar}, A., {et~al.} 2023, arXiv e-prints, arXiv:2311.04949, \dodoi{10.48550/arXiv.2311.04949}

\bibitem[{{Braatz} \& {Gugliucci}(2008)}]{braatz2008}
{Braatz}, J.~A., \& {Gugliucci}, N.~E. 2008, \apj, 678, 96, \dodoi{10.1086/529538}

\bibitem[{{Braatz} {et~al.}(2004){Braatz}, {Henkel}, {Greenhill}, {Moran}, \& {Wilson}}]{braatz2004}
{Braatz}, J.~A., {Henkel}, C., {Greenhill}, L.~J., {Moran}, J.~M., \& {Wilson}, A.~S. 2004, \apjl, 617, L29, \dodoi{10.1086/427185}

\bibitem[{{Braatz} {et~al.}(1996){Braatz}, {Wilson}, \& {Henkel}}]{braatz1996}
{Braatz}, J.~A., {Wilson}, A.~S., \& {Henkel}, C. 1996, \apjs, 106, 51, \dodoi{10.1086/192328}

\bibitem[{{Buchner} \& {Bauer}(2017)}]{buchner2017}
{Buchner}, J., \& {Bauer}, F.~E. 2017, \mnras, 465, 4348, \dodoi{10.1093/mnras/stw2955}

\bibitem[{{Buchner} {et~al.}(2021){Buchner}, {Brightman}, {Balokovi{\'c}}, {Wada}, {Bauer}, \& {Nandra}}]{buchner2021}
{Buchner}, J., {Brightman}, M., {Balokovi{\'c}}, M., {et~al.} 2021, A$\&$A, 651, A58, \dodoi{10.1051/0004-6361/201834963}

\bibitem[{Buchner {et~al.}(2019)Buchner, Brightman, Nandra, Nikutta, \& Bauer}]{Buchner_2019}
Buchner, J., Brightman, M., Nandra, K., Nikutta, R., \& Bauer, F.~E. 2019, Astronomy \& Astrophysics, 629, A16, \dodoi{10.1051/0004-6361/201834771}

\bibitem[{{Buchner} {et~al.}(2019){Buchner}, {Brightman}, {Nandra}, {Nikutta}, \& {Bauer}}]{Buchner2019}
{Buchner}, J., {Brightman}, M., {Nandra}, K., {Nikutta}, R., \& {Bauer}, F.~E. 2019, \aap, 629, A16, \dodoi{10.1051/0004-6361/201834771}

\bibitem[{{Castangia} {et~al.}(2013){Castangia}, {Panessa}, {Henkel}, {Kadler}, \& {Tarchi}}]{castangia2013}
{Castangia}, P., {Panessa}, F., {Henkel}, C., {Kadler}, M., \& {Tarchi}, A. 2013, \mnras, 436, 3388, \dodoi{10.1093/mnras/stt1824}

\bibitem[{{Comastri}(2004)}]{comastri2004}
{Comastri}, A. 2004, in Astrophysics and Space Science Library, Vol. 308, Supermassive Black Holes in the Distant Universe., ed. A.~J. {Barger}, 245

\bibitem[{{Combes} {et~al.}(2019){Combes}, {Garc{\'\i}a-Burillo}, {Audibert}, {Hunt}, {Eckart}, {Aalto}, {Casasola}, {Boone}, {Krips}, {Viti}, {Sakamoto}, {Muller}, {Dasyra}, {van der Werf}, \& {Martin}}]{Combes2019}
{Combes}, F., {Garc{\'\i}a-Burillo}, S., {Audibert}, A., {et~al.} 2019, A$\&$A, 623, A79, \dodoi{10.1051/0004-6361/201834560}

\bibitem[{{Fruscione} {et~al.}(2006){Fruscione}, {McDowell}, {Allen}, {Brickhouse}, {Burke}, {Davis}, {Durham}, {Elvis}, {Galle}, {Harris}, {Huenemoerder}, {Houck}, {Ishibashi}, {Karovska}, {Nicastro}, {Noble}, {Nowak}, {Primini}, {Siemiginowska}, {Smith}, \& {Wise}}]{fruscione2006}
{Fruscione}, A., {McDowell}, J.~C., {Allen}, G.~E., {et~al.} 2006, in \procspie, Vol. 6270, Society of Photo-Optical Instrumentation Engineers (SPIE) Conference Series, 62701V

\bibitem[{{Furui} {et~al.}(2016){Furui}, {Fukazawa}, {Odaka}, {Kawaguchi}, {Ohno}, \& {Hayashi}}]{furui2016}
{Furui}, S., {Fukazawa}, Y., {Odaka}, H., {et~al.} 2016, ApJ, 818, 164, \dodoi{10.3847/0004-637X/818/2/164}

\bibitem[{{Gabriel} {et~al.}(2004){Gabriel}, {Denby}, {Fyfe}, {Hoar}, {Ibarra}, {Ojero}, {Osborne}, {Saxton}, {Lammers}, \& {Vacanti}}]{gabriel2004_SAS}
{Gabriel}, C., {Denby}, M., {Fyfe}, D.~J., {et~al.} 2004, in Astronomical Society of the Pacific Conference Series, Vol. 314, Astronomical Data Analysis Software and Systems (ADASS) XIII, ed. F.~{Ochsenbein}, M.~G. {Allen}, \& D.~{Egret}, 759

\bibitem[{{Garc{\'\i}a-Burillo} {et~al.}(2021){Garc{\'\i}a-Burillo}, {Alonso-Herrero}, {Ramos Almeida}, {Gonz{\'a}lez-Mart{\'\i}n}, {Combes}, {Usero}, {H{\"o}nig}, {Querejeta}, {Hicks}, {Hunt}, {Rosario}, {Davies}, {Boorman}, {Bunker}, {Burtscher}, {Colina}, {D{\'\i}az-Santos}, {Gandhi}, {Garc{\'\i}a-Bernete}, {Garc{\'\i}a-Lorenzo}, {Ichikawa}, {Imanishi}, {Izumi}, {Labiano}, {Levenson}, {L{\'o}pez-Rodr{\'\i}guez}, {Packham}, {Pereira-Santaella}, {Ricci}, {Rigopoulou}, {Rouan}, {Shimizu}, {Stalevski}, {Wada}, \& {Williamson}}]{garcia_burillo2021}
{Garc{\'\i}a-Burillo}, S., {Alonso-Herrero}, A., {Ramos Almeida}, C., {et~al.} 2021, A$\&$A, 652, A98, \dodoi{10.1051/0004-6361/202141075}

\bibitem[{{Gianolli} {et~al.}(2023){Gianolli}, {Kim}, {Bianchi}, {Ag{\'\i}s-Gonz{\'a}lez}, {Madejski}, {Marin}, {Marinucci}, {Matt}, {Middei}, {Petrucci}, {Soffitta}, {Tagliacozzo}, {Tombesi}, {Ursini}, {Barnouin}, {De Rosa}, {Di Gesu}, {Ingram}, {Loktev}, {Panagiotou}, {Podgorny}, {Poutanen}, {Puccetti}, {Ratheesh}, {Veledina}, {Zhang}, {Agudo}, {Antonelli}, {Bachetti}, {Baldini}, {Baumgartner}, {Bellazzini}, {Bongiorno}, {Bonino}, {Brez}, {Bucciantini}, {Capitanio}, {Castellano}, {Cavazzuti}, {Chen}, {Ciprini}, {Costa}, {Del Monte}, {Di Lalla}, {Di Marco}, {Donnarumma}, {Doroshenko}, {Dov{\v{c}}iak}, {Ehlert}, {Enoto}, {Evangelista}, {Fabiani}, {Ferrazzoli}, {Garc{\'\i}a}, {Gunji}, {Heyl}, {Iwakiri}, {Jorstad}, {Kaaret}, {Karas}, {Kislat}, {Kitaguchi}, {Kolodziejczak}, {Krawczynski}, {La Monaca}, {Latronico}, {Liodakis}, {Maldera}, {Manfreda}, {Marscher}, {Marshall}, {Massaro}, {Mitsuishi}, {Mizuno}, {Muleri}, {Negro}, {Ng}, {O'Dell}, {Omodei}, {Oppedisano}, {Papitto}, {Pavlov}, {Peirson}, {Perri},
  {Pesce-Rollins}, {Pilia}, {Possenti}, {Ramsey}, {Rankin}, {Roberts}, {Romani}, {Sgr{\`o}}, {Slane}, {Spandre}, {Swartz}, {Tamagawa}, {Tavecchio}, {Taverna}, {Tawara}, {Tennant}, {Thomas}, {Trois}, {Tsygankov}, {Turolla}, {Vink}, {Weisskopf}, {Wu}, {Xie}, \& {Zane}}]{gianolli2023}
{Gianolli}, V.~E., {Kim}, D.~E., {Bianchi}, S., {et~al.} 2023, MNRAS, 523, 4468, \dodoi{10.1093/mnras/stad1697}

\bibitem[{{Greenhill} {et~al.}(2003{\natexlab{a}}){Greenhill}, {Kondratko}, {Lovell}, {Kuiper}, {Moran}, {Jauncey}, \& {Baines}}]{greenhill2003}
{Greenhill}, L.~J., {Kondratko}, P.~T., {Lovell}, J.~E.~J., {et~al.} 2003{\natexlab{a}}, \apjl, 582, L11, \dodoi{10.1086/367602}

\bibitem[{{Greenhill} {et~al.}(2009){Greenhill}, {Kondratko}, {Moran}, \& {Tilak}}]{greenhill2009}
{Greenhill}, L.~J., {Kondratko}, P.~T., {Moran}, J.~M., \& {Tilak}, A. 2009, \apj, 707, 787, \dodoi{10.1088/0004-637X/707/1/787}

\bibitem[{{Greenhill} {et~al.}(2008){Greenhill}, {Tilak}, \& {Madejski}}]{Greenhill2008}
{Greenhill}, L.~J., {Tilak}, A., \& {Madejski}, G. 2008, \apjl, 686, L13, \dodoi{10.1086/592782}

\bibitem[{{Greenhill} {et~al.}(2003{\natexlab{b}}){Greenhill}, {Booth}, {Ellingsen}, {Herrnstein}, {Jauncey}, {McCulloch}, {Moran}, {Norris}, {Reynolds}, \& {Tzioumis}}]{greenhill2003b}
{Greenhill}, L.~J., {Booth}, R.~S., {Ellingsen}, S.~P., {et~al.} 2003{\natexlab{b}}, \apj, 590, 162, \dodoi{10.1086/374862}

\bibitem[{{Haardt} {et~al.}(1994){Haardt}, {Maraschi}, \& {Ghisellini}}]{haardt1994}
{Haardt}, F., {Maraschi}, L., \& {Ghisellini}, G. 1994, ApJl, 432, L95, \dodoi{10.1086/187520}

\bibitem[{{Hern{\'a}ndez-Garc{\'\i}a} {et~al.}(2015){Hern{\'a}ndez-Garc{\'\i}a}, {Masegosa}, {Gonz{\'a}lez-Mart{\'\i}n}, \& {M{\'a}rquez}}]{hernandez2015}
{Hern{\'a}ndez-Garc{\'\i}a}, L., {Masegosa}, J., {Gonz{\'a}lez-Mart{\'\i}n}, O., \& {M{\'a}rquez}, I. 2015, \aap, 579, A90, \dodoi{10.1051/0004-6361/201526127}

\bibitem[{{Herrnstein} {et~al.}(2005){Herrnstein}, {Moran}, {Greenhill}, \& {Trotter}}]{Herrnstein2005}
{Herrnstein}, J.~R., {Moran}, J.~M., {Greenhill}, L.~J., \& {Trotter}, A.~S. 2005, \apj, 629, 719, \dodoi{10.1086/431421}

\bibitem[{{H{\"o}nig}(2019)}]{honig2019}
{H{\"o}nig}, S.~F. 2019, ApJ, 884, 171, \dodoi{10.3847/1538-4357/ab4591}

\bibitem[{{Jansen} {et~al.}(2001){Jansen}, {Lumb}, {Altieri}, {Clavel}, {Ehle}, {Erd}, {Gabriel}, {Guainazzi}, {Gondoin}, {Much}, {Munoz}, {Santos}, {Schartel}, {Texier}, \& {Vacanti}}]{jansen2001}
{Jansen}, F., {Lumb}, D., {Altieri}, B., {et~al.} 2001, \aap, 365, L1, \dodoi{10.1051/0004-6361:20000036}

\bibitem[{{Kalberla} {et~al.}(2005){Kalberla}, {Burton}, {Hartmann}, {Arnal}, {Bajaja}, {Morras}, \& {P{\"o}ppel}}]{Kalberla2005}
{Kalberla}, P.~M.~W., {Burton}, W.~B., {Hartmann}, D., {et~al.} 2005, A$\&$A, 440, 775, \dodoi{10.1051/0004-6361:20041864}

\bibitem[{{Koss} {et~al.}(2022){Koss}, {Ricci}, {Trakhtenbrot}, {Oh}, {den Brok}, {Mej{\'\i}a-Restrepo}, {Stern}, {Privon}, {Treister}, {Powell}, {Mushotzky}, {Bauer}, {Ananna}, {Balokovi{\'c}}, {B{\"a}r}, {Becker}, {Bessiere}, {Burtscher}, {Caglar}, {Congiu}, {Evans}, {Harrison}, {Heida}, {Ichikawa}, {Kamraj}, {Lamperti}, {Pacucci}, {Ricci}, {Riffel}, {Rojas}, {Schawinski}, {Temple}, {Urry}, {Veilleux}, \& {Williams}}]{koss2022}
{Koss}, M.~J., {Ricci}, C., {Trakhtenbrot}, B., {et~al.} 2022, \apjs, 261, 2, \dodoi{10.3847/1538-4365/ac6c05}

\bibitem[{{Kuo} {et~al.}(2011){Kuo}, {Braatz}, {Condon}, {Impellizzeri}, {Lo}, {Zaw}, {Schenker}, {Henkel}, {Reid}, \& {Greene}}]{kuo2011}
{Kuo}, C.~Y., {Braatz}, J.~A., {Condon}, J.~J., {et~al.} 2011, \apj, 727, 20, \dodoi{10.1088/0004-637X/727/1/20}

\bibitem[{{Laha} {et~al.}(2020){Laha}, {Markowitz}, {Krumpe}, {Nikutta}, {Rothschild}, \& {Saha}}]{Laha2020}
{Laha}, S., {Markowitz}, A.~G., {Krumpe}, M., {et~al.} 2020, \apj, 897, 66, \dodoi{10.3847/1538-4357/ab92ab}

\bibitem[{{Lawrence} \& {Elvis}(2010)}]{lawrence2010}
{Lawrence}, A., \& {Elvis}, M. 2010, \apj, 714, 561, \dodoi{10.1088/0004-637X/714/1/561}

\bibitem[{{Marchese} {et~al.}(2012){Marchese}, {Braito}, {Della Ceca}, {Caccianiga}, \& {Severgnini}}]{marchese2012}
{Marchese}, E., {Braito}, V., {Della Ceca}, R., {Caccianiga}, A., \& {Severgnini}, P. 2012, \mnras, 421, 1803, \dodoi{10.1111/j.1365-2966.2012.20445.x}

\bibitem[{{Marchese} {et~al.}(2014){Marchese}, {Braito}, {Reeves}, {Della Ceca}, {Caccianiga}, {Markowitz}, {Risaliti}, {Severgnini}, \& {Turner}}]{marchese2014}
{Marchese}, E., {Braito}, V., {Reeves}, J.~N., {et~al.} 2014, \mnras, 437, 2806, \dodoi{10.1093/mnras/stt2101}

\bibitem[{{Marchesi} {et~al.}(2022){Marchesi}, {Zhao}, {Torres-Alb{\`a}}, {Ajello}, {Gaspari}, {Pizzetti}, {Buchner}, {Bertola}, {Comastri}, {Feltre}, {Gilli}, {Lanzuisi}, {Matzeu}, {Pozzi}, {Salvestrini}, {Sengupta}, {Silver}, {Tombesi}, {Traina}, {Vignali}, \& {Zappacosta}}]{marchesi2022}
{Marchesi}, S., {Zhao}, X., {Torres-Alb{\`a}}, N., {et~al.} 2022, \apj, 935, 114, \dodoi{10.3847/1538-4357/ac80be}

\bibitem[{{Markowitz} {et~al.}(2014){Markowitz}, {Krumpe}, \& {Nikutta}}]{Markowitz_2014}
{Markowitz}, A.~G., {Krumpe}, M., \& {Nikutta}, R. 2014, \mnras, 439, 1403, \dodoi{10.1093/mnras/stt2492}

\bibitem[{{Masini} {et~al.}(2016){Masini}, {Comastri}, {Balokovi{\'c}}, {Zaw}, {Puccetti}, {Ballantyne}, {Bauer}, {Boggs}, {Brandt}, {Brightman}, {Christensen}, {Craig}, {Gandhi}, {Hailey}, {Harrison}, {Koss}, {Madejski}, {Ricci}, {Rivers}, {Stern}, \& {Zhang}}]{Masini2016}
{Masini}, A., {Comastri}, A., {Balokovi{\'c}}, M., {et~al.} 2016, \aap, 589, A59, \dodoi{10.1051/0004-6361/201527689}

\bibitem[{{Murphy} \& {Yaqoob}(2009)}]{Murphy2009}
{Murphy}, K.~D., \& {Yaqoob}, T. 2009, MNRAS, 397, 1549, \dodoi{10.1111/j.1365-2966.2009.15025.x}

\bibitem[{{Nenkova} {et~al.}(2008{\natexlab{a}}){Nenkova}, {Sirocky}, {Ivezi{\'c}}, \& {Elitzur}}]{nenkova2008a}
{Nenkova}, M., {Sirocky}, M.~M., {Ivezi{\'c}}, {\v{Z}}., \& {Elitzur}, M. 2008{\natexlab{a}}, \apj, 685, 147, \dodoi{10.1086/590482}

\bibitem[{{Nenkova} {et~al.}(2008{\natexlab{b}}){Nenkova}, {Sirocky}, {Nikutta}, {Ivezi{\'c}}, \& {Elitzur}}]{nenkova2008b}
{Nenkova}, M., {Sirocky}, M.~M., {Nikutta}, R., {Ivezi{\'c}}, {\v{Z}}., \& {Elitzur}, M. 2008{\natexlab{b}}, ApJ, 685, 160, \dodoi{10.1086/590483}

\bibitem[{{Oh} {et~al.}(2018){Oh}, {Koss}, {Markwardt}, {Schawinski}, {Baumgartner}, {Barthelmy}, {Cenko}, {Gehrels}, {Mushotzky}, {Petulante}, {Ricci}, {Lien}, \& {Trakhtenbrot}}]{Oh2018}
{Oh}, K., {Koss}, M., {Markwardt}, C.~B., {et~al.} 2018, \apjs, 235, 4, \dodoi{10.3847/1538-4365/aaa7fd}

\bibitem[{{Panessa} \& {Giroletti}(2013)}]{panessa2013}
{Panessa}, F., \& {Giroletti}, M. 2013, \mnras, 432, 1138, \dodoi{10.1093/mnras/stt547}

\bibitem[{{Peck} {et~al.}(2003){Peck}, {Henkel}, {Ulvestad}, {Brunthaler}, {Falcke}, {Elitzur}, {Menten}, \& {Gallimore}}]{peck2003}
{Peck}, A.~B., {Henkel}, C., {Ulvestad}, J.~S., {et~al.} 2003, \apj, 590, 149, \dodoi{10.1086/374924}

\bibitem[{{Petterson}(1977)}]{petterson1997}
{Petterson}, J.~A. 1977, \apj, 214, 550, \dodoi{10.1086/155280}

\bibitem[{{Pizzetti} {et~al.}(2022){Pizzetti}, {Torres-Alb{\`a}}, {Marchesi}, {Ajello}, {Silver}, \& {Zhao}}]{pizzetti2022}
{Pizzetti}, A., {Torres-Alb{\`a}}, N., {Marchesi}, S., {et~al.} 2022, ApJ, 936, 149, \dodoi{10.3847/1538-4357/ac86c6}

\bibitem[{{Ramos Almeida} \& {Ricci}(2017)}]{ramosailmeida_ricci_2017}
{Ramos Almeida}, C., \& {Ricci}, C. 2017, Nature Astronomy, 1, 679, \dodoi{10.1038/s41550-017-0232-z}

\bibitem[{{Risaliti} {et~al.}(2005){Risaliti}, {Elvis}, {Fabbiano}, {Baldi}, \& {Zezas}}]{risaliti_2005}
{Risaliti}, G., {Elvis}, M., {Fabbiano}, G., {Baldi}, A., \& {Zezas}, A. 2005, \apjl, 623, L93, \dodoi{10.1086/430252}

\bibitem[{{Risaliti} {et~al.}(2007){Risaliti}, {Elvis}, {Fabbiano}, {Baldi}, {Zezas}, \& {Salvati}}]{risaliti2007}
{Risaliti}, G., {Elvis}, M., {Fabbiano}, G., {et~al.} 2007, \apjl, 659, L111, \dodoi{10.1086/517884}

\bibitem[{{Risaliti} {et~al.}(2002){Risaliti}, {Elvis}, \& {Nicastro}}]{Risaliti2002}
{Risaliti}, G., {Elvis}, M., \& {Nicastro}, F. 2002, ApJ, 571, 234, \dodoi{10.1086/324146}

\bibitem[{{Ross}(1979)}]{ross1979}
{Ross}, R.~R. 1979, \apj, 233, 334, \dodoi{10.1086/157394}

\bibitem[{{Sanfrutos} {et~al.}(2013){Sanfrutos}, {Miniutti}, {Ag{\'\i}s-Gonz{\'a}lez}, {Fabian}, {Miller}, {Panessa}, \& {Zoghbi}}]{Sanfrutos2013}
{Sanfrutos}, M., {Miniutti}, G., {Ag{\'\i}s-Gonz{\'a}lez}, B., {et~al.} 2013, MNRAS, 436, 1588, \dodoi{10.1093/mnras/stt1675}

\bibitem[{{Schartmann} {et~al.}(2005){Schartmann}, {Meisenheimer}, {Camenzind}, {Wolf}, \& {Henning}}]{Schartmann2005}
{Schartmann}, M., {Meisenheimer}, K., {Camenzind}, M., {Wolf}, S., \& {Henning}, T. 2005, \aap, 437, 861, \dodoi{10.1051/0004-6361:20042363}

\bibitem[{{Siebenmorgen} {et~al.}(2015){Siebenmorgen}, {Heymann}, \& {Efstathiou}}]{Siebenmorgen2015}
{Siebenmorgen}, R., {Heymann}, F., \& {Efstathiou}, A. 2015, \aap, 583, A120, \dodoi{10.1051/0004-6361/201526034}

\bibitem[{{Simpson}(2005)}]{simpson2005}
{Simpson}, C. 2005, \mnras, 360, 565, \dodoi{10.1111/j.1365-2966.2005.09043.x}

\bibitem[{{Smith} {et~al.}(2001){Smith}, {Brickhouse}, {Liedahl}, \& {Raymond}}]{smith2001}
{Smith}, R.~K., {Brickhouse}, N.~S., {Liedahl}, D.~A., \& {Raymond}, J.~C. 2001, \apjl, 556, L91, \dodoi{10.1086/322992}

\bibitem[{{Stalevski} {et~al.}(2011){Stalevski}, {Fritz}, {Baes}, {Nakos}, \& {Popovi{\'c}}}]{stakevski2011}
{Stalevski}, M., {Fritz}, J., {Baes}, M., {Nakos}, T., \& {Popovi{\'c}}, L.~{\v{C}}. 2011, Baltic Astronomy, 20, 490, \dodoi{10.1515/astro-2017-0326}

\bibitem[{Stephens(1974)}]{Stephens1974_andersondarlingtest}
Stephens, M.~A. 1974, Journal of the American Statistical Association, 69, 730, \dodoi{10.1080/01621459.1974.10480196}

\bibitem[{{Tanimoto} {et~al.}(2019){Tanimoto}, {Ueda}, {Odaka}, {Kawaguchi}, {Fukazawa}, \& {Kawamuro}}]{Tanimoto2019}
{Tanimoto}, A., {Ueda}, Y., {Odaka}, H., {et~al.} 2019, \apj, 877, 95, \dodoi{10.3847/1538-4357/ab1b20}

\bibitem[{{Torres-Alb{\`a}} {et~al.}(2023){Torres-Alb{\`a}}, {Marchesi}, {Zhao}, {Cox}, {Pizzetti}, {Ajello}, \& {Silver}}]{nuria2023}
{Torres-Alb{\`a}}, N., {Marchesi}, S., {Zhao}, X., {et~al.} 2023, arXiv e-prints, arXiv:2301.07138, \dodoi{10.48550/arXiv.2301.07138}

\bibitem[{{Torres-Alb{\`a}} {et~al.}(2018){Torres-Alb{\`a}}, {Iwasawa}, {D{\'\i}az-Santos}, {Charmandaris}, {Ricci}, {Chu}, {Sanders}, {Armus}, {Barcos-Mu{\~n}oz}, {Evans}, {Howell}, {Inami}, {Linden}, {Medling}, {Privon}, {U}, \& {Yoon}}]{torres-alba_2018}
{Torres-Alb{\`a}}, N., {Iwasawa}, K., {D{\'\i}az-Santos}, T., {et~al.} 2018, \aap, 620, A140, \dodoi{10.1051/0004-6361/201834105}

\bibitem[{{Urry} \& {Padovani}(1995)}]{urry_padovani}
{Urry}, C.~M., \& {Padovani}, P. 1995, \pasp, 107, 803, \dodoi{10.1086/133630}

\bibitem[{{Vasudevan} {et~al.}(2010){Vasudevan}, {Fabian}, {Gandhi}, {Winter}, \& {Mushotzky}}]{vasudevan2010}
{Vasudevan}, R.~V., {Fabian}, A.~C., {Gandhi}, P., {Winter}, L.~M., \& {Mushotzky}, R.~F. 2010, MNRAS, 402, 1081, \dodoi{10.1111/j.1365-2966.2009.15936.x}

\bibitem[{{Yaqoob}(2012)}]{Yaqoob2012}
{Yaqoob}, T. 2012, MNRAS, 423, 3360, \dodoi{10.1111/j.1365-2966.2012.21129.x}

\bibitem[{{Yaqoob} {et~al.}(2015){Yaqoob}, {Tatum}, {Scholtes}, {Gottlieb}, \& {Turner}}]{yaqoob2015}
{Yaqoob}, T., {Tatum}, M.~M., {Scholtes}, A., {Gottlieb}, A., \& {Turner}, T.~J. 2015, MNRAS, 454, 973, \dodoi{10.1093/mnras/stv2021}

\bibitem[{{Zhao} {et~al.}(2021){Zhao}, {Marchesi}, {Ajello}, {Cole}, {Hu}, {Silver}, \& {Torres-Alb{\`a}}}]{zhao_2021}
{Zhao}, X., {Marchesi}, S., {Ajello}, M., {et~al.} 2021, \aap, 650, A57, \dodoi{10.1051/0004-6361/202140297}

\end{thebibliography}

\clearpage
 \appendix


 \section{X-ray fitting results}\label{appendix:xray_fitting_allsources}
The best-fit values tables of the simultaneous multi-epoch analysis of the sources analyzed in this work are listed in this Appendix. 

\begin{table*}
\centering
\begin{threeparttable}
\caption{X-ray fitting results for NGC 454E}
\label{tab:NGC454e_fitting}
\renewcommand*{\arraystretch}{1.2}
\begin{tabular}{llccc}
  \textbf{Parameter} && \textbf{\borus}  & \textbf{\mytorus\ dec} & \textbf{\uxclumpy}\\

 \hline\hline
     stat/d.o.f       &         & 973.14/859 & 944.00/857  &   -- \\ 
    
    red stat      &         &      1.13 &    1.10     &  -- \\

       T      &  &   $3.89 $   &    $2.97 $   &  -- \\
       p-value  &&  $6.13\times10^{-16}$  & $1.95\times10^{-19}$  & -- \\
    \hline\hline

  kT     &    & $0.67_{-0.11}^{+0.08}$       &  $0.68_{-0.10}^{+0.08}$   & --\\

  apec norm $(\times 10^{-6})$     &  & $3.20_{-0.43}^{+0.80}$       &  $3.43_{-0.83}^{+0.83}$   & --\\

    \hline

    $\Gamma$     &     & $1.65_{-0.05}^{+0.04}$       &  $1.66_{-0.04}^{+0.04}$   & --\\  
 
    N$_{\rm H,av}\times$10$^{24}$\,cm$^{-2}$  &&  $0.056_{-0.007}^{+0.005}$  &  $0.11_{-0.01}^{+0.02}$   & --  \\

    C$_F$    &       & $0.75_{-0.07}^{+0.11}$      & $\dots$       & -- \\

    cos$(\theta_i)$                &    & $<0.33$    & $\dots$  & --        \\
    
    $\theta_i$              &    & $\dots$  & $\dots$     & --      \\
    
    CTKcover &  & $\dots$ &  $\dots$ &  --\\

    TOR$\sigma$ &  & $\dots$ & $\dots$ & --\\

    $A_{S,90}$         &          &  $\dots$      & $0.38_{-0.08}^{+0.13}$   & --  \\

    $A_{S,0}$          &  &  $\dots$   & $1.55_{-0.19}^{+0.21}$    & --  \\

     $F_s (\times 10^{-3})$      & & $14.39_{-4.38}^{+3.26}$       &  $14.05_{-3.37}^{+4.36}$   & --\\

     Norm ($10^{-3}$)       && $1.84_{-0.39}^{+0.67}$       &  $2.00_{-0.49}^{+0.41}$   & --\\
    
    \hline
     & \suzaku\     & $1.64_{-0.43}^{+0.43}$       &  $1.48_{-0.29}^{+0.45}$   & --\\
    
    &\xmm\             & $0.78_{-0.20}^{+0.19}$       &  $0.72_{-0.14}^{+0.21}$   & --\\

    $C_{\rm AGN}$ &     \chandra\ - 1  & $0.66_{-0.20}^{+0.23}$       &  $0.62_{-0.17}^{+0.25}$   & --\\
    
    & \nustar\ - 1  & 1       & 1   & --\\
    
    & \nustar\ - 2          & $0.70_{-0.21}^{+0.20}$       &  $0.63_{-0.15}^{+0.21}$   & --\\

     &     \chandra\ - 2  & =\nustar\ - 2       &  =\nustar\ - 2   & --\\

     \hline

     & \suzaku\     & $123.13_{-7.86}^{+8.71}$       &  $136.07_{-11.98}^{+16.68}$   & --\\

      & \xmm\         & $26.82_{-1.34}^{+1.43}$       &  $26.66_{-1.39}^{+1.45}$   & --\\
      
      N$_{\rm H,los}\times$10$^{22}$\,cm$^{-2}$     &     \chandra\ - 1  & $19.02_{-3.87}^{+4.62}$       &  $19.00_{-3.91}^{+4.69}$   & --\\

    & \nustar\ - 1 & $96.32_{-16.16}^{+24.59}$       &  $109.81_{-23.07}^{+25.06}$   & --\\ 
 
   &   \nustar\ - 2         & $101.07_{-14.93}^{+12.36}$       &  $104.29_{-14.61}^{+15.70}$   & --\\
   
   &     \chandra\ - 2  & =\nustar\ - 2       &  = \nustar\ - 2  & --\\
 
   \hline
       \hline
    
   log L$_{NuSTAR, 2-10\,keV}$ & & $42.38_{-0.09}^{+0.04}$       &  $42.43_{-0.12}^{+0.11}$   & --\\
   log L$_{NuSTAR, 10-40\,keV}$ & & $42.02_{-0.06}^{+0.05}$      &  $41.87_{-0.06}^{+0.06}$   & --\\

   \hline
    \hline
    
   Red stat - No Var. &&  3.87 & 3.70 & -- \\
   T       &  &   $84.62 $   &    $79.62 $   &  --\\

\hline

   Red stat - No C$_{AGN}$Var. && 1.80 & 1.73 & -- \\
      T       &  &   $23.74 $   &    $21.61 $   &  --\\
   \hline

   Red stat - No N$_{\rm H,los}$ Var. && 3.08  & 2.94 & -- \\
     T       &  &   $61.21 $   &    $57.06 $   &  -- \\
   \hline\hline
\end{tabular}

\begin{tablenotes}
      \footnotesize
     \textbf{Notes:} Same as Table \ref{tab:2MASXJ06411806+3249313_fitting}.
    \end{tablenotes}
  
    \end{threeparttable}

\end{table*}

\begin{table*}
\centering
\begin{threeparttable}
\caption{X-ray fitting results for  MRK 348}
\label{table:mkn348_fitting}
\renewcommand*{\arraystretch}{1.2}
\begin{tabular}{llccc}
  \textbf{Parameter} && \textbf{\borus}  & \textbf{\mytorus\ dec} & \textbf{\uxclumpy}\\              
 \hline\hline
   $\chi^2$/d.o.f         &         & 3978.12/3168 & 3820.06/3170 &  3983.08/3172  \\ 
    
    red $\chi^2$       &         &  1.25    &     1.20    & 1.25 \\

       T       &  &   $14.39 $   &    $11.54 $   &  $14.40 $ \\
   p-value  && $4.43\times10^{-55}$& $5.31\times10^{-319}$& $6.01\times10^{-59}$ \\
    \hline\hline

  $kT$    &     & $0.78_{-0.05}^{+0.04}$       &  $0.78_{-0.04}^{+0.03}$  & $0.79_{-0.05}^{+0.05}$\\

  & \xmm\ - 1   & $<0.84$       &  $5.41_{-1.92}^{+1.90}$  & = XMM - 2\\ 
  
   apec norm$(\times 10^{-6} )$   & \xmm\ - 2   & $16.22_{-2.23}^{+2.24}$       &  $18.29_{-2.22}^{+2.23}$  & $15.95_{-2.24}^{+2.22}$\\ 
  
     & \suzaku   & $21.34_{-2.09}^{+2.07}$       &  $27.93_{-2.07}^{+2.07}$  & $20.62_{-2.24}^{+2.06}$\\
    \hline

    $\Gamma$   &     & $1.51_{-0.01}^{+0.02}$       &  $1.61_{-0.01}^{+0.01}$  & $1.69_{-0.01}^{+0.01}$\\  
 
    N$_{\rm H,av}\times$10$^{24}$\,cm$^{-2}$  &&$0.22_{-0.03}^{+0.10}$       &  $0.30_{-0.02}^{+0.02}$   & $\dots$  \\

    C$_F$   &       & $0.38_{-0.07}^{+0.04}$      & $\dots$       & $\dots$  \\

    cos$(\theta_i)$        &    & $0.35_{-0.07}^{+0.02}$    & $\dots$  & $\dots$        \\
    
    $\theta_i$                &    & $\dots$  & $\dots$     & $70^*$       \\
    
    CTKcover&  & $\dots$ &  $\dots$ &  $0^*$\\

    TOR$\sigma$ &  & $\dots$ & $\dots$ & $27.99_{-0.59}^{+2.58}$\\

    $A_{S,90}$          &          &  $\dots$      & $0.14_{-0.02}^{+0.02}$   & $\dots$   \\

    $A_{S,0}$   &  &  $\dots$   & $4.68_{-0.42}^{+0.48}$    & $\dots$  \\

     $F_s (\times 10^{-3} )$    & & $5.67_{-0.32}^{+0.29}$ & $2.92_{-0.61}^{+0.50}$   & $14.61_{-1.70}^{+0.82}$\\

     Norm ($10^{-3}$)      && $9.72_{-0.26}^{+0.29}$ & $10.39_{-0.30}^{+0.31}$  & $18.77_{-0.72}^{+0.65}$\\
    
    \hline

    &\xmm\ - 1           & $1.16_{-0.02}^{+0.02}$  &$1.09_{-0.02}^{+0.02}$ & $1.06_{-0.02}^{+0.03}$\\
       
    & \suzaku\     & $1.06_{-0.01}^{+0.01}$  &$1.02_{-0.01}^{+0.01}$ & $0.99_{-0.02}^{+0.02}$\\
      
    $C_{\rm AGN}$ &    \xmm\ - 2  & $0.33_{-0.01}^{+0.01}$ & $0.31_{-0.01}^{+0.01}$  & $0.30_{-0.01}^{+0.01}$\\
    
    & \nustar\ - 1  & 1       & 1   & 1\\
    
    & \nustar\ - 2         & $0.90_{-0.01}^{+0.01}$  &$0.85_{-0.01}^{+0.01}$ & $0.86_{-0.01}^{+0.01}$\\

     \hline

      & \xmm\ -1          & $13.08_{-0.24}^{+0.20}$ & $13.50_{-0.24}^{+0.27}$  &$12.73_{-0.17}^{+0.17}$ \\

    & \suzaku\     & $7.58_{-0.07}^{+0.07}$ & $7.46_{-0.07}^{+0.08}$  &$7.52_{-0.08}^{+0.08}$ \\
      
      N$_{\rm H,los}\times$10$^{22}$\,cm$^{-2}$  &       \xmm\ - 2  & $12.61_{-0.48}^{+0.51}$ & $12.91_{-0.56}^{+0.62}$  &$12.30_{-0.42}^{+0.42}$ \\
  
    & \nustar\ - 1 & $6.16_{-0.34}^{+0.41}$ & $5.74_{-0.40}^{+0.41}$  &$6.72_{-0.35}^{+0.34}$ \\ 
 
   &   \nustar\ - 2         & $7.43_{-0.30}^{+0.31}$ & $7.13_{-0.32}^{+0.34}$  &$7.52_{-0.28}^{+0.27}$\\

   \hline
       \hline
    
   log L$_{NuSTAR, 2-10\,keV}$ & & $43.42_{-0.05}^{+0.04}$       &  $43.374_{-0.008}^{+0.008}$   & $43.58$\\
   log L$_{NuSTAR, 10-40\,keV}$ & & $43.59_{-0.05}^{+0.04}$       &  $43.204_{-0.007}^{+0.007}$   & $43.69$\\

   \hline
    \hline
    
   Red stat - No Var. && 19.50 & 19.78 &  19.85\\
     T       &  &   1042.74    &   1058.92     &  1063.43  \\

\hline

   Red stat - No C$_{AGN}$Var. && 1.83 &1.63 &  2.02\\
    T       &  &   47.20    &   35.83     &  57.69  \\
   \hline

   Red stat - No N$_{\rm H,los}$ Var. && 2.32&2.29 & 2.80 \\
     T       &  &   74.57    &   72.87    & 101.78  \\
   \hline
    \hline 

\end{tabular}

\begin{tablenotes}
      \footnotesize
     \textbf{Notes:} Same as Table \ref{tab:2MASXJ06411806+3249313_fitting}. $\theta_{i,uxclumpy}=arccos(cos(\theta_{i,borus02}))$. CTKcover is frozen as it is compatible with the hard limit of the available range.
    \end{tablenotes}
  
    \end{threeparttable}

\end{table*}

\begin{table*}
\centering
\begin{threeparttable}
\caption{X-ray fitting results for NGC 4992}
\renewcommand*{\arraystretch}{1.2}
\begin{tabular}{llccc}
  \textbf{Parameter} && \textbf{\borus}  & \textbf{\mytorus\ dec} & \textbf{\uxclumpy}\\  
\hline\hline
  stat/d.o.f         &         & 529.95/530 & 538.43/532  &   543.71/530 \\ 
    
    red stat       &         &     0.99 &     1.01    &  1.02\\

       T       &  &   $0.002 $   &    $0.30 $   &  $0.60 $ \\

       p-value  && 0.58& 0.38& 0.74 \\
    \hline
\hline

  $kT$    &    & $0.23_{-0.03}^{+0.04}$       &  $0.24_{-0.03}^{+0.03}$   & $0.24_{-0.03}^{+0.04}$\\

  apec norm$(\times 10^{-4} )$    &  & $0.92_{-0.29}^{+0.46}$       &  $0.91_{-0.28}^{+0.45}$   & $0.85_{-0.25}^{+0.43}$\\

    \hline

    $\Gamma$ &     & $1.68_{-0.12}^{+0.09}$       &  $1.68_{-0.07}^{+0.06}$   & $1.66_{-0.06}^{+0.11}$\\  
 
    N$_{\rm H,av}\times$10$^{24}$\,cm$^{-2}$  &&$1.86_{-0.34}^{+0.48}$       &  $2.00_{-0.32}^{+0.53}$   & $\dots$ \\

    C$_F$  &       & $0.67_{-0.20}^{+0.12}$     & $\dots$       & $\dots$  \\

    cos$(\theta_i)$              &    & $0.59_{-0.16}^{+0.55}$     & $\dots$  & $\dots$        \\
    
    $\theta_i$               &    & $\dots$  & $\dots$     & $>26.19$       \\
    
    CTKcover&  & $\dots$ &  $\dots$ &  $<0.30$\\

    TOR$\sigma$&  & $\dots$ & $\dots$ & $28.18_{-4.11}^{+18.32}$\\

    $A_{S,90}$         &          &  $\dots$      & $0.85_{-0.27}^{+0.69}$   & $\dots$   \\

    $A_{S,0}$           &  &  $\dots$   & $0.71_{-0.37}^{+0.35}$    & $\dots$  \\

     $F_s (\times 10^{-3} )$    & & $4.00_{-1.14}^{+1.85}$       &  $3.37_{-0.95}^{+1.26}$   & $10.73_{-6.39}^{+10.38}$\\

     Norm ($10^{-3}$)    && $3.34_{-0.81}^{+0.90}$       &  $6.63_{-0.70}^{+0.83}$   & $4.36_{-1.39}^{+2.00}$\\
    
    \hline

& \chandra\ - a        &  $1.09_{-0.37}^{+0.51}$   & $1.11_{-0.39}^{+0.55}$         & $1.04_{-0.34}^{+0.56}$\\
     
    &\suzaku\             & $1.12_{-0.13}^{+0.12}$       &  $1.17_{-0.13}^{+0.15}$   & $1.14_{-0.12}^{+0.19}$\\

    $C_{\rm AGN}$ & \xmm\ & $1.06_{-0.14}^{+0.08}$       &  $1.08_{-0.14}^{+0.16}$   & $1.01_{-0.15}^{+0.18}$ \\

    &  \nustar\        & 1       & 1   & 1\\
    
    & \chandra\ - b         & $0.92_{-0.20}^{+0.24}$       &  $0.92_{-0.20}^{+0.26}$   & $0.89_{-0.17}^{+0.21}$\\

     \hline
     
     & \chandra\ - a          & $61.74_{-14.17}^{+16.80}$       &  $64.03_{-15.03}^{+17.79}$   & $63.98_{-14.88}^{+22.90}$\\
        
         &  \suzaku\        & $52.04_{-2.65}^{+3.68}$       &  $54.72_{-2.74}^{+2.85}$   & $55.06_{-2.17}^{+9.87}$\\

    N$_{\rm H,los}\times$10$^{22}$\,cm$^{-2}$  &  \xmm\        & $58.24_{-4.17}^{+3.94}$       &  $60.41_{-4.22}^{+4.62}$   & $59.78_{-4.27}^{+7.40}$\\

 &  \nustar\        & $41.26_{-3.41}^{+4.87}$       &  $42.27_{-4.17}^{+4.02}$   & $45.14_{-7.15}^{+6.18}$\\
 
   & \chandra\ - b         & $39.86_{-5.45}^{+5.81}$       &  $38.87_{-5.45}^{+5.81}$   & $39.84_{-4.23}^{+6.24}$\\
 
   \hline
       \hline
    
   log L$_{NuSTAR, 2-10\,keV}$ & & $43.24_{-0.04}^{+0.01}$       &  $43.28_{-0.06}^{+0.07}$   & $43.50$\\
   log L$_{NuSTAR, 10-40\,keV}$ & & $43.22_{-0.06}^{+0.06}$       &  $43.18_{-0.05}^{+0.04}$   & $43.63$\\

   \hline
    \hline
    
   Red stat - No Var. &&1.08 & 1.31& 1.29 \\
     T       &  &   $2.17 $   &    $7.34 $   &  $6.83 $ \\
\hline

   Red stat - No C$_{AGN}$Var. && 0.99&1.02 &  1.03\\
    T       &  &   $0.01 $   &    $0.37 $   &  $0.73 $ \\
   \hline

   Red stat - No N$_{\rm H,los}$ Var. && 1.06& 1.07& 1.07 \\
    T       &  &   $1.35 $   &    $1.71 $   &  $1.61 $ \\
\hline
 \hline 

\end{tabular}

\begin{tablenotes}
      \footnotesize
     \textbf{Notes:} Same as Table \ref{tab:2MASXJ06411806+3249313_fitting}.
    \end{tablenotes}
  
    \end{threeparttable}
\label{table:ngc4992_fitting}
\end{table*}

\begin{table*}
\centering
\begin{threeparttable}
\caption{X-ray fitting results for ESO 383-18}
\renewcommand*{\arraystretch}{1.2}
\begin{tabular}{llccc}
  \textbf{Parameter} && \textbf{\borus}  & \textbf{\mytorus\ dec} & \textbf{\uxclumpy}\\  
 \hline\hline
   stat/d.o.f         && 1763.60/1477 &  1647.21/1474 &  1817.30/1477  \\ 
    
    red stat       &  &   1.19    &     1.11    &  1.23\\

    T       &  &  $7.45 $   &    $4.51 $   &  $8.85 $ \\

    p-value  && 0.43& 0.11& 0.80 \\
\hline
\hline

  $kT$   &  & $0.17_{-0.03}^{+0.03}$       &  $0.68_{-0.12}^{+0.10}$   & $0.22_{-0.04}^{+0.05}$\\

  apec norm$(\times 10^{-4} )$     &  & $0.11_{-0.06}^{+0.08}$       &  $0.05_{-0.01}^{+0.01}$   & $0.13_{-0.06}^{+0.06}$\\

    \hline

    $\Gamma$  &   & $1.52_{-0.04}^{+0.03}$ &  $1.79_{-0.03}^{+0.03}$   & $1.46_{-0.01}^{+0.03}$\\  
 
    N$_{\rm H,av}\times$10$^{24}$\,cm$^{-2}$  &&$0.97_{-0.10}^{+0.07}$       &  $1.00_{-0.05}^{+0.03}$   & $\dots$\\

    C$_F$   && $0.55_{-0.12}^{+0.02}$      & $\dots$       & $\dots$  \\

    cos$(\theta_i)$  && $<0.29$    & $\dots$  & $\dots$\\
    
    $\theta_i$  &    & $\dots$  & $\dots$     & $64.55_{-8.56}^{+16.59}$       \\
    
    CTKcover &  & $\dots$ &  $\dots$ &  $0*$\\

    TOR$\sigma$ &  & $\dots$ & $\dots$ & $7.90_{-1.17}^{+3.58}$\\

    $A_{S,90}$ &  &  $\dots$      & $0.06_{-0.02}^{+0.02}$   & $\dots$   \\

    $A_{S,0}$          &  &  $\dots$   & $7.65_{-1.08}^{+0.91}$    & $\dots$  \\

     $F_s (\times 10^{-3} )$  & & $8.90_{-1.31}^{+1.10}$ &  $2.42_{-1.10}^{+1.04}$   & $12.37_{-0.98}^{+0.98}$\\

     Norm ($10^{-3}$)      && $2.16_{-0.24}^{+0.16}$       &  $2.26_{-0.17}^{+0.17}$   & $2.63_{-0.30}^{+0.31}$\\
    
    \hline
     
    &\xmm\    & $1.27_{-0.04}^{+0.06}$       &  $1.35_{-0.07}^{+0.06}$   & $1.24_{-0.05}^{+0.06}$\\
 
    $C_{\rm AGN}$ & \nustar\--1 & 1       & 1   & 1\\
    
    & \nustar\--2        & $0.93_{-0.02}^{+0.02}$       &  $0.92_{-0.03}^{+0.03}$   & $0.91_{-0.02}^{+0.04}$\\
    
    & \chandra\          & $0.42_{-0.10}^{+0.08}$       &  $0.44_{-0.14}^{+0.32}$   & $0.41_{-0.14}^{+0.21}$\\

     \hline
         &  \xmm\        & $18.82_{-0.43}^{+0.94}$       &  $19.95_{-0.95}^{+0.88}$   & $17.76_{-0.71}^{+0.68}$\\

    N$_{\rm H,los}\times$10$^{22}$\,cm$^{-2}$  &\nustar\--1 & $16.03_{-1.01}^{+1.76}$       &  $13.79_{-1.53}^{+1.65}$   & $14.79_{-1.38}^{+1.51}$\\ 
 
 & \nustar\--2          & $17.17_{-0.77}^{+0.99}$       &  $15.14_{-0.95}^{+0.91}$   & $16.06_{-0.94}^{+0.96}$\\
 
   & \chandra\          & $18.72_{-3.03}^{+3.56}$       &  $19.95_{-4.17}^{+5.69}$   & $17.61_{-3.61}^{+4.48}$\\
 
   \hline
    \hline
    
   log L$_{NuSTAR, 2-10\,keV}$ & & $42.58_{-0.01}^{+0.03}$       &  $42.46_{-0.03}^{+0.03}$   & $42.69$\\
   log L$_{NuSTAR, 10-40\,keV}$ & & $42.74_{-0.02}^{+0.02}$       &  $41.81_{-0.05}^{+0.05}$   & $42.95$\\

   \hline
    \hline
    
   Red stat - No Var. && 1.31&1.38 & 1.49 \\
     T       &  &   $12.18 $   &    $14.79 $   &  $18.94 $ \\

\hline

   Red stat - No C$_{AGN}$Var. &&1.24 &1.25 &  1.35\\
      T       &  &   $9.36 $   &    $9.88 $   &  $13.43 $ \\
   \hline

   Red stat - No N$_{\rm H,los}$ Var. && 1.23& 1.18& 1.23 \\
    T       &  &   $9.04 $   &    $7.18 $   &  $9.11 $ \\
   \hline
\hline
\end{tabular}

\begin{tablenotes}
      \footnotesize
     \textbf{Notes:} Same as Table \ref{tab:2MASXJ06411806+3249313_fitting}.
    \end{tablenotes}
  
    \end{threeparttable}
\label{table:eso383_fitting}
\end{table*}

\begin{table*}
\centering
\begin{threeparttable}
\caption{X-ray fitting results for  MRK 417}
\renewcommand*{\arraystretch}{1.2}
\begin{tabular}{llccc}
  \textbf{Parameter} && \textbf{\borus}  & \textbf{\mytorus\ dec} & \textbf{\uxclumpy}\\  
\hline\hline
   stat/d.o.f         &   & 571.20/550  & 568.64/552  &   564.47/552 \\ 
    
    red stat       &         &    1.04   &    1.03     & 1.02 \\

   T       &  &   $0.90 $   &    $0.71 $   &  $0.53 $ \\

   p-value  && 0.07& 0.11& 0.07 \\
    
    \hline
\hline

  $kT$     &    & $0.79_{-0.09}^{+0.08}$       &  $0.79_{-0.09}^{+0.08}$   & $0.79_{-0.10}^{+0.08}$\\

  apec norm$(\times 10^{-4} )$      &  & $0.07_{-0.02}^{+0.02}$       &  $0.07_{-0.01}^{+0.02}$   & $0.06_{-0.02}^{+0.02}$\\

    \hline

    $\Gamma$    &     & $1.53_{-0.07}^{+0.06}$       &  $1.64_{-0.09}^{+0.07}$   & $1.64_{-0.08}^{+0.11}$\\  
 
    N$_{\rm H,av}\times$10$^{24}$\,cm$^{-2}$  &&$0.12_{-0.03}^{+0.06}$       &  $0.69_{-0.25}^{+0.32}$   &  $\dots$\\

    C$_F$    &       & $0.44_{-0.12}^{+0.12}$      & $\dots$       & $\dots$  \\

    cos$(\theta_i)$               &    & $<0.37$    & $\dots$  & $\dots$        \\
    
    $\theta_i$                &    & $\dots$  & $\dots$     & $56.06_{-16.60}^{+10.45}$       \\
    
    CTKcover &  & $\dots$ &  $\dots$ &  $0*$\\

    TOR$\sigma$ &  & $\dots$ & $\dots$ & $9.39_{-2.92}^{+6.77}$\\

    $A_{S,90}$         &          &  $\dots$      & $0.22_{-0.06}^{+1.04}$   & $\dots$   \\

    $A_{S,0}$   &  &  $\dots$   & $1.93_{-1.60}^{+0.87}$    & $\dots$  \\

     $F_s (\times 10^{-3} )$      & & $9.53_{-1.66}^{+3.11}$       &  $9.20_{-2.39}^{+3.03}$   & $14.34_{-8.12}^{+7.28}$\\

     Norm ($10^{-3}$)      && $2.11_{-0.42}^{+0.43}$       &  $2.13_{-0.42}^{+0.42}$   & $3.14_{-0.63}^{+1.27}$\\
    
    \hline
     
    &\xmm\             & $0.74_{-0.12}^{+0.15}$       &  $0.70_{-0.11}^{+0.12}$   & $0.54_{-0.06}^{+0.11}$\\

    $C_{\rm AGN}$& \suzaku\ & $1.54_{-0.20}^{+0.18}$       &  $1.47_{-0.17}^{+0.22}$   & $1.26_{-0.14}^{+0.12}$\\

    & \nustar\ & 1       & 1   & 1\\
    
    & \chandra\          & $0.80_{-0.18}^{+0.25}$       &  $0.81_{-0.17}^{+0.22}$   & $0.78_{-0.16}^{+0.20}$\\

     \hline
         &  \xmm\        & $55.18_{-5.34}^{+5.43}$       &  $55.10_{-6.40}^{+7.58}$   & $56.22_{-4.79}^{+4.82}$\\

    N$_{\rm H,los}\times$10$^{22}$\,cm$^{-2}$  & \suzaku\ & $49.33_{-3.20}^{+3.13}$       &  $47.56_{-3.85}^{+5.69}$   & $44.00_{-2.22}^{+2.30}$\\ 

& \nustar\          & $32.90_{-3.80}^{+3.44}$       &  $30.38_{-3.32}^{+3.92}$   & $30.48_{-3.94}^{+5.01}$\\
 
   & \chandra\          & $30.95_{-5.57}^{+6.50}$       &  $29.23_{-5.21}^{+6.48}$   & $27.99_{-4.73}^{+5.72}$\\
 
   \hline
       \hline
    
  log L$_{NuSTAR, 2-10\,keV}$ & & $43.42_{-0.08}^{+0.08}$       &  $43.32_{-0.07}^{+0.08}$   & $43.50$\\
   
  log L$_{NuSTAR, 10-40\,keV}$ & & $43.48_{-0.04}^{+0.03}$       &  $42.25_{-0.07}^{+0.09}$   & $43.65$\\

   \hline
    \hline
    
   Red stat - No Var. && 2.58& 2.58 &  2.57\\
    T       &  &   $37.23 $   &    $37.29 $   &  $37.06 $ \\

\hline

   Red stat - No C$_{AGN}$Var. && 1.22&1.23 &  1.21\\
     T       &  &   $5.11 $   &    $5.57 $   &  $4.98 $ \\
   \hline

   Red stat - No N$_{\rm H,los}$ Var. && 1.11&1.10 &1.10  \\
      T       &  &   $2.64 $   &    $2.57 $   &  $2.57 $ \\
\hline
\hline
\end{tabular}

\begin{tablenotes}
      \footnotesize
     \textbf{Notes:} Same as Table \ref{tab:2MASXJ06411806+3249313_fitting}.
    \end{tablenotes}
  
    \end{threeparttable}
\label{table:mkn417_fitting}
\end{table*}

\begin{table*}
\centering
\begin{threeparttable}
\caption{X-ray fitting results for MCG-01-05-047}
\renewcommand*{\arraystretch}{1.2}
\begin{tabular}{llccc}
  \textbf{Parameter} && \textbf{\borus}  & \textbf{\mytorus\ dec} & \textbf{\uxclumpy}\\  
\hline\hline
     $\chi^2$/d.o.f         &         & 624.05/529 & 565.25/531  &  642.27/530  \\ 
    
    red $\chi^2$       &         &    1.18   &     1.06   & 1.21 \\

   T       &  &   $4.13 $   &    $1.48 $   &  $4.87 $ \\

   p-value  && 0.003 & $3.01\times10^{-274}$ &  0.046\\
    
    \hline\hline

      $kT$      &    & --      &  --  & -- \\ 

    apec norm    &    & --      &  --  & --\\

    \hline
    
    $\Gamma$     &     & $1.61_{-0.03}^{+0.15}$       &  $1.66_{-0.15}^{+0.18}$   & $1.37_{-0.11}^{+0.07}$\\  
 
    N$_{\rm H,av}\times$10$^{24}$\,cm$^{-2}$  &&$>1.41$       &  $0.21_{-0.03}^{+0.05}$   & $\dots$\\

    C$_F$   &       & $0.69_{-0.40}^{+0.01}$      & $\dots$       & $\dots$  \\

    cos$(\theta_i)$            &    & $0.75_{-0.18}^{+0.04}$    & $\dots$  & $\dots$        \\
    
    $\theta_i$             &    & $\dots$  & $\dots$     & $0*$       \\
    
    CTKcover &  & $\dots$ &  $\dots$ &  $>0.21$\\

    TOR$\sigma$ &  & $\dots$ & $\dots$ & $27.23_{-2.52}^{+16.78}$\\

    $A_{S,90}$          &          &  $\dots$      & $0.15_{-0.03}^{+0.03}$   & $\dots$   \\

    $A_{S,0}$           &  &  $\dots$   & $4.77_{-1.96}^{+2.22}$    & $\dots$  \\
  
     $F_s (\times 10^{-3} )$      & & $19.61_{-9.76}^{+8.64}$       &  $3.55_{-3.47}^{+2.92}$   & $>31.83$\\

     Norm ($10^{-3}$)      && $1.08_{-0.10}^{+0.52}$       &  $4.81_{-0.88}^{+1.32}$   & $0.80_{-0.23}^{+0.21}$\\
    
    \hline
     
                    &\suzaku\     & $0.75_{-0.10}^{+0.10}$       &  $0.84_{-0.05}^{+0.05}$   & $0.86_{-0.08}^{+0.12}$\\

    $C_{\rm AGN}$   & \xmm\ & $1.47_{-0.20}^{+0.20}$       &  $0.35_{-0.04}^{+0.04}$   & $1.72_{-0.16}^{+0.22}$ \\

                    & \nustar\          & 1       & 1   & 1\\
    
                    & \chandra\          & $1.86_{-0.30}^{+0.33}$       &  $0.46_{-0.06}^{+0.11}$   & $2.20_{-0.26}^{+0.41}$\\

     \hline
                                            &  \suzaku\        & $11.24_{-1.36}^{+0.46}$       &  $>188.52$   & $9.13_{-0.72}^{+0.63}$\\

    N$_{\rm H,los}\times$10$^{22}$\,cm$^{-2}$   & \xmm\ &$14.15_{-1.45}^{+0.60}$       &  $15.39_{-1.55}^{+1.85}$   & $12.24_{-1.36}^{+1.11}$\\ 

                                            & \nustar\          & $14.71_{-4.24}^{+4.28}$       &  $234.25_{-42.94}^{+53.58}$   & $9.68_{-1.73}^{+2.59}$\\
 
                                             & \chandra\          & $16.30_{-1.78}^{+0.97}$       &  $18.66_{-2.31}^{+3.12}$   & $14.14_{-0.84}^{+2.00}$\\
 
   \hline
       \hline
    
   log L$_{NuSTAR, 2-10\,keV}$ & & $42.53_{-0.08}^{+0.06}$       &  $43.14_{-0.17}^{+0.29}$   & $42.54$\\
  log log L$_{NuSTAR, 10-40\,keV}$ & & $42.29_{-0.06}^{+0.05}$       &  $41.87_{-0.06}^{+0.12}$   & $42.85$\\

   \hline
    \hline
    
   Red stat - No Var. &&2.63 & 2.13&  2.12\\
     T       &  &   $37.72 $   &    $26.24 $   &  $26.00 $ \\

\hline

   Red stat - No C$_{AGN}$Var. &&1.79 & 1.31&  1.62\\
     T       &  &   $18.33 $   &    $7.28 $   &  $14.43 $ \\
   \hline

   Red stat - No N$_{\rm H,los}$ Var. && 1.42& 1.37&  1.27\\
    T       &  &   $9.84 $   &    $8.62 $   &  $6.32 $ \\
\hline
 \hline 
\end{tabular}

\begin{tablenotes}
      \footnotesize
     \textbf{Notes:} Same as Table \ref{tab:2MASXJ06411806+3249313_fitting}. \uxclumpy\ angle $\theta_i$ is frozen as it is compatible with the hard limit of the available range.
    \end{tablenotes}
  
    \end{threeparttable}
\label{table:mcg_fitting}
\end{table*}

\begin{table*}
\centering
\begin{threeparttable}
\caption{X-ray fitting results for ESO 103-35}
\renewcommand*{\arraystretch}{1.2}
\begin{tabular}{llccc}
  \textbf{Parameter} && \textbf{\borus}  & \textbf{\mytorus\ dec} & \textbf{\uxclumpy}\\  
\hline\hline
 
     $\chi^2$/d.o.f    &   & 3961.46/3868 & 3926.45/3869  &  3947.50/3870  \\ 
    
    red $\chi^2$     &         &    1.02   &    1.01     & 1.02 \\

       T       &  &   $0.92 $   &    $1.50 $   &  $1.24 $ \\

       p-value  && 0.07& 0.06& 0.02 \\
    \hline
\hline

    $kT$    &    & $0.73_{-0.05}^{+0.04}$       &  $0.74_{-0.04}^{+0.04}$   & $0.73_{-0.05}^{+0.04}$\\

    apec norm$(\times 10^{-4} )$    & \xmm\ & $0.03_{-0.02}^{+0.02}$       &  $0.07_{-0.02}^{+0.02}$   & $0.03_{-0.02}^{+0.02}$\\

            &\suzaku\ & $0.33_{-0.03}^{+0.03}$       &  $0.37_{-0.03}^{+0.02}$   & $0.33_{-0.03}^{+0.03}$\\
    
    \hline
    
    $\Gamma$     &     & $1.81_{-0.01}^{+0.02}$       &  $1.95_{-0.02}^{+0.02}$   & $1.81_{-0.02}^{+0.03}$\\  
 
    N$_{\rm H,av}\times$10$^{24}$\,cm$^{-2}$  &&$2.13_{-0.28}^{+0.29}$       &  $0.60_{-0.21}^{+0.29}$   & $\dots$\\

    C$_F$   &       & $0.56_{-0.03}^{+0.02}$      & $\dots$       & $\dots$  \\

    cos$(\theta_i)$              &    & $0.49_{-0.02}^{+0.02}$    & $\dots$  & $\dots$     \\
    
    $\theta_i$              &    & $\dots$  & $\dots$     & $79.81_{-10.73}^{+8.30}$       \\
    
    CTKcover&  & $\dots$ &  $\dots$ &  $0.34_{-0.03}^{+0.05}$\\

    TOR$\sigma$ &  & $\dots$ & $\dots$ & $>81.83$\\

    $A_{S,90}$       &          &  $\dots$      & $0.30_{-0.03}^{+0.04}$   & $\dots$   \\

    $A_{S,0}$          &  &  $\dots$   & $1.67_{-0.12}^{+0.14}$    & $\dots$  \\

     $F_s (\times 10^{-3} )$      & & $1.06_{-0.19}^{+0.26}$       &  $<0.10$   & $10.07_{-16.373}^{+4.66}$\\

     Norm ($10^{-3}$)       && $14.20_{-0.55}^{+0.81}$       &  $17.75_{-1.21}^{+1.29}$   & $18.03_{-0.85}^{+0.11}$\\
    
    \hline
     
        &\xmm\             & $1.37_{-0.04}^{+0.04}$       &  $1.37_{-0.04}^{+0.04}$   & $1.39_{-0.04}^{+0.04}$\\

    $C_{\rm AGN}$ &\suzaku\     & $1.42_{-0.03}^{+0.03}$       &   $1.43_{-0.03}^{+0.03}$  & $1.44_{-0.03}^{+0.04}$\\
    
        & \nustar\ 1  & 1       & 1   & 1\\
    
        & \nustar\ 2          & $1.02_{-0.02}^{+0.02}$       &  $1.01_{-0.02}^{+0.02}$   & $1.02_{-0.02}^{+0.02}$\\

     \hline
     
        & \xmm\         & $19.97_{-0.39}^{+0.40}$       &  $20.84_{-0.44}^{+0.46}$   & $18.84_{-0.32}^{+0.31}$\\
      
    N$_{\rm H,los}\times$10$^{22}$\,cm$^{-2}$     &     \suzaku\     & $20.04_{-0.27}^{+0.29}$       &  $20.91_{-0.34}^{+0.35}$   & $18.76_{-0.21}^{+0.22}$\\

        & \nustar\ 1 & $17.56_{-0.58}^{+0.69}$       &  $18.04_{-0.67}^{+0.73}$   & $16.37_{-0.61}^{+0.59}$\\ 
 
        &   \nustar\ 2         & $18.91_{-0.55}^{+0.65}$       &  $19.39_{-0.64}^{+0.70}$   & $17.62_{-0.54}^{+0.52}$\\

   \hline
       \hline
    
   log L$_{NuSTAR, 2-10\,keV}$ & & $43.26_{-0.001}^{+0.001}$       &  $43.41_{-0.02}^{+0.01}$   & $43.47$\\
   
   log L$_{NuSTAR, 10-40\,keV}$ & & $43.20_{-0.01}^{+0.01}$       &  $43.01_{-0.01}^{+0.01}$   & $43.38$\\

   \hline
    \hline
    
   Red stat - No Var. && 1.56&1.63 & 1.65 \\
    T       &  &   $34.96 $   &    $39.25 $   &  $40.76 $ \\

\hline

   Red stat - No C$_{AGN}$Var. &&1.17 &1.17 &  1.19\\
     T       &  &   $10.80 $   &    $10.88 $   &  $12.18 $ \\
   \hline

   Red stat - No N$_{\rm H,los}$ Var. &&1.08 &1.03 &  1.09\\
     T       &  &   $4.94 $   &    $1.80 $   &  $5.58 $ \\

\hline
\hline
\end{tabular}

\begin{tablenotes}
      \footnotesize
     \textbf{Notes:} Same as Table \ref{tab:2MASXJ06411806+3249313_fitting}.
    \end{tablenotes}
  
    \end{threeparttable}
\label{table:eso103_fitting}
\end{table*}

\begin{table*}
\centering
\begin{threeparttable}
\caption{X-ray fitting results for NGC 1142}
\renewcommand*{\arraystretch}{1.2}
\begin{tabular}{llccc}
  \textbf{Parameter} && \textbf{\borus}  & \textbf{\mytorus\ dec} & \textbf{\uxclumpy}\\  
\hline\hline
   $\chi^2$/d.o.f         &         &  724.33/680 &  720.40/680 &    723.91/680 \\ 
    
    red $\chi^2$       &         &  1.06      &   1.06       & 1.06 \\

   T       &  &   $1.69 $   &    $1.55 $   &  $1.68 $ \\

   p-value  && $1.04\times10^{-9}$& $1.28\times10^{-15}$& $7.88\times10^{-45}$ \\
    
    \hline
\hline

  $kT$    &    & $0.75_{-0.03}^{+0.03}$      & $0.75_{-0.03}^{+0.03}$     & $0.76_{-0.03}^{+0.04}$      \\   

    &\xmm &   $0.11_{-0.02}^{+0.02}$     &    $0.11_{-0.02}^{+0.02}$     & $0.09_{-0.02}^{+0.01}$\\
     
    apec norm$(\times 10^{-4} )$      & \suzaku\ -- 2   &   $0.27_{-0.02}^{+0.02}$      &  $0.27_{-0.02}^{+0.02}$      & $0.26_{-0.02}^{+0.02}$\\
     
     &\suzaku\ -- 1        & $0.24_{-0.03}^{+0.03}$         &   $0.25_{-0.03}^{+0.03}$     & $0.23_{-0.03}^{+0.04}$\\

    \hline

    $\Gamma$    &     & $1.64_{-0.08}^{+0.08}$      & $1.65_{-0.05}^{+0.09}$    & $2.08_{-0.15}^{+0.07}$   \\   
 
   N$_{\rm H,av}\times$10$^{24}$\,cm$^{-2}$  & & $0.29_{-0.02}^{+0.03}$       & $0.22_{-0.01}^{+0.02}$     & $\dots$\\

    C$_F$  &       & $0.40_{-0.05}^{+0.03}$      & $\dots$       & $\dots$  \\

    cos$(\theta_i)$             &    & $<0.22$    & $\dots$  & $\dots$        \\

     $\theta_i$              &    & $\dots$  & $\dots$     & $>77.29$       \\
    
    CTKcover &  & $\dots$ &  $\dots$ & $<0.26$\\

    TOR$\sigma$ &  & $\dots$ & $\dots$ & $25.71_{-8.10}^{+31.17}$\\

    $A_{S,90}$         &          &  $\dots$      & $1.09_{-0.18}^{+0.08}$   & $\dots$   \\

    $A_{S,0}$          &  &  $\dots$   & $<0.15$    & $\dots$  \\
 
     $F_s (\times 10^{-3} )$    &  &  $5.29_{-0.99}^{+1.14}$ & $3.72_{-0.88}^{+1.17}$ & $6.96_{-4.39}^{+13.08}$  \\

     Norm ($10^{-3}$)       &&  $4.08_{-0.86}^{+1.37}$ &  $4.55_{-1.11}^{+1.51}$  &  $16.70_{-2.97}^{+4.17}$ \\
    
    \hline
     
    &\xmm\             & $1.15_{-0.15}^{+0.19}$        & $1.15_{-0.16}^{+0.20}$      & $0.83_{-0.19}^{+0.22}$    \\

    $C_{\rm AGN}$ &   \suzaku\ -- 1         & $1.87_{-0.22}^{+0.28}$        & $1.88_{-0.23}^{+0.30}$   & $1.32_{-0.26}^{+0.36}$\\
    
    & \suzaku\ -- 2         & $1.93_{-0.23}^{+0.30}$       & $1.93_{-0.24}^{+0.31}$   & $1.44_{-0.27}^{+0.32}$\\

    &   \nustar\           & $1$        & $1$     & $1$         \\

     \hline
         &  \xmm\        & $55.63_{-3.55}^{+3.83}$  &$56.06_{-3.11}^{+3.16}$ & $56.15_{-1.89}^{+1.95}$\\

    N$_{\rm H,los}\times$10$^{22}$\,cm$^{-2}$  & \suzaku\ -- 1    &$67.30_{-3.58}^{+3.72}$     & $68.19_{-3.08}^{+2.56}$   & $66.45_{-2.74}^{+2.29}$\\

    & \suzaku\ -- 2    & $92.42_{-5.39}^{+5.62}$   &$92.16_{-2.44}^{+5.09}$ & $90.90_{-5.85}^{+5.04}$\\ 
    
    &  \nustar\  & $146.84_{-16.24}^{+15.72}$   &$147.22_{-13.08}^{+11.81}$ & $174.37_{-26.01}^{+27.91}$\\

   \hline
       \hline
    
   log L$_{NuSTAR, 2-10\,keV}$ & & $43.51_{-0.01}^{+0.01}$       &  $43.57_{-0.10}^{+0.08}$   & $43.82$\\
   log L$_{NuSTAR, 10-40\,keV}$ & & $42.96_{-0.08}^{+0.07}$       &  $43.20_{-0.06}^{+0.06}$   & $43.69$\\

   \hline
    \hline
    
   Red stat - No Var. && 7.12& 7.24& 7.39 \\
      T       &  &   $160.52 $   &    $163.70 $   &  $167.41 $ \\

\hline

   Red stat - No C$_{AGN}$Var. && 1.36& 1.35&  1.25\\
     T       &  &   $9.55 $   &    $9.06 $   &  $6.47 $ \\
   \hline

   Red stat - No N$_{\rm H,los}$ Var. && 1.36& 1.35&  1.38\\
      T       &  &   $9.57 $   &    $9.17 $   &  $10.02 $ \\
\hline
 \hline 

\end{tabular}

\begin{tablenotes}
      \footnotesize
     \textbf{Notes:} Same as Table \ref{tab:2MASXJ06411806+3249313_fitting}.
    \end{tablenotes}
  
    \end{threeparttable}
\label{table:NGC 1142_fitting}
\end{table*}

\begin{table*}
\centering
\begin{threeparttable}
\caption{X-ray fitting results for IRAS 16288+3929}
\renewcommand*{\arraystretch}{1.2}
\begin{tabular}{llccc}
  \textbf{Parameter} && \textbf{\borus}  & \textbf{\mytorus\ dec} & \textbf{\uxclumpy}\\  
\hline\hline
  
    $\chi^2$/d.o.f         &         &  276.17/265 & 277.37/267  &  294.25/265  \\ 
    
    red $\chi^2$      &         &  1.04     &     1.04   &  1.11\\

       T       &  &   $0.68 $   &    $0.63 $   &  $1.79 $ \\

 p-value  && $2.77\times10^{-09}$& $2.02\times10^{-12}$& $4.60\times10^{-10}$ \\
    \hline
\hline

  $kT$      &    & $0.71_{-0.09}^{+0.07}$       &  $0.72_{-0.08}^{+0.07}$   & $0.71_{-0.08}^{+0.07}$\\

  apec norm$(\times 10^{-6} )$    &  & $4.31_{-0.75}^{+0.75}$       &  $4.21_{-0.93}^{+0.87}$   & $4.48_{-0.85}^{+0.79}$\\

    \hline

    $\Gamma$   &     & $2.20_{-0.08}^{+0.17}$       &  $2.19_{-0.11}^{+0.11}$   & $2.33_{-0.14}^{+0.23}$\\  
 
    N$_{\rm H,av}\times$10$^{24}$\,cm$^{-2}$  &&$0.29_{-0.06}^{+0.07}$       &  $0.29_{-0.06}^{+0.06}$   & $\dots$ \\

    C$_F$    &       & $0.20_{-0.09}^{+0.37}$      & $\dots$       & $\dots$  \\

    cos$(\theta_i)$               &    & $0.05$    & $\dots$  & $\dots$        \\
    
    $\theta_i$               &    & $\dots$  & $\dots$     & $0*$       \\
    
    CTKcover &  & $\dots$ &  $\dots$ &  $>0.57$\\

    TOR$\sigma$ &  & $\dots$ & $\dots$ & $70.00_{-5.94}^{+7.99}$\\

    $A_{S,90}$         &          &  $\dots$      & $0.29_{-0.12}^{+0.09}$   & $\dots$   \\

    $A_{S,0}$          &  &  $\dots$   & $<0.11$    & $\dots$  \\

     $F_s (\times 10^{-3} )$     & & $0.007_{-0.002}^{+0.003}$       &  $0.82_{-0.37}^{+0.47}$   & $11.57_{-5.54}^{+14.02}$\\

     Norm ($10^{-3}$)      && $5.74_{-1.92}^{+2.61}$       &  $9.76_{-3.17}^{+4.81}$   & $12.43_{-3.92}^{+15.42}$\\
    
    \hline
     
    &\suzaku           & $5.65_{-0.91}^{+1.03}$       &  $5.59_{-1.65}^{+2.08}$   & $4.03_{-0.87}^{+0.78}$  \\

    $C_{\rm AGN}$ & \xmm\ & $0.77_{-0.12}^{+0.12}$       &  $0.78_{-0.18}^{+0.23}$   & $0.81_{-0.28}^{+0.15}$ \\
    
    & \nustar          & 1       & 1   & 1\\

     \hline
         &  \suzaku       & $155.61_{-15.63}^{+22.63}$       &  $151.73_{-16.85}^{+19.00}$   & $131.18_{-16.03}^{+13.29}$\\

    N$_{\rm H,los}\times$10$^{22}$\,cm$^{-2}$  &\xmm\ & $62.22_{-4.79}^{+4.88}$       &  $61.04_{-3.30}^{+3.84}$   & $ 56.20_{-1.67}^{+3.59}$\\ 
 
   & \nustar               &  $85.12_{-7.65}^{+8.47}$ & $83.36_{-10.23}^{+12.19}$  & $77.19_{-7.22}^{+20.68}$\\
 
   \hline
       \hline
    
   log L$_{NuSTAR, 2-10\,keV}$ & & $43.57_{-0.09}^{+0.13}$       &  $43.58_{-0.13}^{+0.14}$   & $43.61$\\
   log L$_{NuSTAR, 10-40\,keV}$ & & $43.12_{-0.48}^{+0.09}$       &  $43.36_{-0.11}^{+0.11}$   & $43.30$\\

   \hline
    \hline
    
   Red stat - No Var. &&1.58 &1.62 & 1.59 \\
    T       &  &   $9.50 $   &    $10.16 $   &  $9.64 $ \\

\hline

   Red stat - No C$_{AGN}$Var. && 1.36& 1.36&  1.32\\
      T       &  &   $5.89 $   &    $5.88 $   &  $5.20 $ \\
   \hline

   Red stat - No N$_{\rm H,los}$ Var. &&1.30 & 1.31&  1.29\\
      T       &  &   $4.95 $   &    $5.12 $   &  $4.75 $ \\
\hline
 \hline 

\end{tabular}

\begin{tablenotes}
      \footnotesize
     \textbf{Notes:} Same as Table \ref{tab:2MASXJ06411806+3249313_fitting}.
    \end{tablenotes}
  
    \end{threeparttable}
\label{table:iras_fitting}
\end{table*}

\begin{table*}
\centering
\begin{threeparttable}
\caption{X-ray fitting results for ESO 263-13}
\renewcommand*{\arraystretch}{1.2}
\begin{tabular}{llccc}
  \textbf{Parameter} && \textbf{\borus}  & \textbf{\mytorus\ dec} & \textbf{\uxclumpy}\\  
\hline\hline
 
     $\chi^2$/d.o.f         &         & 862.24/685 & 859.47/686  &  797.37/684  \\ 
    
    red $\chi^2$      &         &   1.31    &    1.25     & 1.16 \\

   T       &  &   $6.77 $   &    $6.62 $   &  $4.33 $ \\

      p-value  && $2.87\times10^{-42}$& $8.10\times10^{-51}$& $6.89\times10^{-89}$ \\
    
    \hline
\hline

  $kT$    &    & $0.78_{-0.06}^{+0.05}$       &  $0.78_{-0.06}^{+0.06}$   & $0.78_{-0.05}^{+0.05}$\\

  apec norm$(\times 10^{-6} )$      &  & $8.98_{-1.65}^{+1.66}$       &  $8.74_{-1.57}^{+1.58}$   & $10.42_{-1.13}^{+1.10}$\\

    \hline

    $\Gamma$   &     & $1.65_{-0.06}^{+0.06}$       &  $1.69_{-0.05}^{+0.05}$   & $1.59_{-0.08}^{+0.06}$\\  
 
    N$_{\rm H,av}\times$10$^{24}$\,cm$^{-2}$  &&$0.09_{-0.02}^{+0.02}$       &  $0.08_{-0.02}^{+0.02}$   & $\dots$\\

    C$_F$   &       & $0.04_{-0.08}^{+0.14}$      & $\dots$       & $\dots$  \\

    cos$(\theta_i)$             &    & $0.05_u^u$    & $\dots$  & $\dots$        \\
    
    $\theta_i$        &    & $\dots$  & $\dots$     & $44.58_{-4.62}^{+3.68}$       \\
    
    CTKcover &  & $\dots$ &  $\dots$ &  $0.25_{-0.02}^{+0.01}$\\

    TOR$\sigma$ &  & $\dots$ & $\dots$ & $7.03_{-0.22}^{+1.19}$\\

    $A_{S,90}$         &          &  $\dots$      & $1.26_{-0.32}^{+0.35}$   & $\dots$   \\

    $A_{S,0}$          &  &  $\dots$   & $0*$    & $\dots$  \\

     $F_s (\times 10^{-3} )$      & & $9.99_{-2.41}^{+2.79}$       &  $9.42_{-2.25}^{+2.82}$   & $15.13_{-0.96}^{+2.10}$\\

     Norm ($10^{-3}$)      && $3.51_{-0.74}^{+0.83}$       &  $3.95_{-0.74}^{+0.86}$   & $3.21_{-0.33}^{+0.41}$\\
    
    \hline
     
    &\suzaku\             & $1.02_{-0.13}^{+0.16}$       &  $1.03_{-0.13}^{+0.15}$   & $1.67_{-0.17}^{+0.32}$\\

    $C_{\rm AGN}$ &\xmm\      & $0.61_{-0.08}^{+0.10}$       &  $0.62_{-0.08}^{+0.10}$   & $0.63_{-0.11}^{+0.11}$\\
    
    & \nustar\ & 1       & 1   & 1\\

     \hline
         &  \suzaku\        & $30.09_{-1.54}^{+1.72}$       &  $30.80_{-1.64}^{+1.66}$   & $26.85_{-0.56}^{+0.40}$\\

    N$_{\rm H,los}\times$10$^{22}$\,cm$^{-2}$  &  \xmm\               &  $26.01_{-0.97}^{+1.18}$ & $25.45_{-1.06}^{+1.07}$  & $20.68_{-0.75}^{+0.75}$\\
 
   & \nustar\          & $81.27_{-8.29}^{+8.99}$       &  $81.08_{-8.08}^{+6.60}$   & $56.04_{-2.12}^{+3.62}$\\
 
   \hline
       \hline
    
   log L$_{NuSTAR, 2-10\,keV}$ & & $43.53_{-0.13}^{+0.14}$       &  $43.56_{-0.11}^{+0.11}$   & $44.22$\\
   log L$_{NuSTAR, 10-40\,keV}$ & & $43.69_{-0.01}^{+0.01}$       &  $43.71_{-0.01}^{+0.01}$   & $44.39$\\

   \hline
    \hline
    
   Red stat - No Var. && 4.52& 4.59 & 4.57 \\
    T       &  &   $92.43 $   &    $94.31 $   &  $93.73v $ \\

\hline

   Red stat - No C$_{AGN}$Var. && 1.42&1.49 & 1.41 \\
      T       &  &   $11.11 $   &    $12.99 $   &  $10.93 $ \\
   \hline

   Red stat - No N$_{\rm H,los}$ Var. && 1.62&1.71 & 1.44 \\
     T       &  &   $16.29 $   &    $18.66 $   &  $11.69 $ \\
\hline
 \hline 

\end{tabular}

\begin{tablenotes}
      \footnotesize
     \textbf{Notes:} Same as Table \ref{tab:2MASXJ06411806+3249313_fitting}.
    \end{tablenotes}
  
    \end{threeparttable}
\label{table:eso263_fitting}
\end{table*}

\begin{table*}
\centering
\begin{threeparttable}
\caption{X-ray fitting results for Fairall 272}
\renewcommand*{\arraystretch}{1.2}
\begin{tabular}{llccc}
  \textbf{Parameter} && \textbf{\borus}  & \textbf{\mytorus\ dec} & \textbf{\uxclumpy}\\  
\hline\hline
    stat/d.o.f         &         & 590.59/616 & 584.99/616  &   593.97/617  \\ 
    
    red stat       &         &    0.96    &     0.95    & 0.96 \\

       T       &  &   $0.96 $   &    $0.95 $   &  $0.93 $ \\

   p-value  && $3.18\times10^{-4}$& $2.61\times10^{-12}$& 0.01 \\
    \hline
\hline

  $kT$     &    & $0.75_{-0.13}^{+0.11}$       & $0.79_{-0.09}^{+0.08}$   & $0.77_{-0.12}^{+0.09}$\\

  apec norm$(\times 10^{-4} )$    &  & $0.06_{-0.01}^{+0.01}$& $0.08_{-0.01}^{+0.01}$    & $0.07_{-0.01}^{+0.02}$\\

    \hline

    $\Gamma$      &     & $<1.52$       & $1.53_{-0.09}^{+0.11}$   & $1.56_{-0.06}^{+0.04}$\\  
 
    N$_{\rm H,av}\times$10$^{24}$\,cm$^{-2}$  && $0.65_{-0.11}^{+0.37}$       & $0.74_{-0.23}^{+0.37}$      & $\dots$\\

    C$_F$  &       & $0.65_{-0.20}^{+0.07}$      & $\dots$       & $\dots$  \\

    cos$(\theta_i)$               &    & $<0.40$    & $\dots$  & $\dots$        \\
    
    $\theta_i$       &    & $\dots$  & $\dots$     & $<49.45$       \\
    
    CTKcover&  & $\dots$ &  $\dots$ & $<0.16$\\

    TOR$\sigma$ &  & $\dots$ & $\dots$ & $21.27_{-5.84}^{+5.99}$\\

    $A_{S,90}$        &          &  $\dots$      & $0.23_{-0.07}^{+0.11}$   & $\dots$   \\

    $A_{S,0}$         &  &  $\dots$   & $3.49_{-1.33}^{+1.68}$    & $\dots$  \\

     $F_s (\times 10^{-3} )$    & & $6.03_{-1.59}^{+1.72}$       & $<4.14$   & $5.84_{-2.06}^{+37.76}$\\

     Norm ($10^{-3}$)      && $1.72_{-0.27}^{+0.02}$       & $1.75_{-0.39}^{+0.66}$   & $3.89_{-0.77}^{+1.00}$\\
    
    \hline
     
    &\xmm\             & $0.78_{-0.07}^{+0.07}$       & $0.82_{-0.07}^{+0.09}$   & $0.78_{-0.09}^{+0.08}$\\

    $C_{\rm AGN}$ & \nustar\ & 1       & 1   & 1\\
    
    & \chandra\          & $0.26_{-0.08}^{+0.13}$       & $0.27_{-0.08}^{+0.12}$   & $0.24_{-0.08}^{+0.12}$\\

     \hline
         &  \xmm\        & $28.09_{-0.98}^{+2.26}$       & $32.02_{-2.94}^{+3.53}$   & $30.39_{-3.06}^{+3.01}$\\

    N$_{\rm H,los}\times$10$^{22}$\,cm$^{-2}$ & \nustar\ & $13.92_{-1.22}^{+1.46}$       & $12.96_{-1.67}^{+1.79}$   & $15.06_{-1.38}^{+1.80}$\\ 
 
   & \chandra\          & $13.42_{-4.22}^{+6.01}$       & $13.11_{-4.46}^{+6.63}$   & $12.90_{-4.11}^{+6.21}$\\
 
   \hline
       \hline
    
   log L$_{NuSTAR, 2-10\,keV}$ & & $43.04_{-0.02}^{+0.01}$       &  $42.92_{-0.07}^{+0.08}$   & $43.32$\\
   log L$_{NuSTAR, 10-40\,keV}$ & & $43.38_{-0.03}^{+0.01}$       &  $42.76_{-0.10}^{+0.14}$   & $43.51$\\

   \hline
    \hline
    
   Red stat - No Var. && 2.38& 2.34&  2.34\\
   T       &  &   $33.39 $   &    $33.36 $   &  $33.54 $ \\

\hline

   Red stat - No C$_{AGN}$Var. &&1.00 & 0.97&  1.01\\
     T       &  &   $0.13 $   &    $0.63 $   &  $0.30 $ \\
   \hline

   Red stat - No N$_{\rm H,los}$ Var. && 1.14& 1.14&  1.12\\
    T       &  &   $3.82 $   &    $3.57 $   &  $3.07 $ \\
\hline
 \hline 

\end{tabular}

\begin{tablenotes}
      \footnotesize
     \textbf{Notes:} Same as Table \ref{tab:2MASXJ06411806+3249313_fitting}.
    \end{tablenotes}
  
    \end{threeparttable}
\label{table:fairall_fitting}
\end{table*}

\begin{table*}
\centering
\begin{threeparttable}
\caption{X-ray fitting results for LEDA 2816387}
\label{table:leda_fitting}
\renewcommand*{\arraystretch}{1.2}
\begin{tabular}{llccc}
  \textbf{Parameter} && \textbf{\borus}  & \textbf{\mytorus\ dec} & \textbf{\uxclumpy}\\  
\hline\hline
  
     $\chi^2$/d.o.f         &         &344.97/305  & 346.23/312  &  345.28/313  \\ 
    
    red $\chi^2$      &         &    1.13   &    1.11     & 1.10  \\

       T       &  &   $2.29 $   &    $1.93 $   &  $1.82 $ \\

 p-value  && 0.42& 0.15& 1\\
    \hline
\hline

  $kT$     &    & $0.90_{-0.11}^{+0.12}$       &  $0.90_{-0.11}^{+0.12}$   & $0.89_{-0.13}^{+0.13}$\\ 

  apec norm$(\times 10^{-6} )$     &  & $6.60_{-1.89}^{+2.08}$       &  $6.59_{-1.59}^{+1.90}$   & $5.51_{-1.63}^{+1.87}$\\

    \hline

    $\Gamma$     &     & $<1.4$       &  $<1.4$   & $1.4*$\\  
 
    N$_{\rm H,av}\times$10$^{24}$\,cm$^{-2}$  &&$0.05_{-0.03}^{+0.03}$       &  $0.09_{-0.05}^{+0.14}$   & $\dots$\\

    C$_F$    &       & $0.25_{-0.08}^{+0.19}$     & $\dots$       & $\dots$  \\

    cos$(\theta_i)$             &    & $<0.05$    & $\dots$  & $\dots$        \\
    
    $\theta_i$             &    & $\dots$  & $\dots$     & $90*$       \\
    
    CTKcover &  & $\dots$ &  $\dots$ &  $<0.27$\\

    TOR$\sigma$ &  & $\dots$ & $\dots$ & $7.03_{-2.23}^{+6.36}$\\

    $A_{S,90}$         &          &  $\dots$      & $0.22_{-0.15}^{+1.66}$   & $\dots$   \\

    $A_{S,0}$         &  &  $\dots$   & $0.34_{-0.4}^{+0.95}$    & $\dots$  \\

     $F_s (\times 10^{-3} )$   & & $5.72_{-2.82}^{+4.59}$ &  $5.13_{-3.22}^{+5.13}$   & $15.48_{-9.07}^{+13.96}$\\

Norm ($10^{-3}$)     && $1.08_{-0.12}^{+0.03}$       &  $1.16_{-0.13}^{+0.03}$   & $1.34_{-0.17}^{+0.41}$\\
    
    \hline
     
    &\nustar\ - 1             & 1       & 1   & 1\\

    $C_{\rm AGN}$ & \xmm\ & $1.43_{-0.16}^{+0.30}$       &  $1.19_{-0.08}^{+0.22}$   & $0.91_{-0.26}^{+0.32}$\\
    
    & \nustar\ - 2          & $0.98_{-0.17}^{+0.21}$       &  $0.99_{-0.16}^{+0.20}$   & $0.98_{-0.15}^{+0.18}$\\

     \hline
         &  \nustar\ - 1        & $66.55_{-7.04}^{+7.71}$       &  $64.28_{-7.07}^{+7.55}$   & $62.97_{-7.90}^{+7.37}$\\

    N$_{\rm H,los}\times$10$^{22}$\,cm$^{-2}$  &\xmm\ & $89.55_{-10.15}^{+12.37}$       &  $97.41_{-9.57}^{+13.59}$   & $75.42_{-11.68}^{+13.48}$\\ 
 
   & \nustar\ - 2          & $76.32_{-9.73}^{+10.75}$       &  $75.46_{-8.98}^{+10.41}$   & $73.98_{-8.66}^{+9.12}$\\
 
       \hline
    
   log L$_{NuSTAR, 2-10\,keV}$ & & $44.28_{-0.09}^{+0.01}$       &  $44.26_{-0.05}^{+0.02}$   & $44.38$\\
   log L$_{NuSTAR, 10-40\,keV}$ & & $44.58_{-0.05}^{+0.05}$       &  $44.57_{-0.05}^{+0.05}$   & $44.68$\\

   \hline
    \hline
    
   Red stat - No Var. && 1.29 &1.26 & 1.24 \\
     T       &  &   $5.08 $   &    $4.66 $   &  $4.26 $ \\

\hline

   Red stat - No C$_{AGN}$Var. && 1.14& 1.24& 1.10 \\
     T       &  &   $2.42 $   &    $4.20 $   &  $1.73 $ \\
   \hline

   Red stat - No N$_{\rm H,los}$ Var. &&1.15 & 1.17& 1.11 \\
      T       &  &   $2.62 $   &    $3.02 $   &  $1.91 $ \\
\hline
 \hline 

\end{tabular}

\begin{tablenotes}
      \footnotesize
     \textbf{Notes:} Same as Table \ref{tab:2MASXJ06411806+3249313_fitting}. $\theta_{i,uxclumpy}=arccos(cos(\theta_{i,borus02}))$
    \end{tablenotes}
    \end{threeparttable}

\end{table*}

\clearpage 

\section{Sources spectra}\label{sec:allspectra}
 In this section, we present the multi-epoch best-fit spectra of the sources analyzed in this work, as modeled by \borus. We chose to show only \borus\ best-fit spectra for clarity purposes. The spectra should be read as follows:
\begin{itemize}
    \item Considering we are simultaneously fitting all the observations available for a single source, all the observations for a single source are shown together, each with a different color. 
    \item All detectors in the same telescope are colored the same for each observation.
    \item \nustar\  observations are colored chronologically, as listed in Table \ref{tab:all_observ}. The color order is as follows, from the first to the last: blue, green.
    \item \chandra\ observations are colored chronologically, as listed in Table \ref{tab:all_observ}. The color order is as follows, from the first to the last: magenta, orange.
    \item \suzaku\ observations are colored chronologically, as listed in Table \ref{tab:all_observ}. The color order is as follows, from the first to the last: red, cyan.
    \item \xmm\ observations are colored chronologically, as listed in Table \ref{tab:all_observ}. The color order is as follows, from the first to the last: purple, gray.
    \item A solid black line highlights the model's best-fit.
    \item The model components have the same color as the observation to which they correspond but different line styles. Reflection: dashed, Line-of-sight: dash-dotted, Scattering: dash-dot-dot-dot, \texttt{apec}: dotted.

\end{itemize}
\clearpage

\begin{figure}[ht]
    \centering
    \includegraphics[ width=0.45\textwidth]{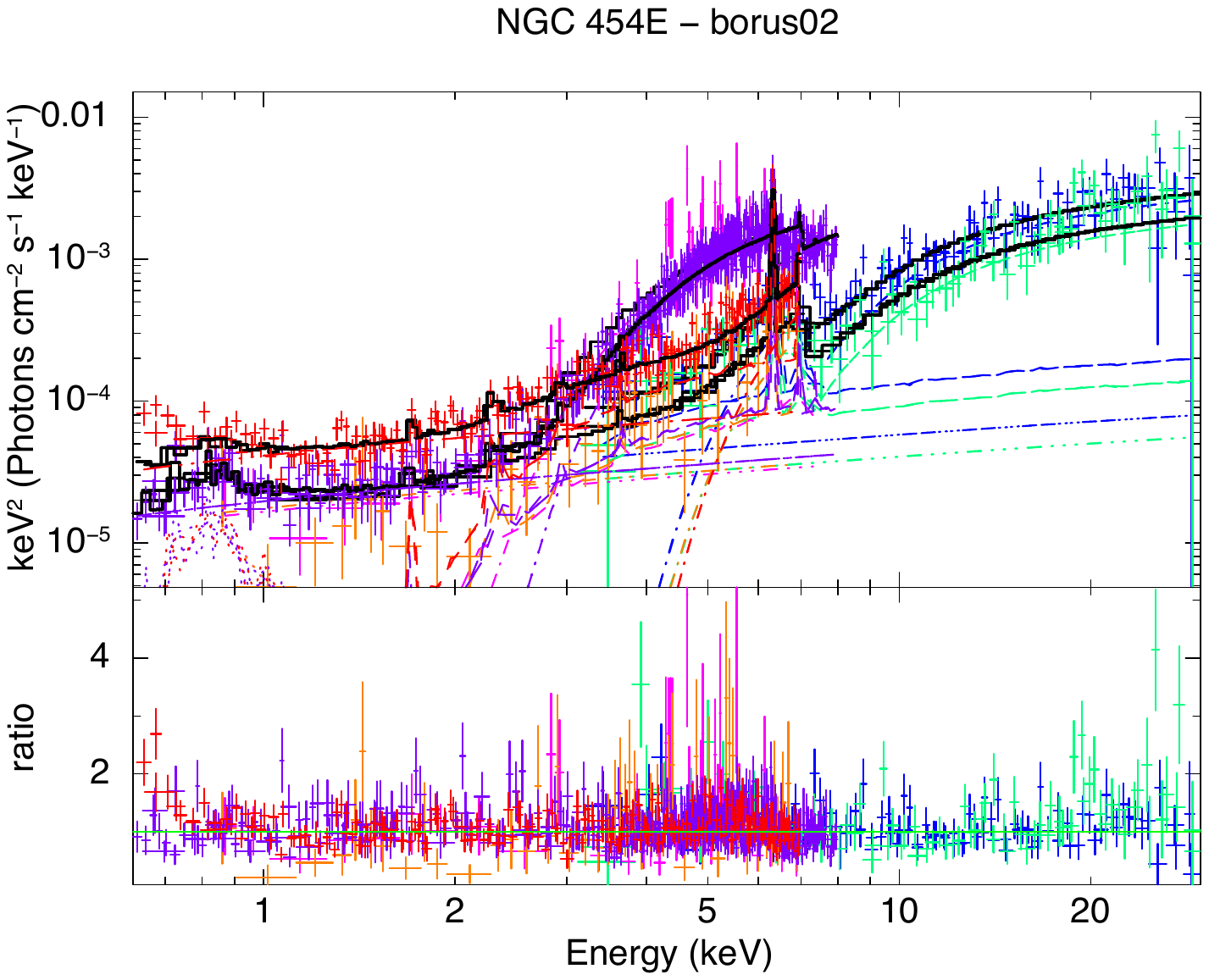}
    \includegraphics[ width=0.45\textwidth]{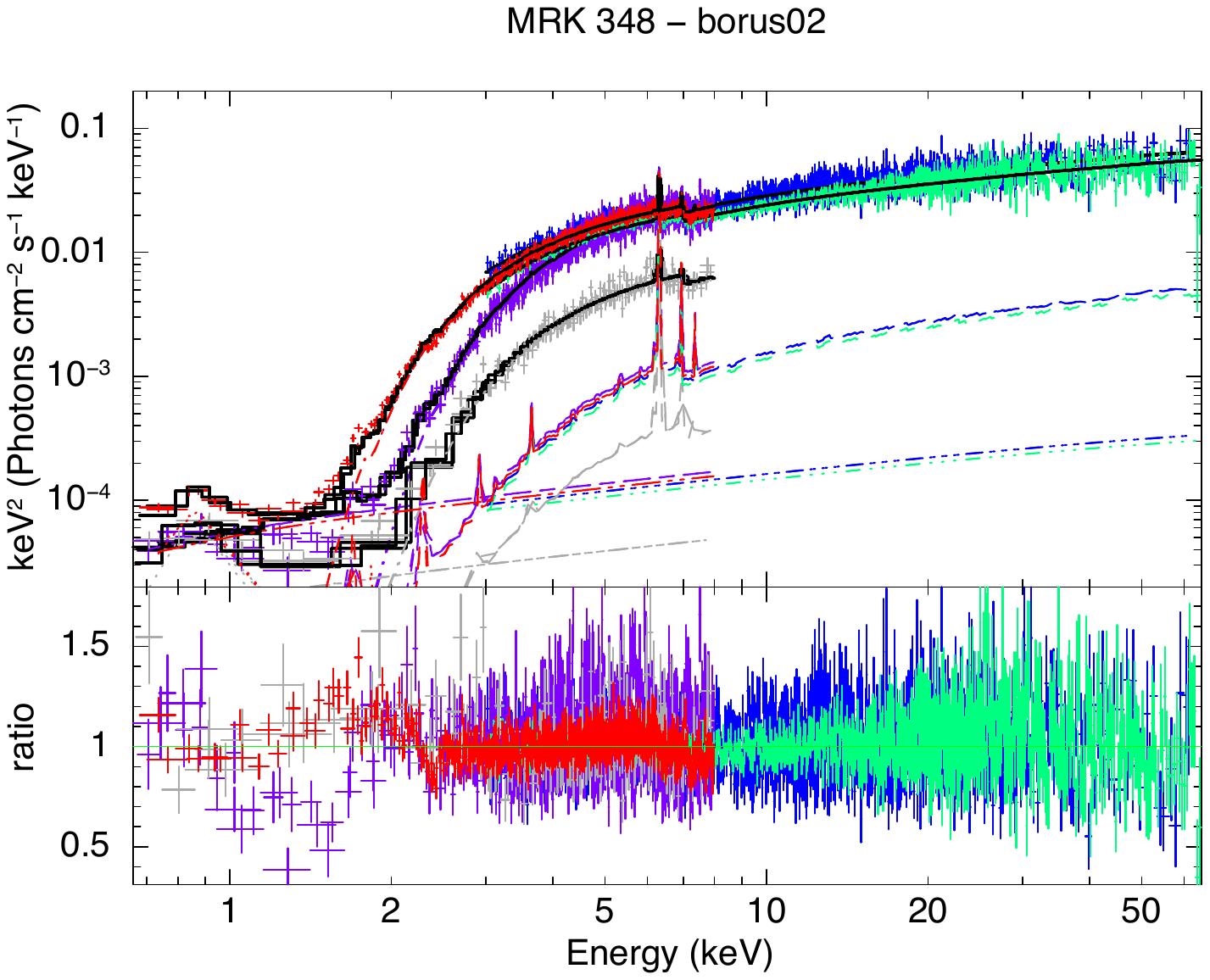}
    \includegraphics[ width=0.45\textwidth]{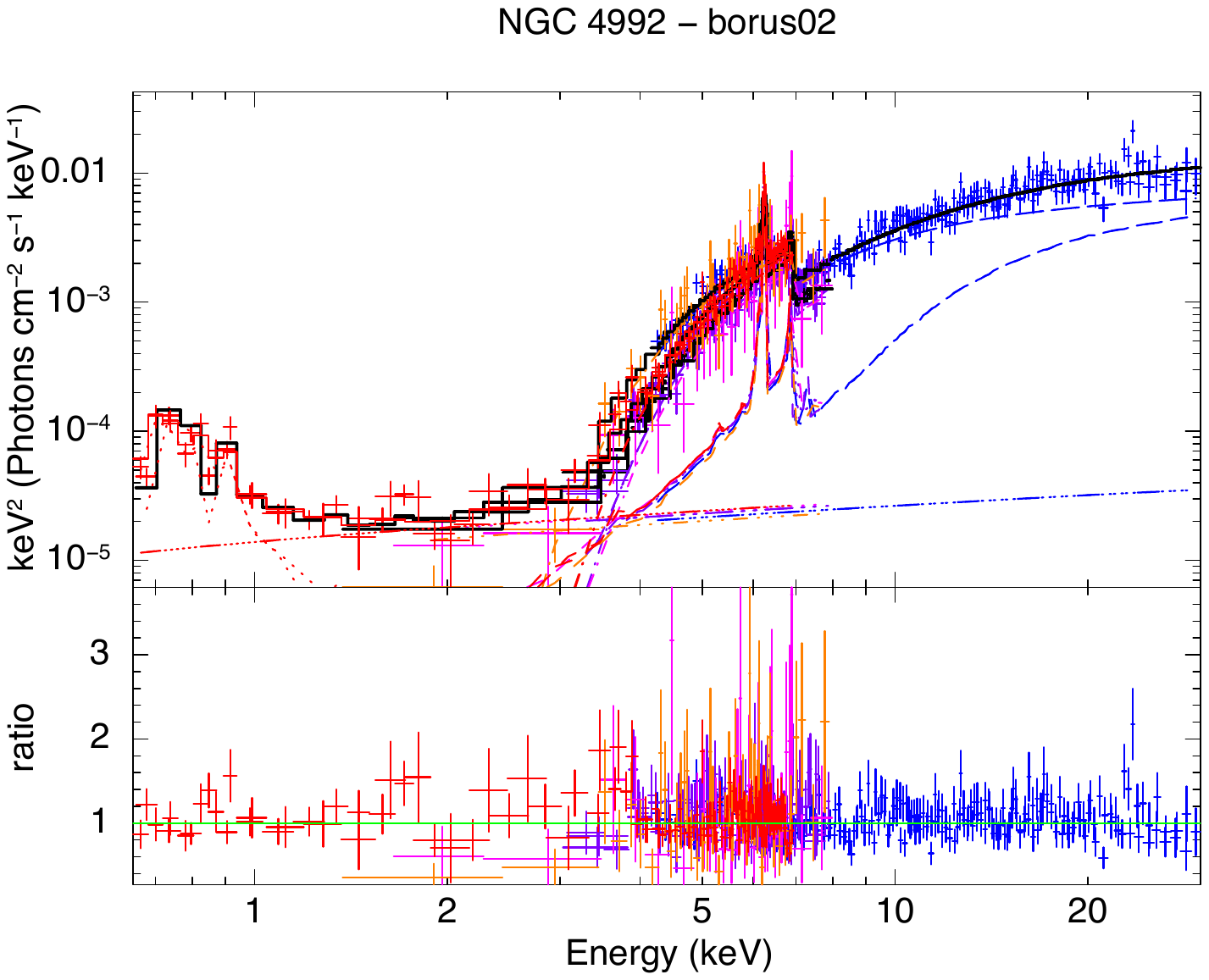}
    \includegraphics[ width=0.45\textwidth]{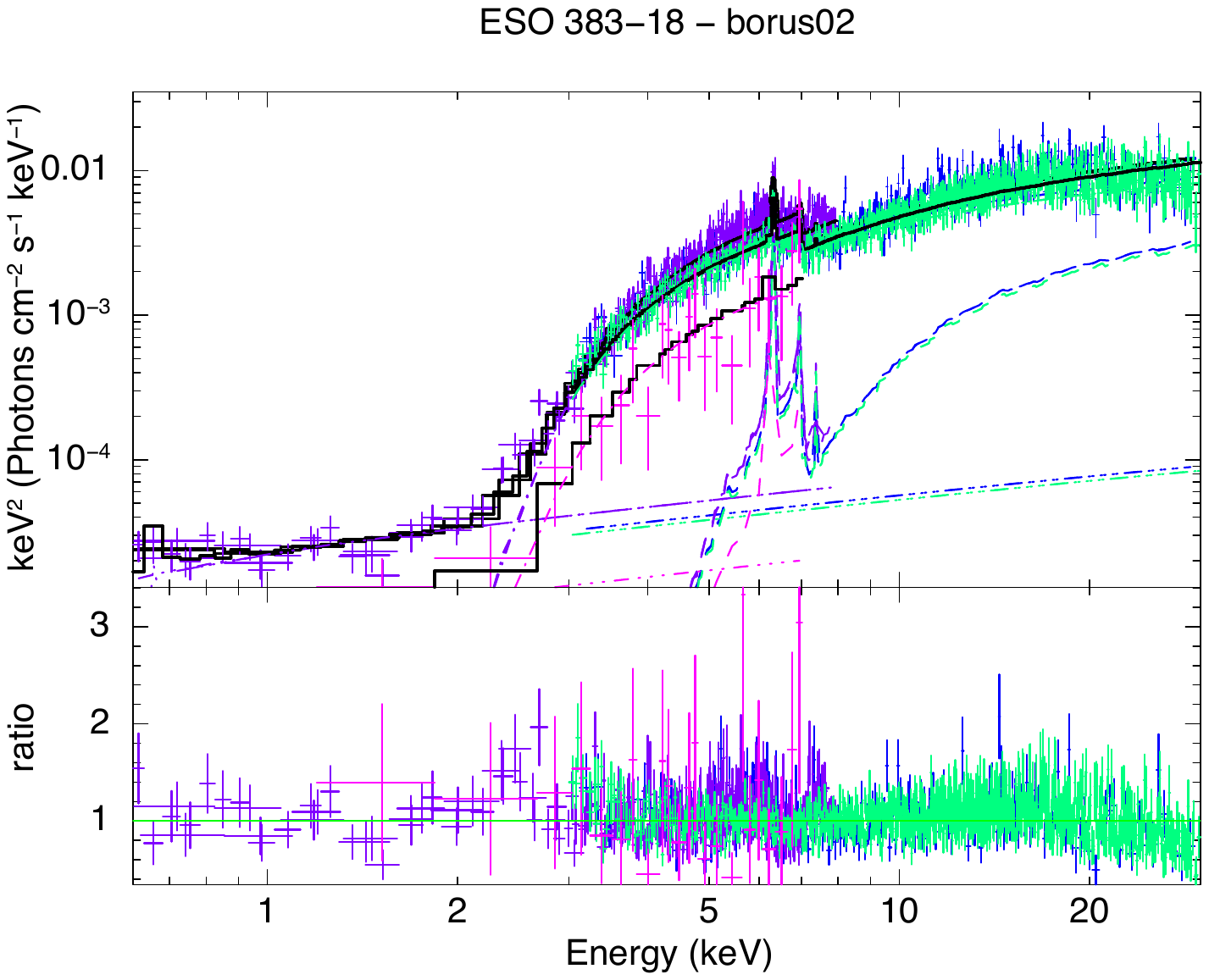}
    \includegraphics[ width=0.45\textwidth]{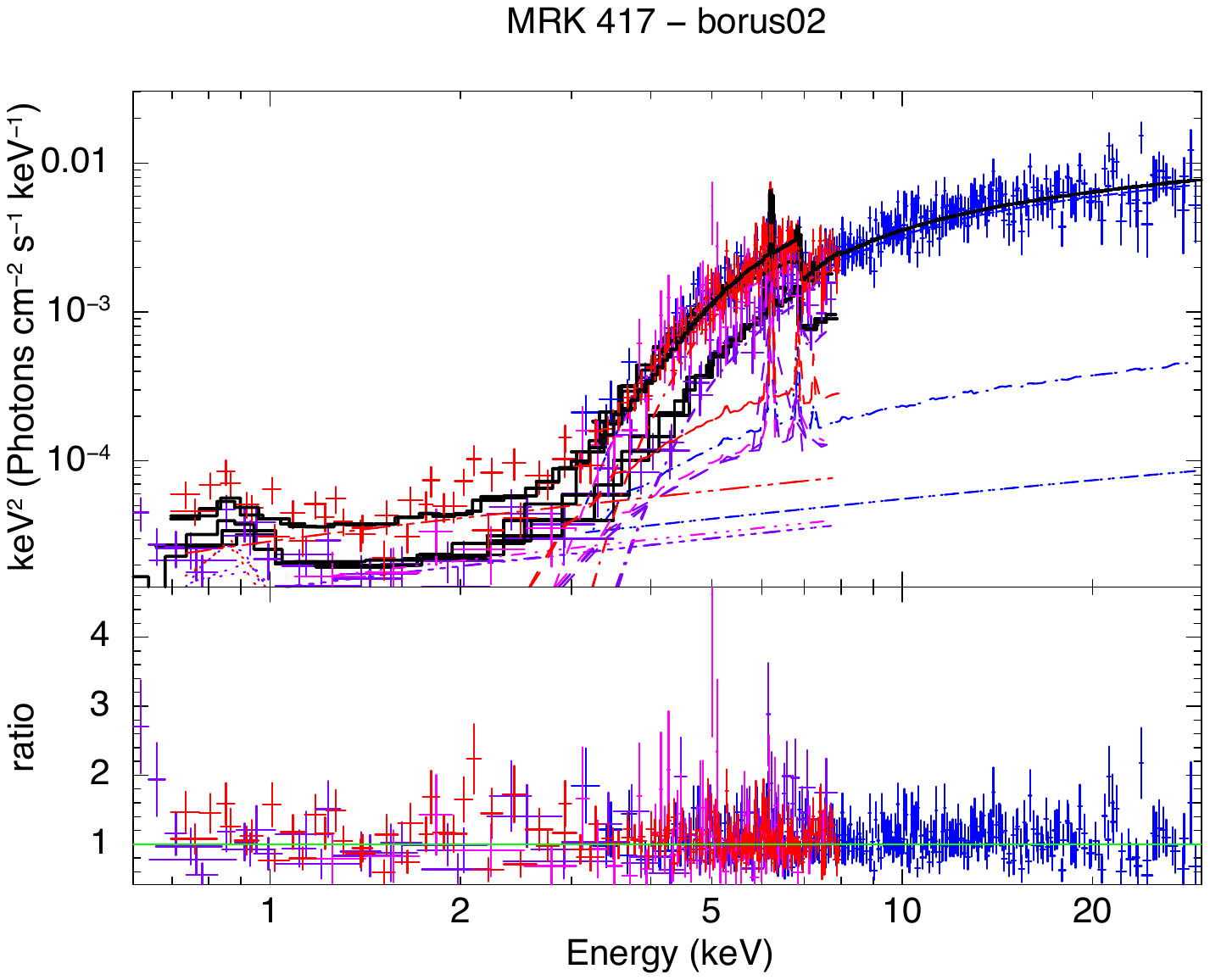}
    \includegraphics[ width=0.45\textwidth]{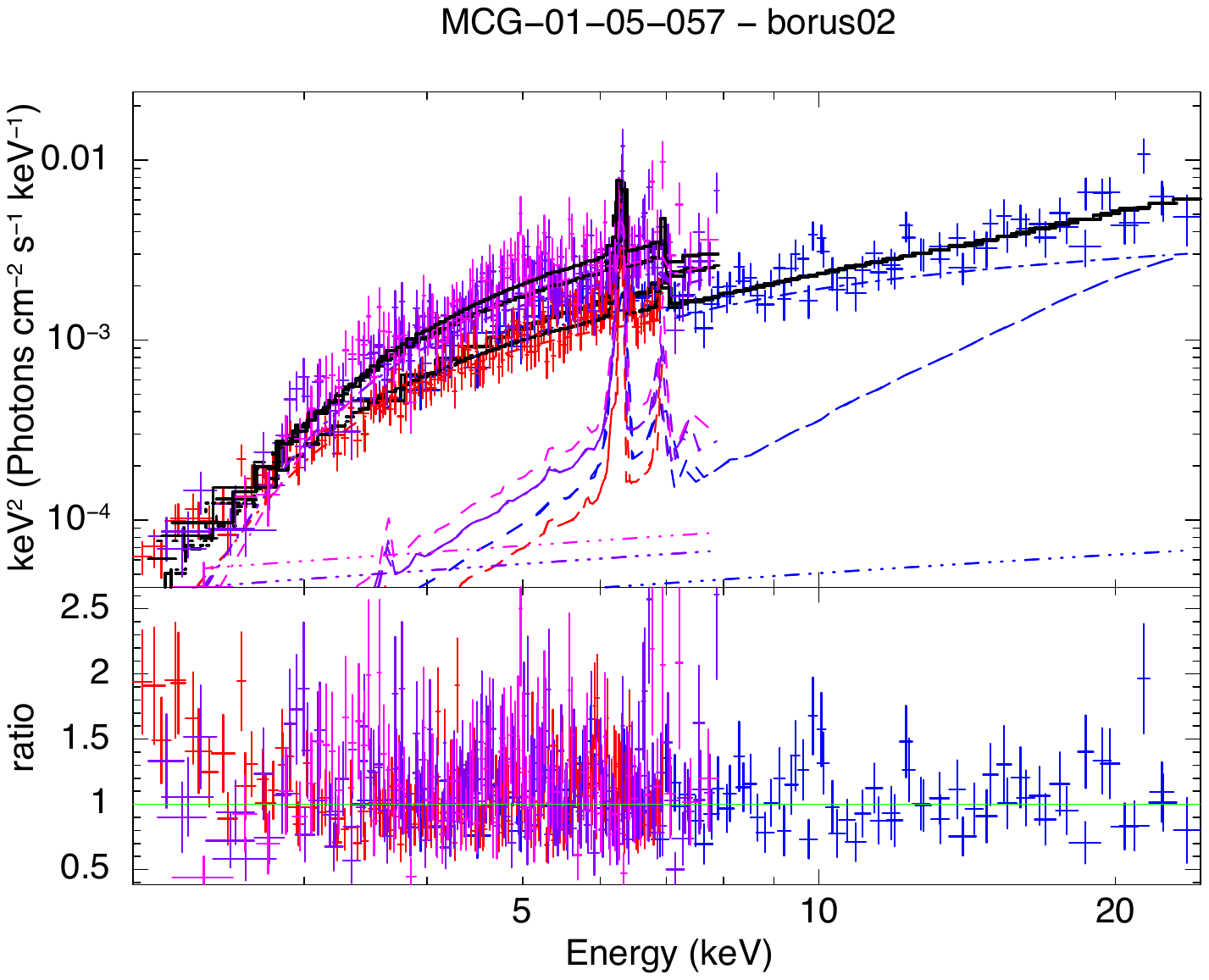}

    \caption{From left to right, top to bottom, unfolded \borus\ spectra of NGC 454e, MRK 348, NGC 4992, ESO 383-18, MRK 417, MCG-01-05-027, respectively. Color code as explained in Appendix \ref{sec:allspectra}.}
    \label{fig:single_epochs}
\end{figure}
\clearpage
   \begin{figure}[ht]
    \centering
    \includegraphics[ width=0.45\textwidth]{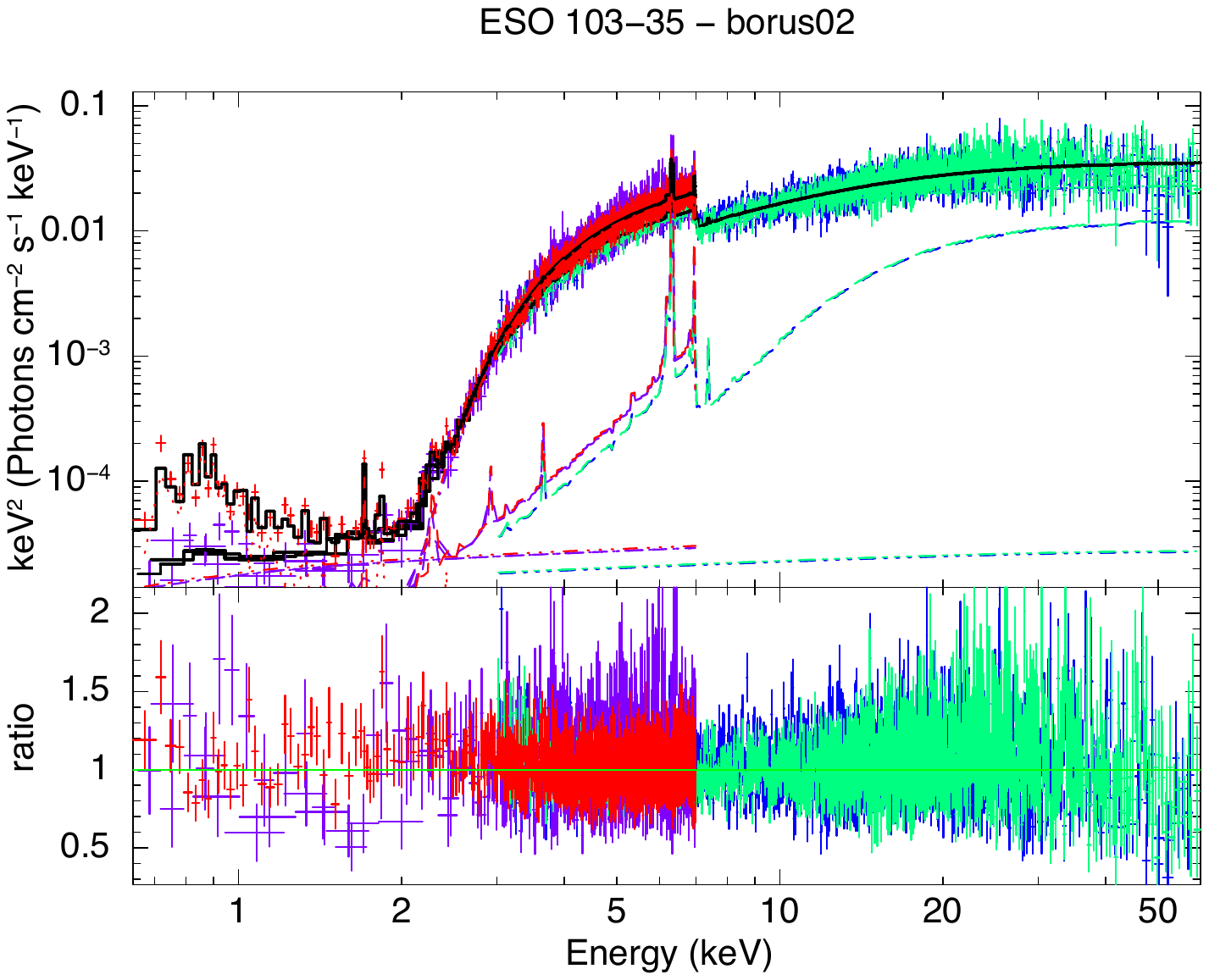}
    \includegraphics[ width=0.45\textwidth]{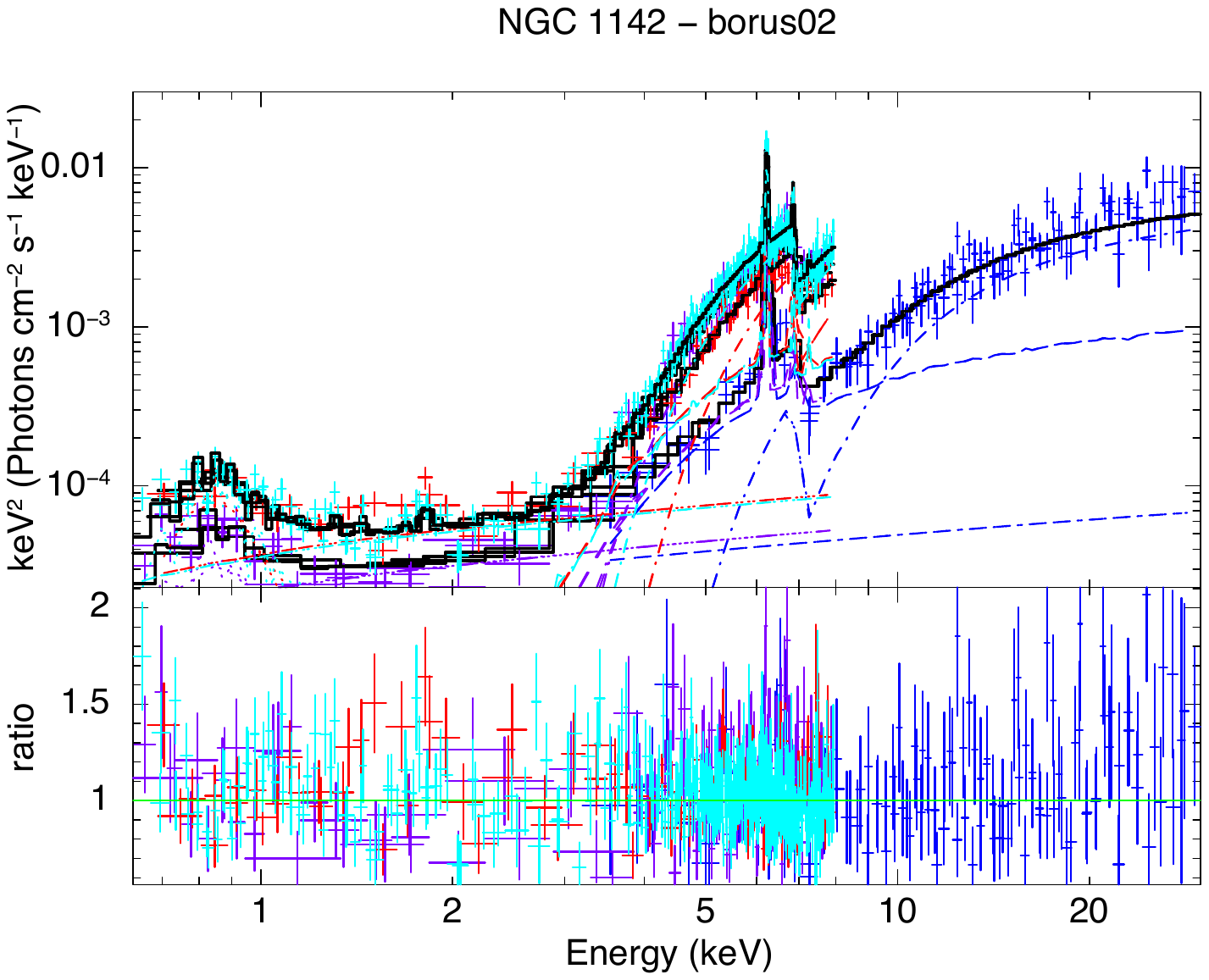}
    \includegraphics[ width=0.45\textwidth]{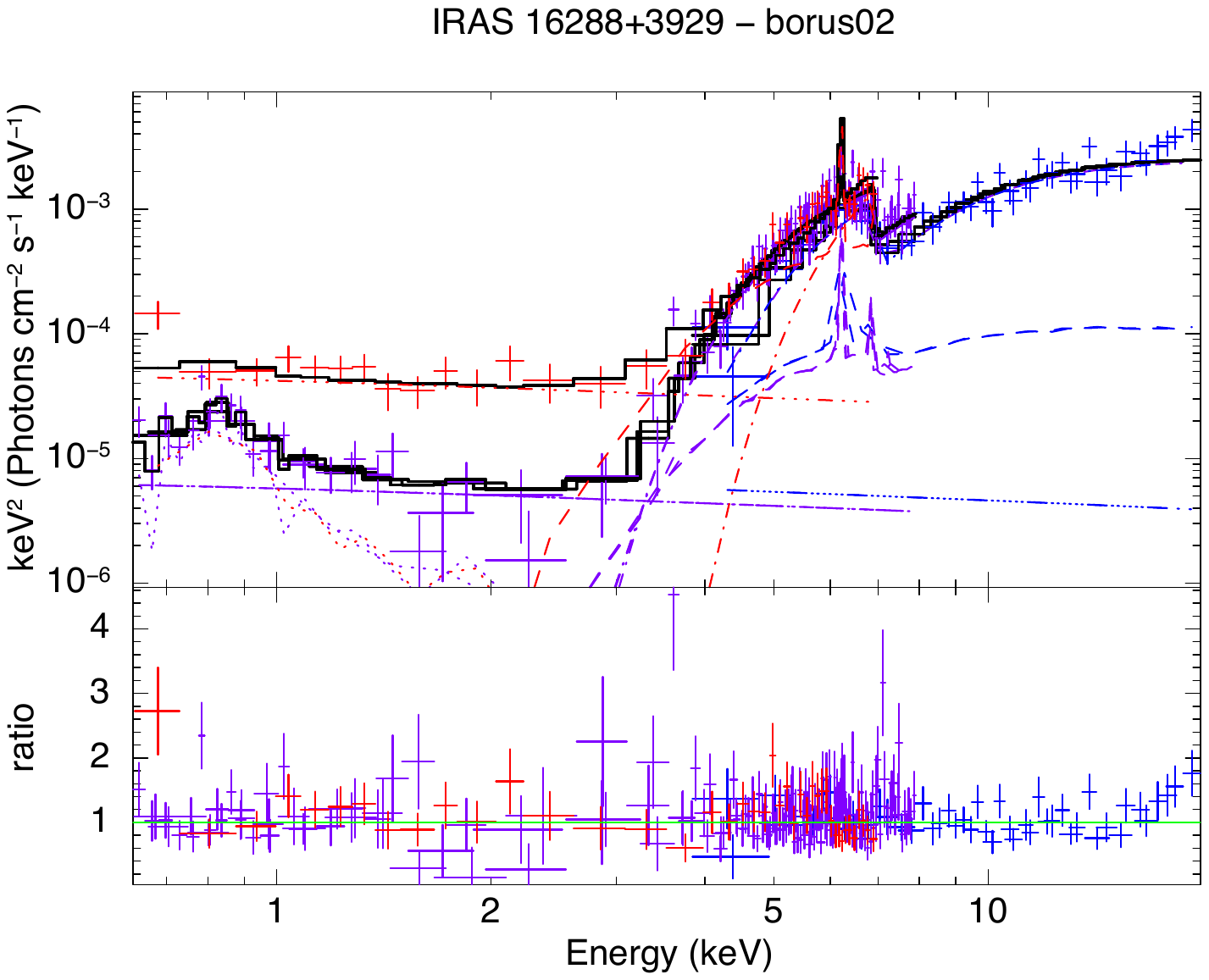}
    \includegraphics[width=0.45\textwidth]{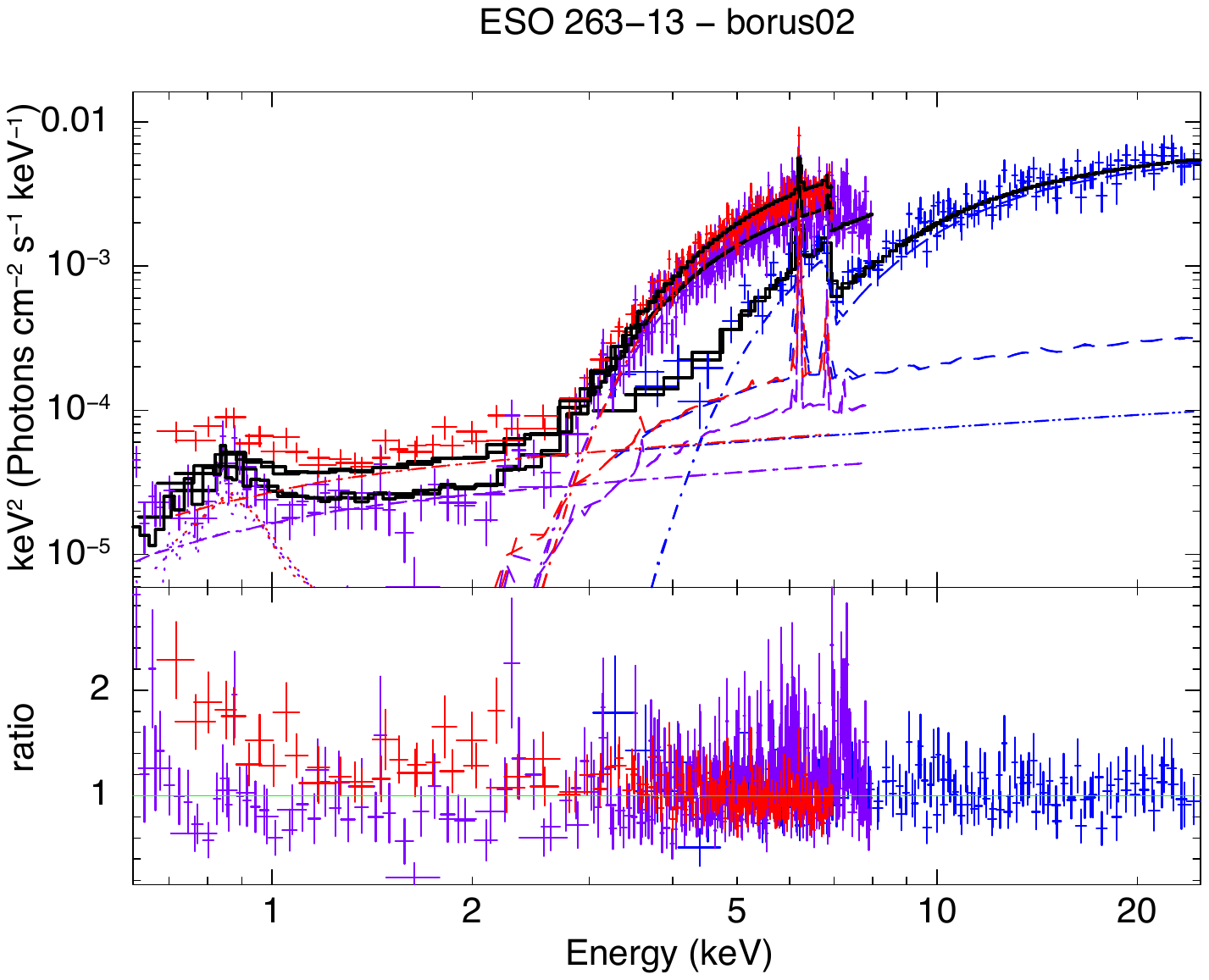}
    \includegraphics[ width=0.45\textwidth]{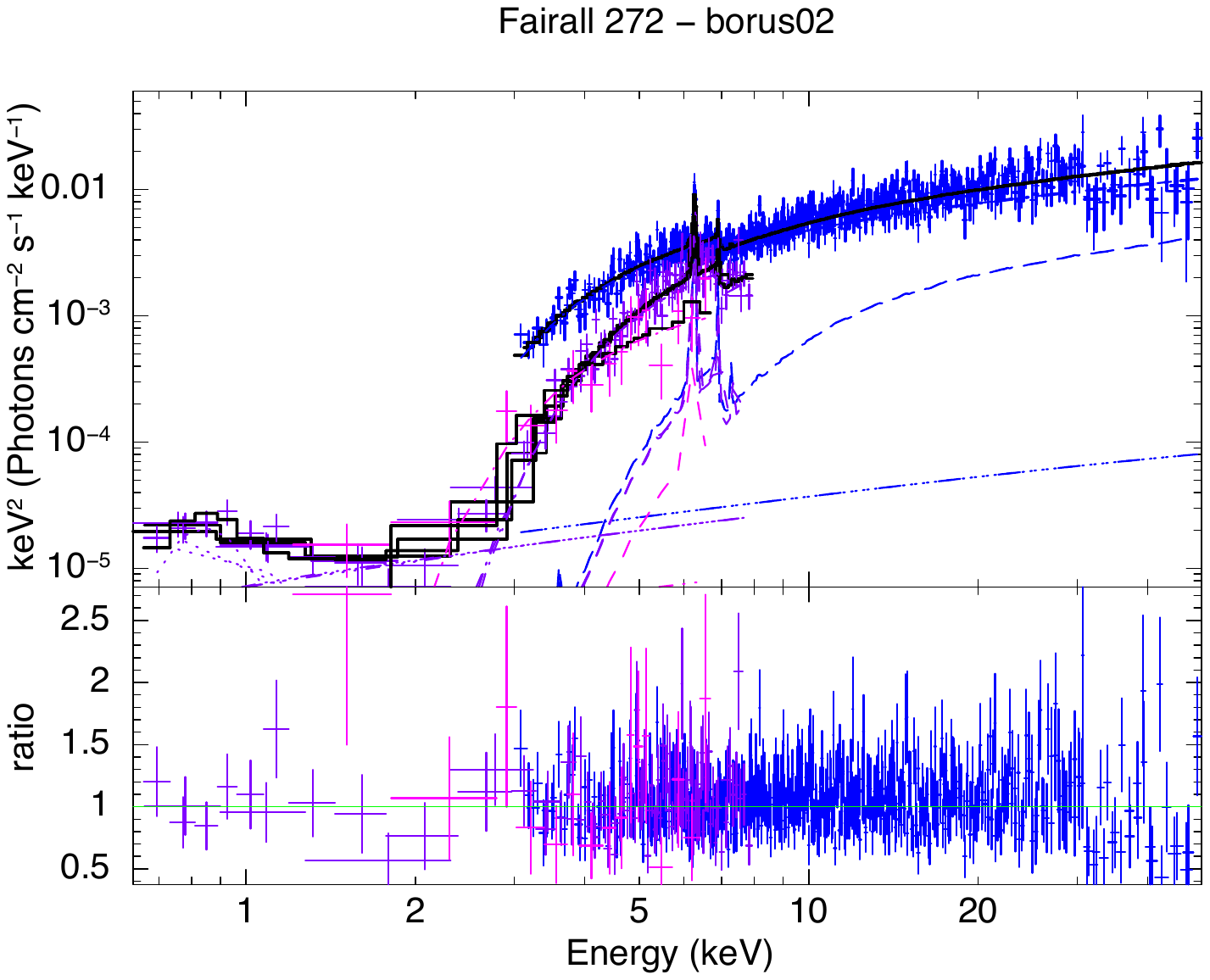}
    \includegraphics[ width=0.45\textwidth]{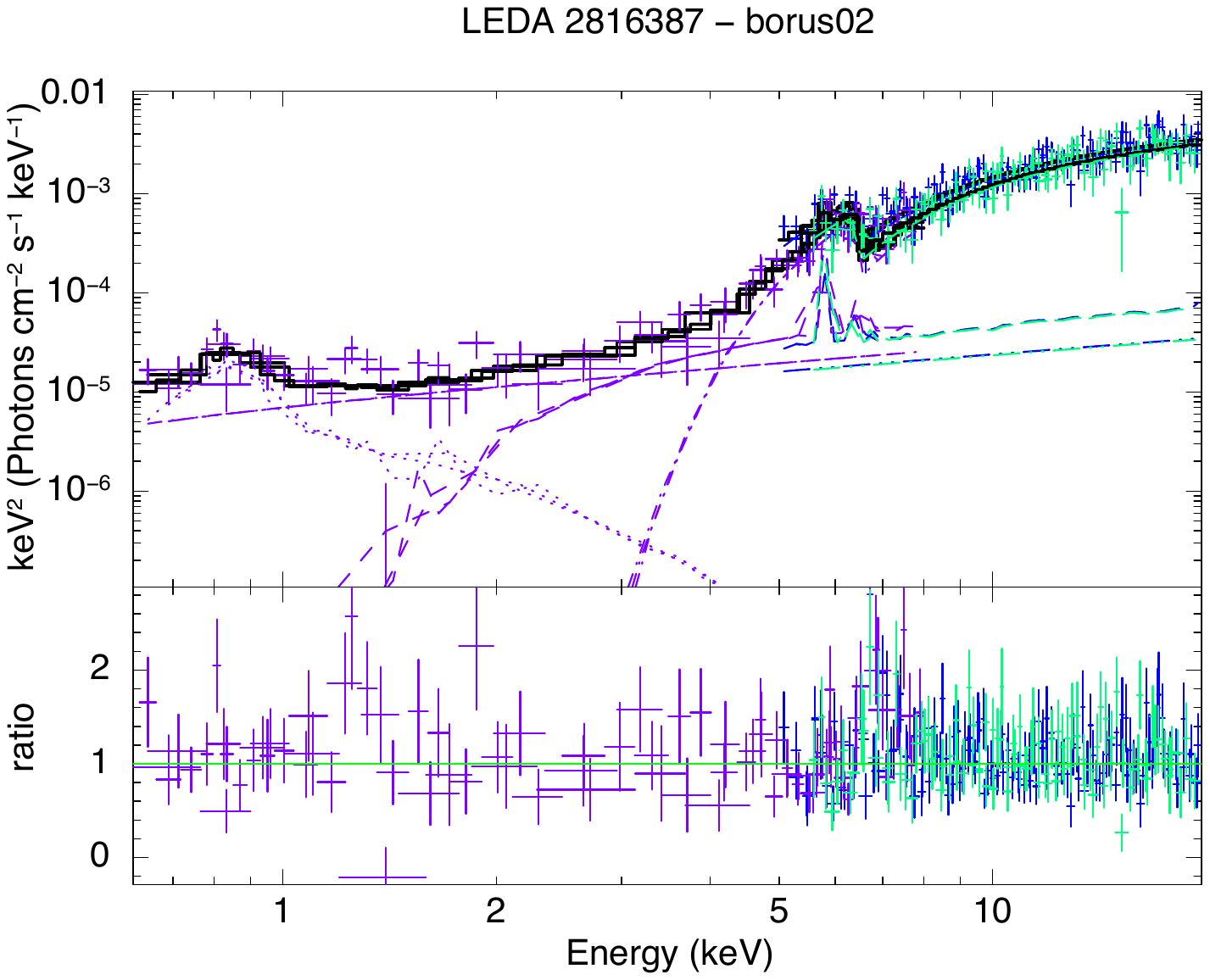}
    
    \caption{From left to right, top to bottom, unfolded \borus\ spectra of ESO 103-35, NGC 1142, IRAS 16288+3929, ESO 263-13, Fairall 272, LEDA 2816387, respectively. Color code as explained in Appendix \ref{sec:allspectra}. }
    \label{fig:single_epochs2}
\end{figure}

\clearpage
\section{\nh variability plots}\label{sec:variability_plots}
This appendix presents the line-of-sight hydrogen column density variability plots of the sources analyzed in this study.
\vspace{-1cm}
\begin{figure*}[ht]
    \centering
    \includegraphics[ width=0.48\textwidth]{ 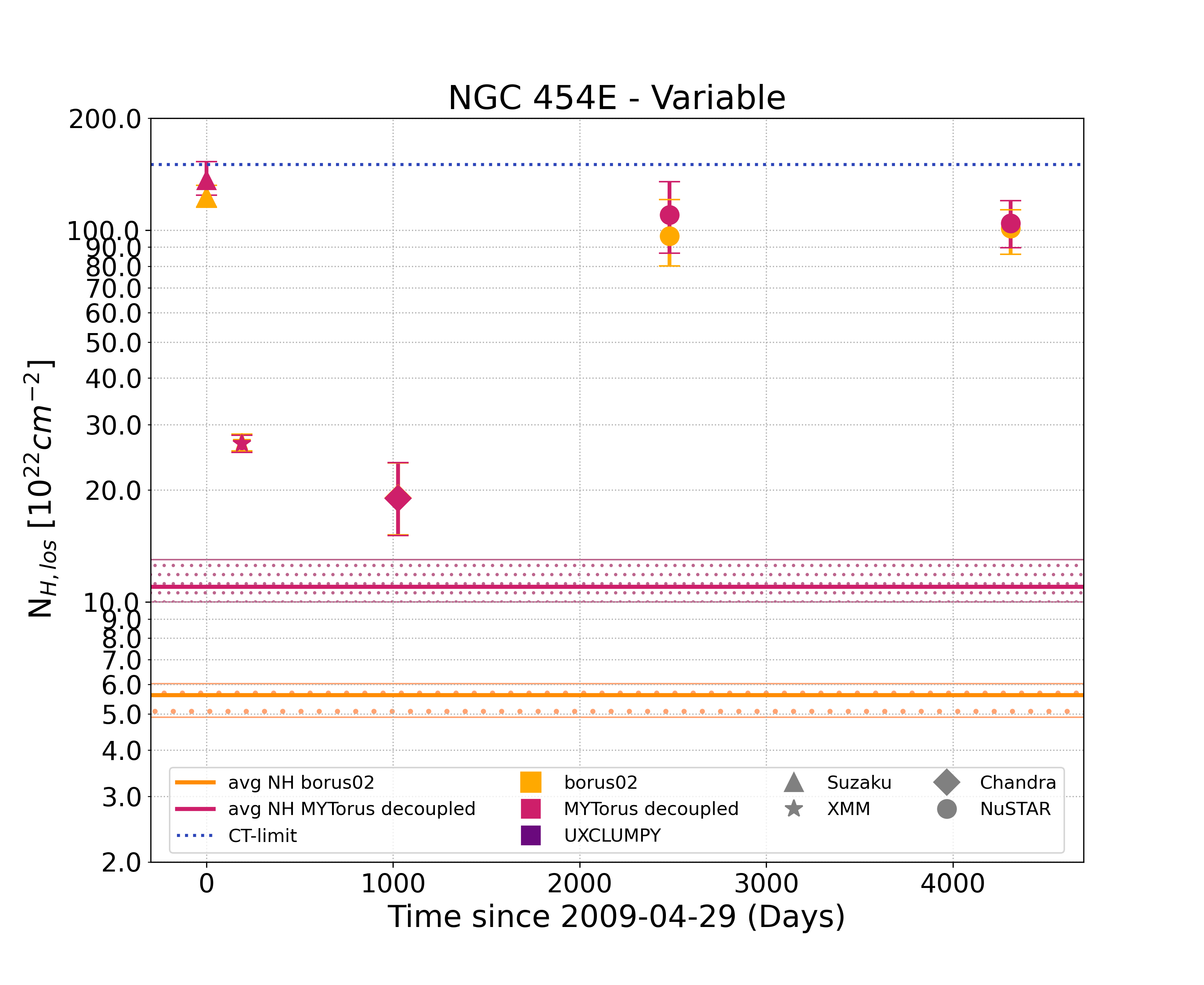}
    \includegraphics[ width=0.48\textwidth]{ 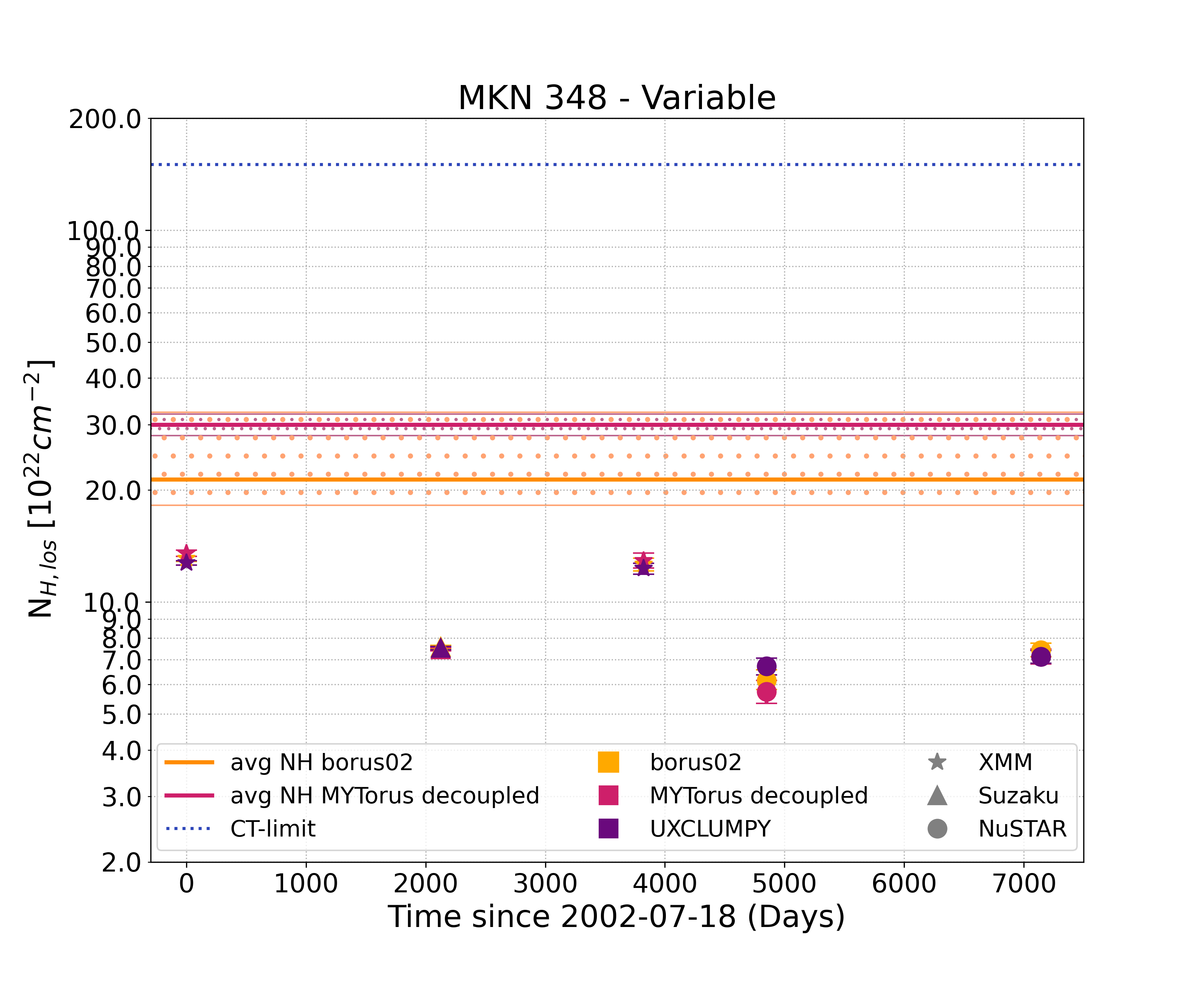}
    \includegraphics[ width=0.48\textwidth]{ 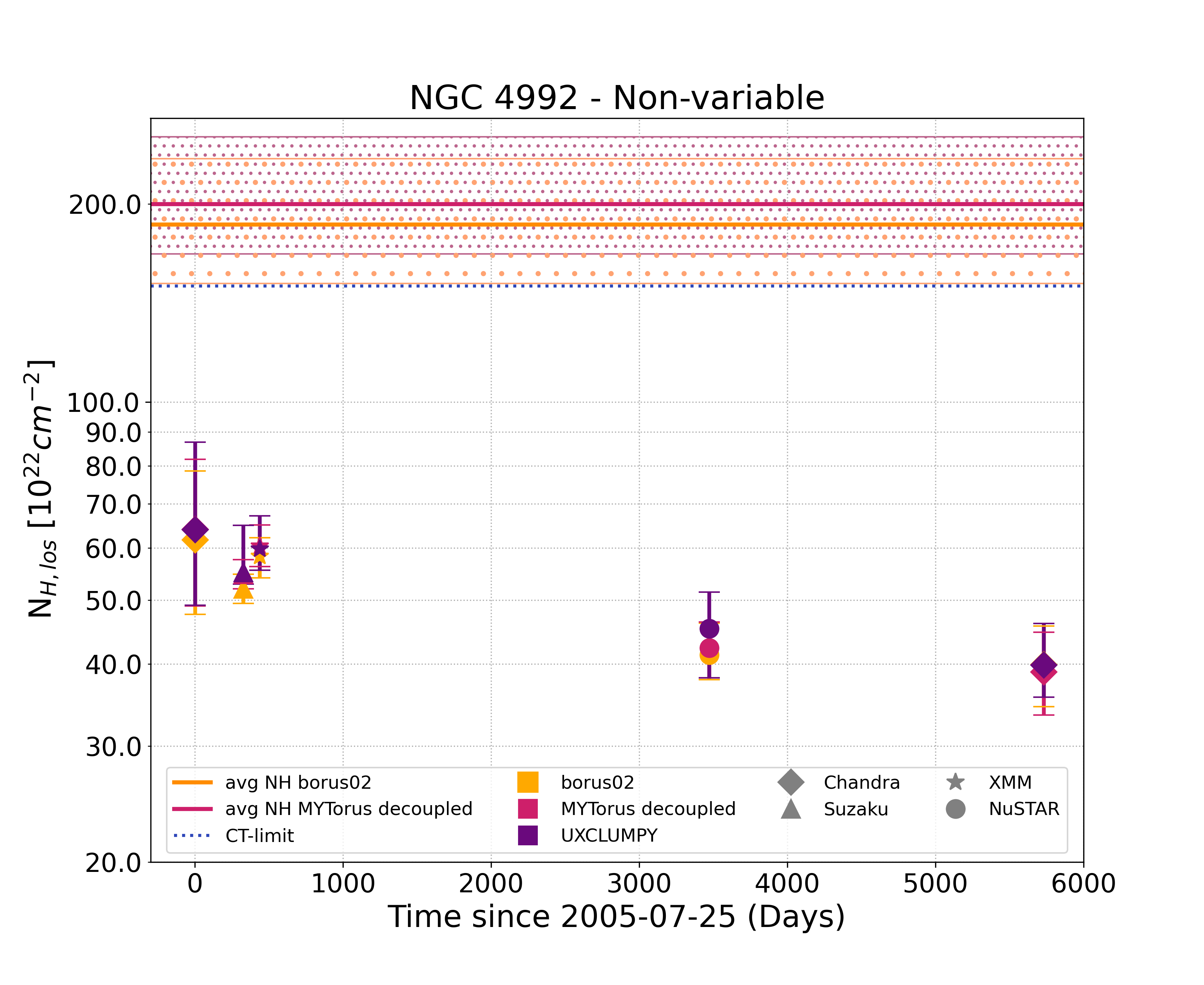}
    \includegraphics[ width=0.48\textwidth]{ 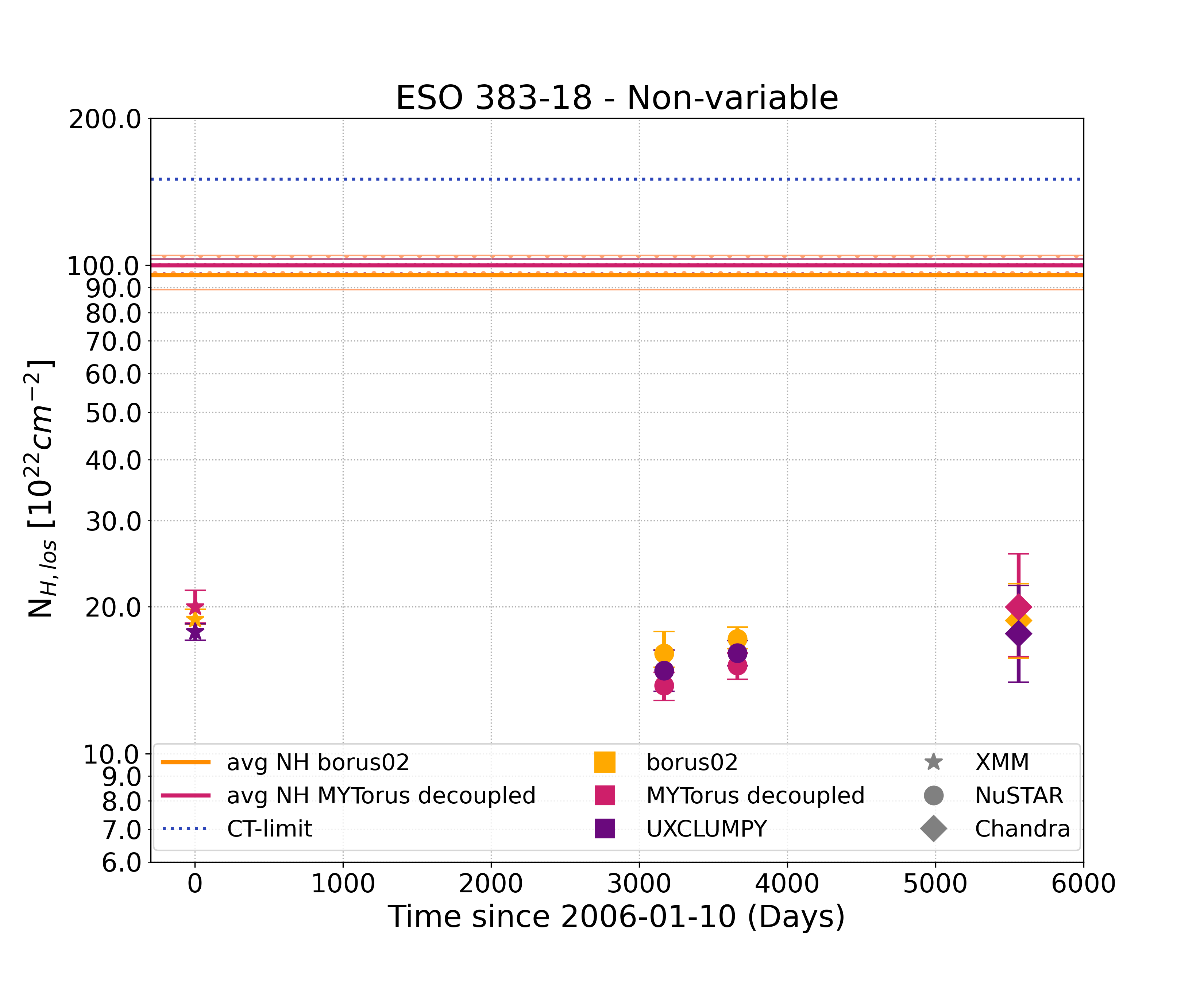}
    \includegraphics[ width=0.48\textwidth]{ 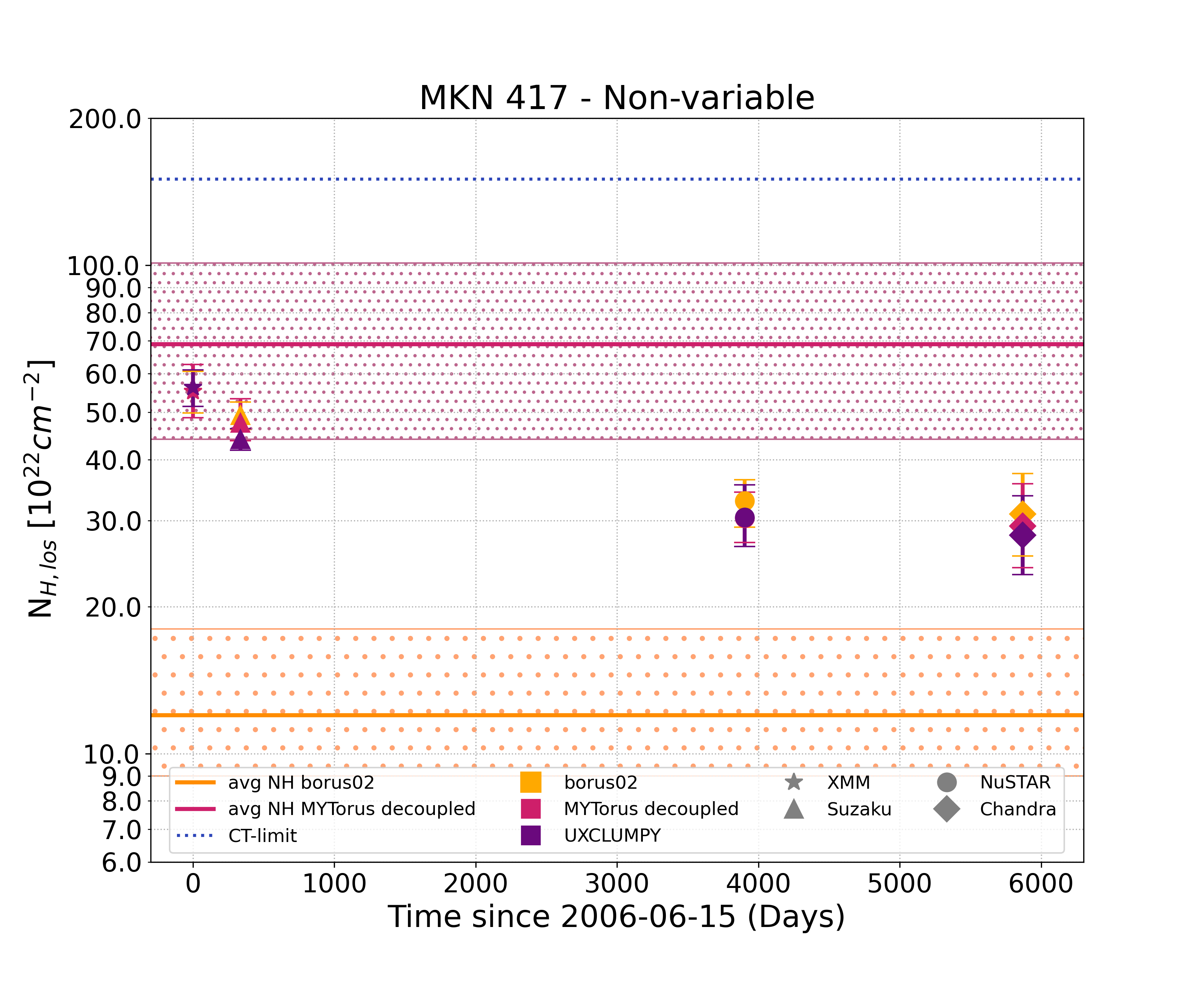}
    \includegraphics[ width=0.48\textwidth]{ 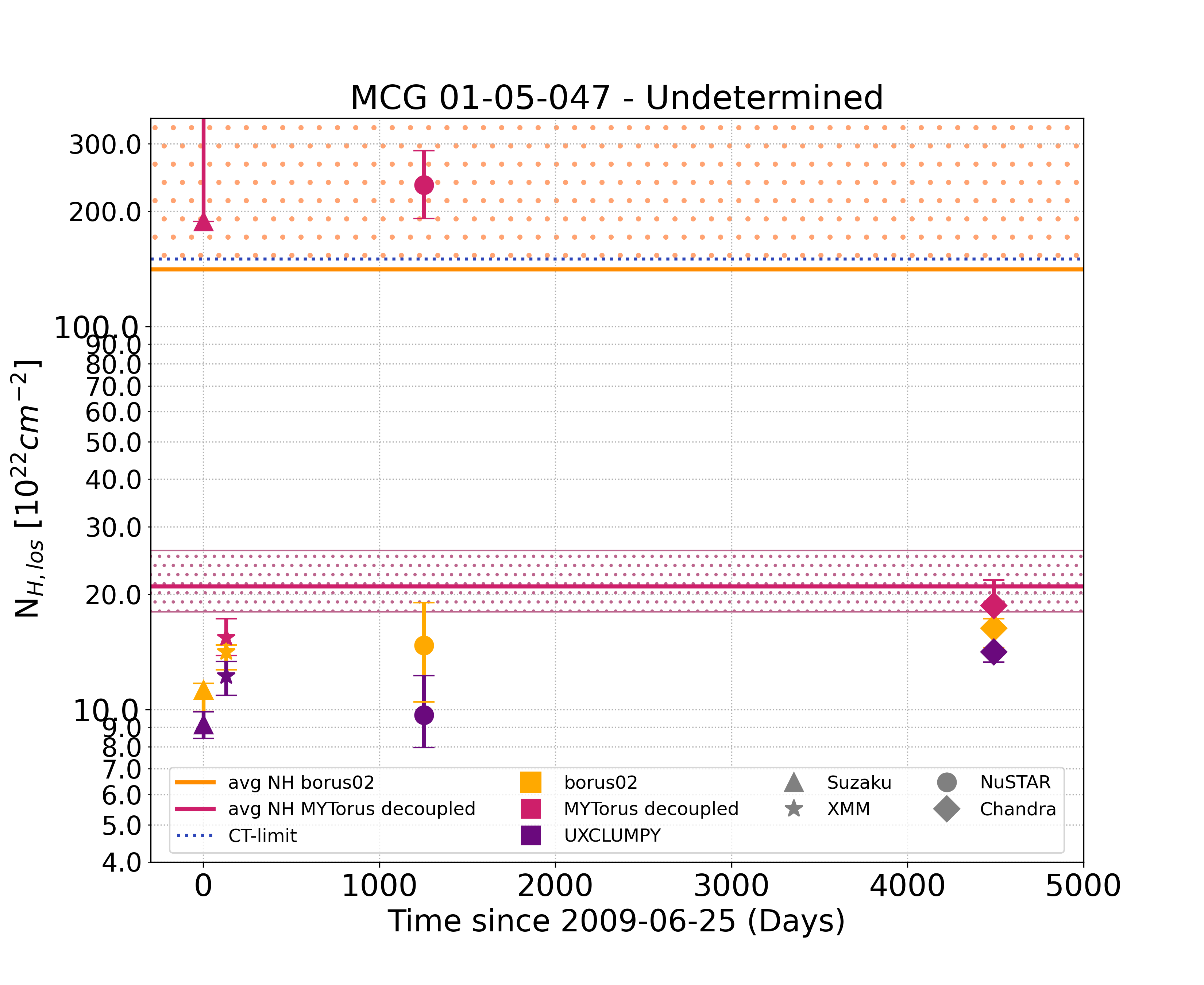}
    
    \caption{From left to right, top to bottom, line-of-sight hydrogen column density variability of NGC 454E, MRK 348, NGC 4992, ESO 383-18, MRK 417, MCG-01-05-047, respectively. Color code as explained in Figure \ref{fig:nhvar_plots3}}
    \label{fig:nhvar_plots}
\end{figure*}

\clearpage
   \begin{figure*}[ht]
    \centering
    \includegraphics[ width=0.48\textwidth]{ 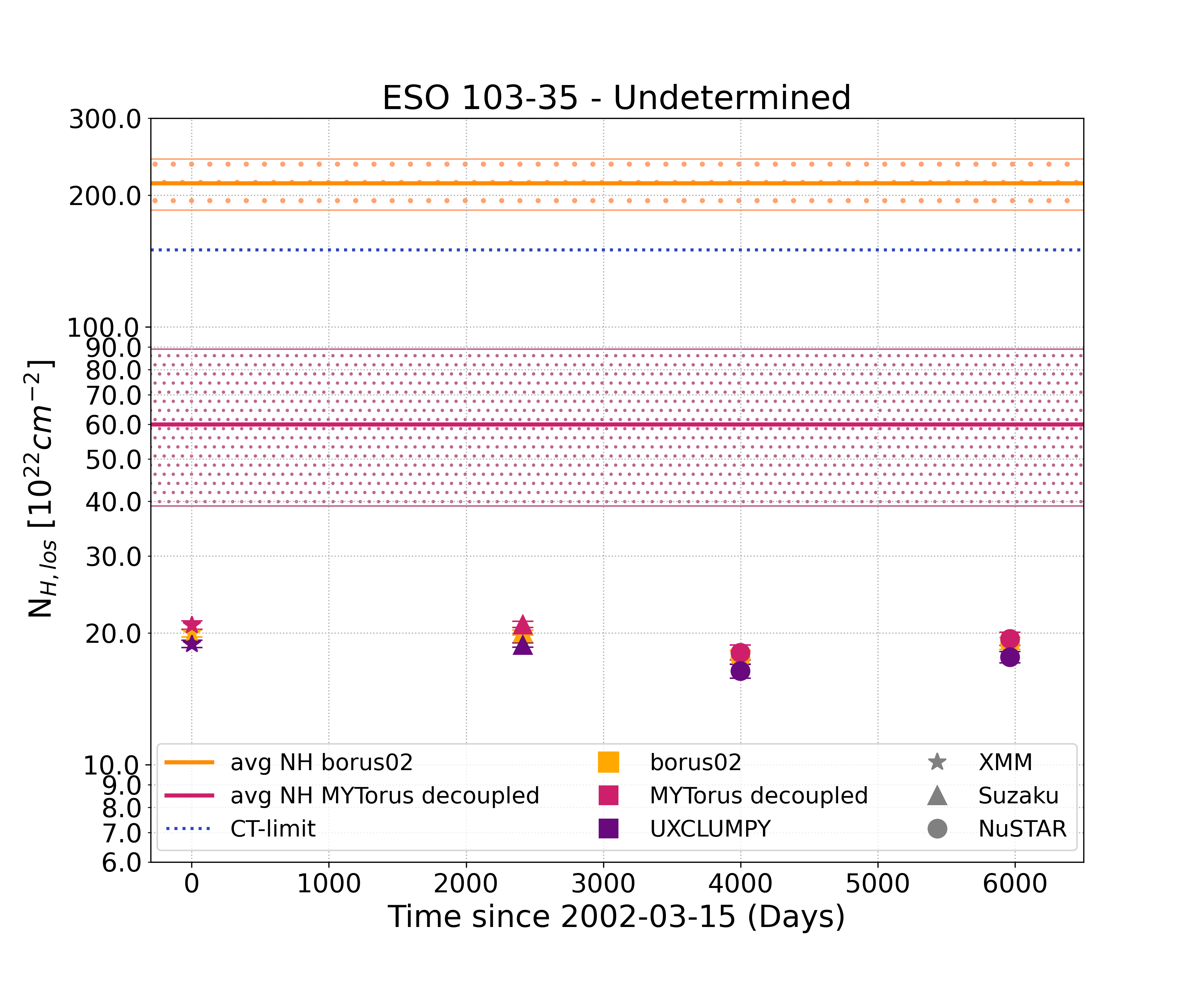}
    \includegraphics[ width=0.48\textwidth]{ 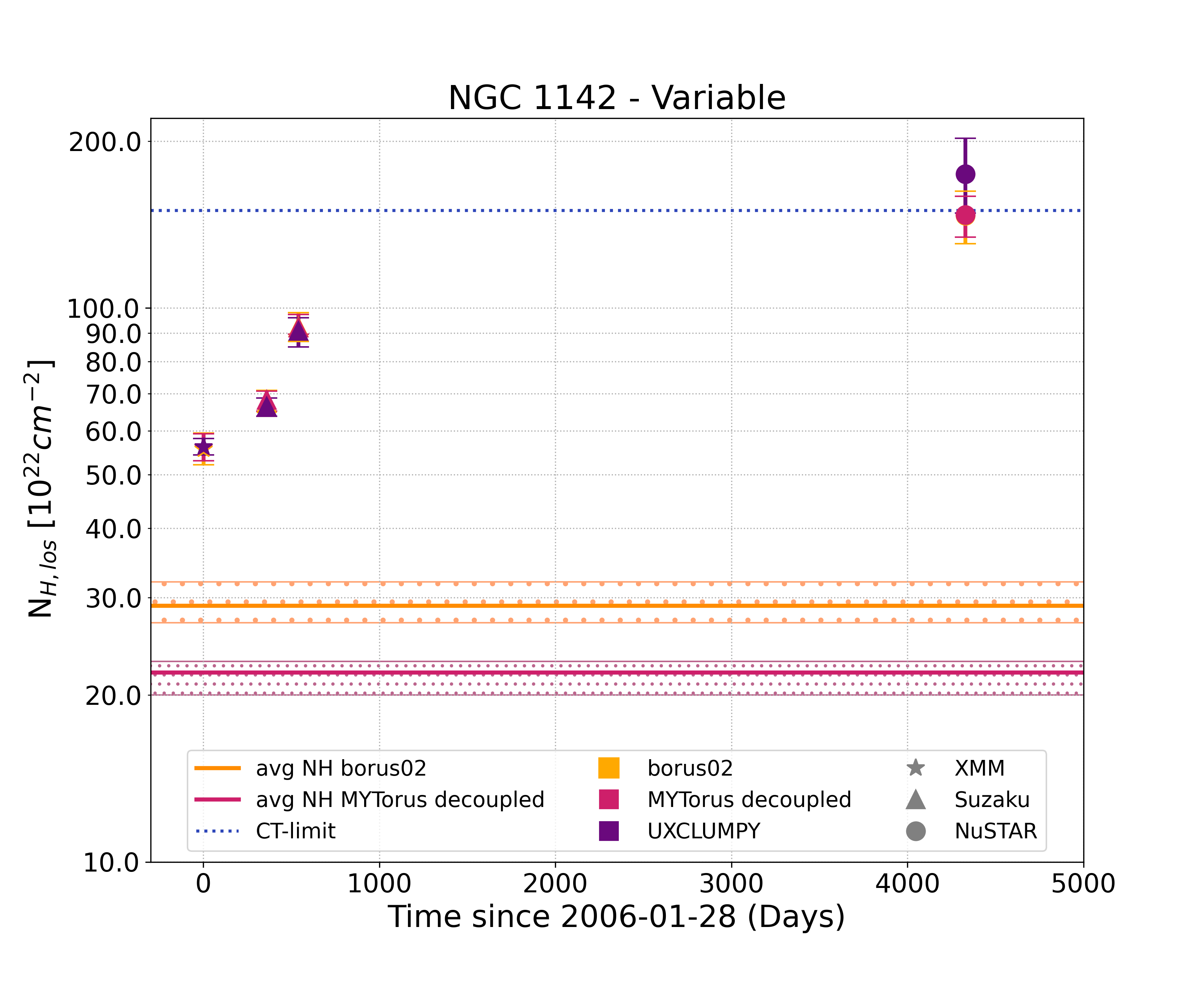}
    \includegraphics[ width=0.48\textwidth]{ 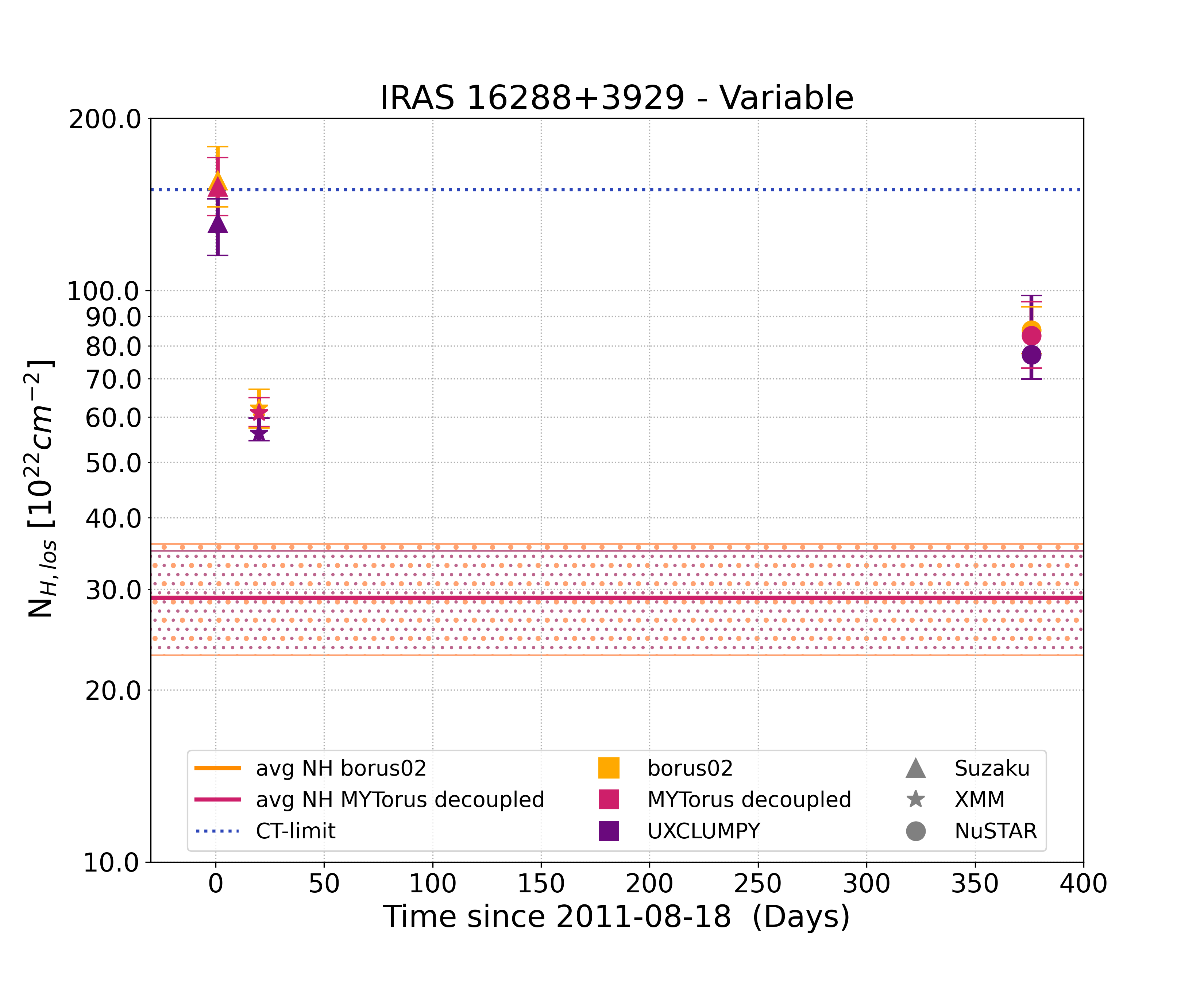}
    \includegraphics[width=0.48\textwidth]{ 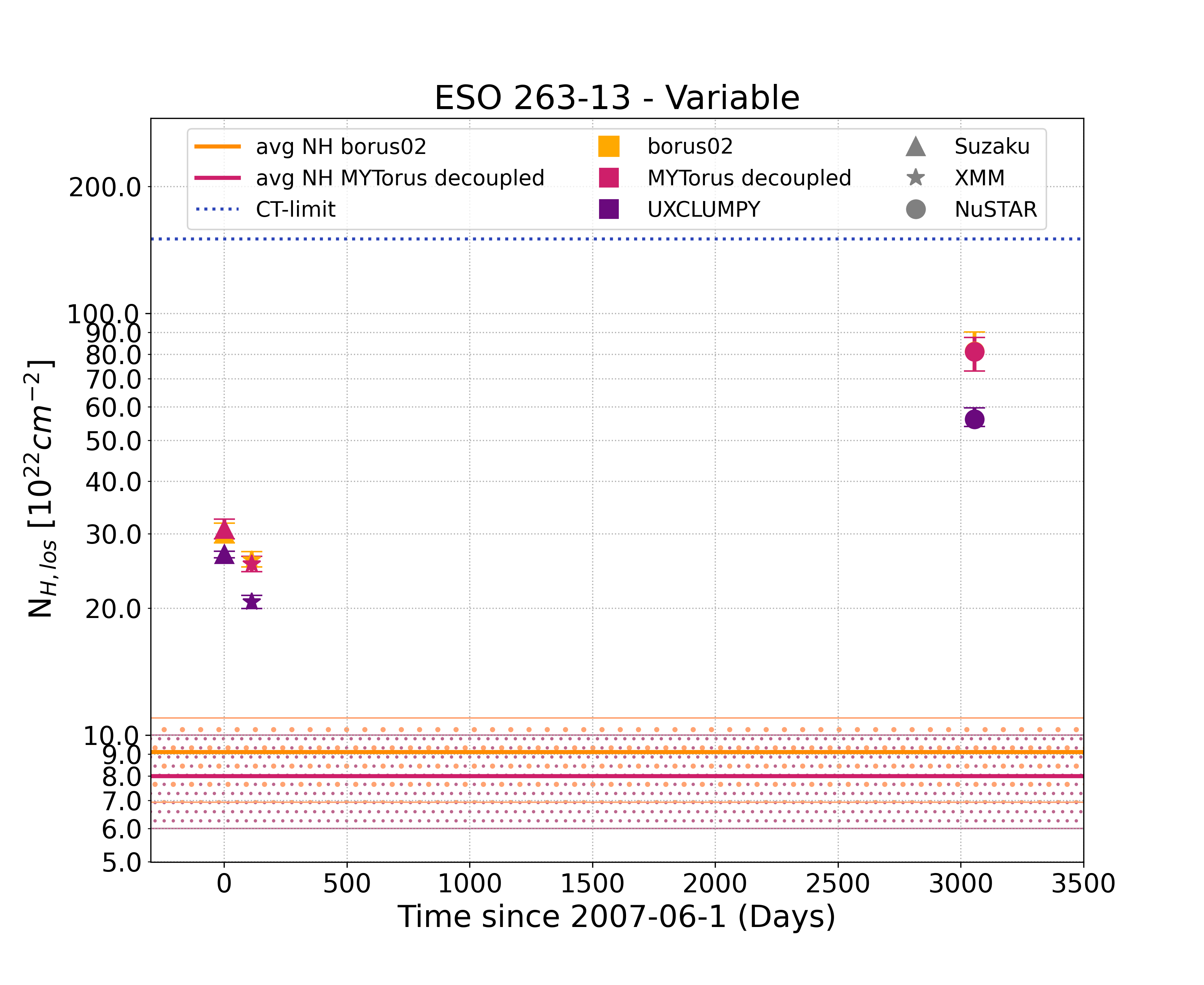}
    \includegraphics[ width=0.48\textwidth]{ 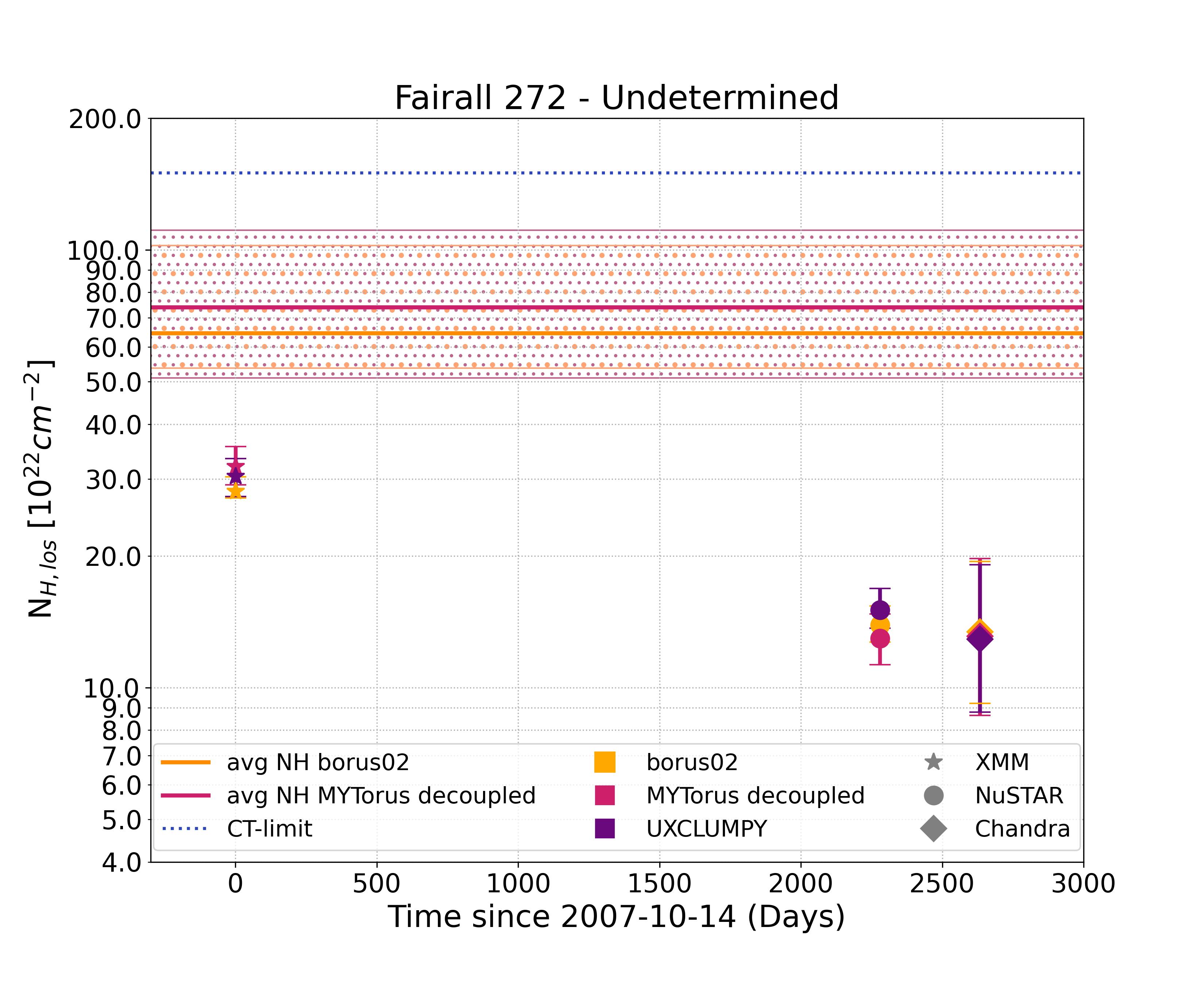}
    \includegraphics[ width=0.48\textwidth]{ 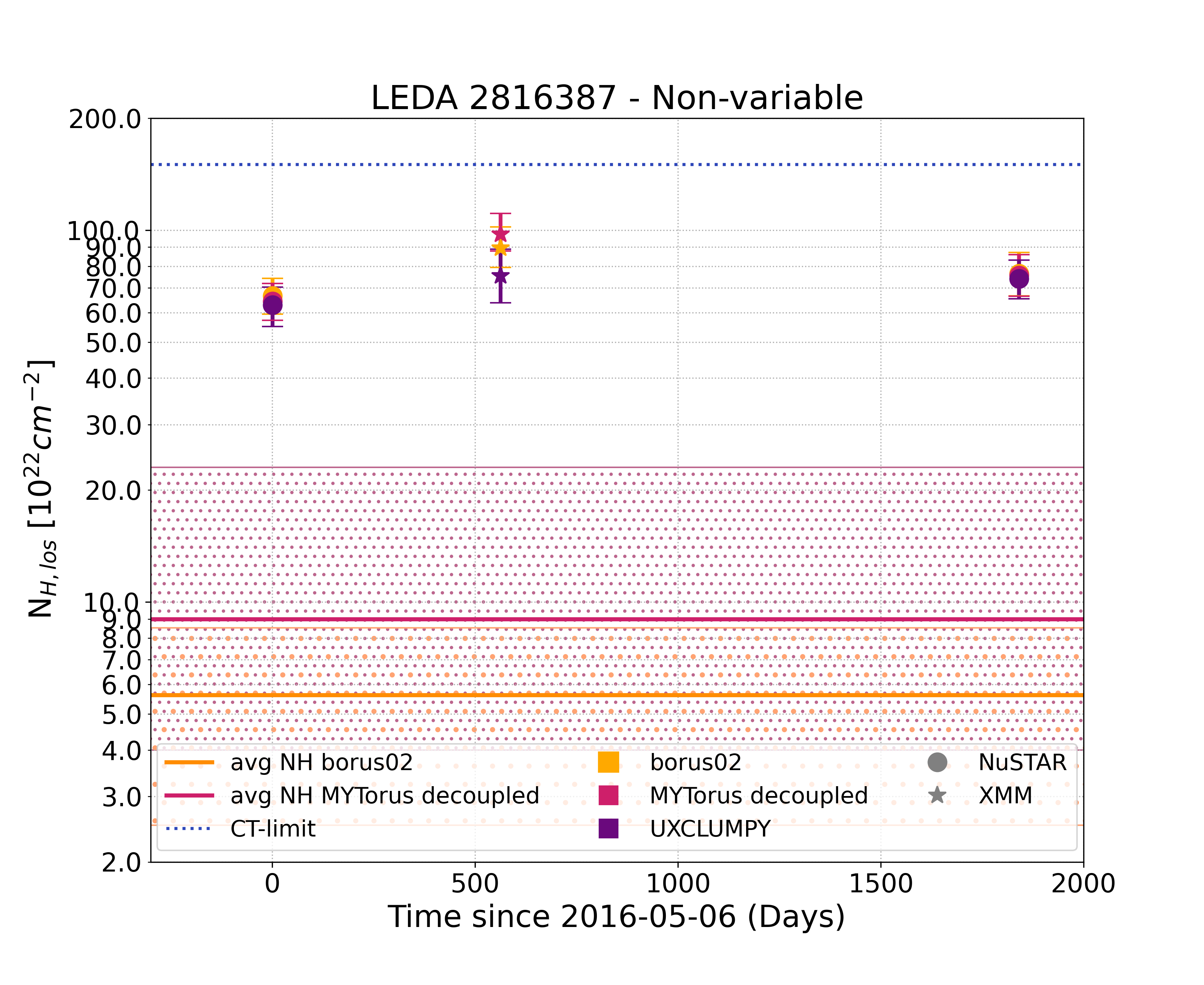}
    \caption{From left to right, top to bottom, line-of-sight hydrogen column density variability of ESO 103-35, NGC 1142, IRAS 16288+3929, ESO 263-13, Fairall 272, LEDA 2816387, respectively. Color code as explained in Figure \ref{fig:nhvar_plots3}}
    \label{fig:nhvar_plots2}
\end{figure*}
\clearpage
\section{Comments on individual sources}\label{appendix:individual_sources}

In this appendix, we provide the specifics of data reduction and fitting procedures for each source when deviating from those outlined in Section \ref{sec:datareduction} and \ref{sec:analysis}. We comment on the results of the X-ray analysis for each individual source examined in the work, focusing on the contrast between the different models used. The comparison with the results obtained by \citetalias{zhao_2021} and other relevant works is also discussed.

\subsection{NGC 454E}\label{sec:ngc454e}
\noindent \textbf{Data reduction and fitting:} For the two \chandra\ observations, the background was extracted from a 15" circle in a uniform region to avoid contamination from a small emitting source located 10" from the nucleus. This second source is neither resolved by \xmm\ nor \suzaku. Due to the data quality of the two \chandra\ observations, the spectra were binned with 5 and 3 cts/bin, respectively, applying C-statistics to the fit. Thus, we used a combination of $\chi^2$ and C-stat as total statistics.
\textbf{Analysis of results:} NGC 454E is part of a merging system with the companion NGC 454W. As reported in \citet{marchese2012}, no X-ray emission is detected in the latter. We note that two observations, \nustar\- 2 and \chandra\- 2, were taken almost simultaneously ($\Delta$t=6h). After checking for possible intrinsic flux and luminosity variability, we opted to couple them in the spectral analysis. Thus, we consider them to be a unique joint observation. Due to the complexity of the six spectra analyzed in this work, as seen in the top left panel of Figure \ref{fig:single_epochs}, we are presenting results only for \borus\ and \mytorus, as \uxclumpy\ failed to provide physically meaningful results. \mytorus\ and \borus\ are in good agreement, matching the results obtained by \citet{marchese2012}, whose analysis was performed only taking into account \suzaku, \xmm\ and \textit{Swift}-XRT observations. \citetalias{zhao_2021} analysis, performed by modeling with \borus\ the first \nustar\ and the one \xmm\ observations, found a thicker ($\gtrsim 10^{24}$\,cm$^{-2}$) average torus density with respect to the one measured by our analysis, in which the \nhav is of the order of $\sim 10^{22}$\,cm$^{-2}$ for \borus, and $\sim 10^{23}$\,cm$^{-2}$ for \mytorus. We measure a $\Delta$N$_{\rm H,los}\simeq 95\times 10^{22}$\,cm$^{-2}$ over a time span of 190 days ($\sim$ six months), which would position the origin of the obscuration around $\sim90$ pc from the central engine, following the approach firstly proposed by \citet{risaliti_2005}. We note that for \borus\, the tension is above $3 $; however, when removing the variability from the analysis, we get much higher tensions ($\rm T>20 $) in favor of the `\nh\ Variable' scenario. This is also supported by the very low derived p-values. We thus classify the source as `\nh\ Variable', in agreement with the variable nature of the source found by \citet{marchese2012} and \citetalias{zhao_2021}.

\subsection{MRK 348}\label{sec:mrk348}
\noindent \textbf{Data reduction and fitting:} \chandra\ observation (ObsID: 12809) was not included in the fitting due to pile-up affecting the spectrum. The data quality of all observations was good enough to bin each spectrum with at least 15 cts/bin; we thus applied the $\chi^2$ statistics to our data.
\textbf{Analysis of results:} The data is well-fit by all models, although we find a significant tension ($\rm T>10 $) in all cases. This may be due to the large number of counts available for this source, suggesting the models we use are too simple to fit the data. Nevertheless, there are no apparent issues observed in the fit residuals that might indicate specific problems. The best-fit data for the torus parameters and the \nh\ are in good agreement between the three models, within errors. To properly model the soft part of the spectrum, we disentangled the \texttt{apec} normalization between the three soft observations (two \xmm\ and one \suzaku). We froze \uxclumpy\ inclination angle to be equal to the one we found using \borus, such that $\theta_{i,uxclumpy}=arccos(cos\theta_{i,borus02})=70^{\circ}$. Despite all the calculated tensions being $\rm T>10 $, the `\nh\ Variable' scenario is the one for which we find the lowest tension values. Considering this and the very low derived p-values, we classify this source as `\nh\ Variable'.
The best-fit results are compatible with the ones found by \citetalias{zhao_2021} and \cite{marchese2014}. Both the aforementioned studies found variability in the absorber of MRK 348, in agreement with our classification of the AGN as `\nh\ Variable'. 

\subsection{NGC 4992}\label{sec:ngc4992}
\noindent \textbf{Data reduction and fitting:} Due to the high particle background, we ignored the 0-3\,keV and 0-4\,keV range for \xmm\ (ObsID 0312192101) and \nustar\ (ObsID 60061239002) spectra, respectively. We fit \chandra\ observations 1 and 2 applying C-statistics, as the data quality forced us to bin them with 2 and 5 cts/bin, respectively. Thus, we used a combination of $\chi^2$ and C-statistics as total statistics to fit the data.
\textbf{Analysis of results:} All models fit the data well and are in agreement with each other. We note that our results agree with the ones found by \citetalias{zhao_2021}, although the latter found a slightly lower photon index with respect to the one measured in this work. We notice that, although the hydrogen column density in the line of sight is Compton-thin ($\sim 10^{23}$\,cm$^{-2}$), the average column density of the torus is Compton-thick ($> 1.5 \times 10^{24}$\,cm$^{-2}$). This is measured both by \borus\ and \mytorus\, and it is validated by the need of the CT \textit{inner ring} in \uxclumpy\ (CTKcover$\sim 0.30$), in a scenario similar to the one found by \cite{pizzetti2022} for NGC 7479 and by \citetalias{nuria2023} for NGC 612 and IC 4518A. The introduction of \nh\ variability is unnecessary to fit the data well, reinforced by a p-value of $>0.01$ for all models. We thus classify the source as `Non-variable in \nh'

\subsection{ESO 383-18}\label{sec:eso383-18}
\noindent \textbf{Data reduction and fitting:} Due to the poor data quality, we grouped the \chandra\ observation with 3 cts/bin, using C-statistic to fit the data. We thus use a mix of C-stat and $\chi^2$ as total statistics.
\textbf{Analysis of results:} The data is well-fit by all models, although they do not strongly agree regarding general torus properties. \mytorus\ favors a face-on scenario while \borus\ and \uxclumpy\ favor an edge-on scenario, with a lower plasma temperature and photon index with respect to \mytorus, in agreement with what was found by \citetalias{zhao_2021}. Within errors, the three models agree in the \nh\ measurements and in the classification of the source as `Non-variable in \nh'; we thus classify it as such. 

\subsection{MRK 417}\label{sec:mrk417}
\noindent \textbf{Data reduction and fitting:} Due to the low exposure time of the \chandra\ observation, we grouped the data with 5 cts/bin and used C-statistic for the fit. 
\textbf{Analysis of results:} The three models fit the data well. However, \mytorus\ and \borus\ disagree on the \nhav\ determination, with \uxclumpy\ favoring a transmission-dominated scenario (CTK=0) rather than a reflection-dominated one as inferred by \borus, in accordance with what was found by \citetalias{zhao_2021}. Nevertheless, the three models agree on the \nh\ measurements. We can not strongly affirm \nh\ variability is needed to fit the data, while the statistic analysis suggests flux variability is needed. We thus classify MRK 417 as `Non-variable in \nh'. 

\subsection{MCG-01-05-047}\label{sec:mcg0105047}
\noindent \textbf{Data reduction and fitting:} We omitted the energy range below 2\,keV due to the complexity of the fitting procedure; thus, the fitting is performed in the 2-25\,keV range. 
\textbf{Analysis of results:} The three models disagree, with \mytorus\ fitting the data with a better statistic than \borus\ and \uxclumpy. While the first model favors a Compton-thin \nhav and a \nh-variable scenario, the latter models support a CT torus, in agreement with \citetalias{zhao_2021}, with non-variable Compton-thin clouds, wherein variability primarily arises from intrinsic flux fluctuations. We classify this source as `Undetermined' due to the model disagreement in discerning the cause of variability.
 
\subsection{ESO 103-35}\label{sec:eso103-35}
\noindent \textbf{Data reduction and fitting:}  We disentangled the \texttt{apec} normalization of the two soft X-ray observations to properly fit the soft part of the spectrum.
\textbf{Analysis of results:} The data is well-fit by the models.  \borus\ and \uxclumpy\ support a reflection-dominated scenario, in which a high \nhav\ is needed to model the hard-energy tail of the data, in agreement with what reported by \citetalias{zhao_2021}; \mytorus\ on the other hand, prefers a low \nhav and low scattering fraction scenario, although the photon index is in agreement with the other two models. As reported by \citet{braatz1996} and mentioned in Section \ref{subsec:inner_ring}, ESO 103-35 hosts a nuclear 22 GHz water megamaser emission, which aligns in inclination with the torus: $\theta_{i,maser}=67^{\circ}$ \citep{bennert2004}, $\theta_{i,torus}\in [60^{\circ}-79.81^{\circ}]$.
All models require intrinsic flux variability to fit the data, while only \borus\ and \uxclumpy\ favor a \nh-variable scenario. Additionally, the p-value is above the threshold for all three models, suggesting the introduction of \nh\ variability is not necessary to reach a proper fit. Although the results convey some sort of variability is needed (either intrinsic flux or \nh), considering the discrepancy between the results, we classify this source as `Undetermined', as we are not able to discern the cause of such variability.

\subsection{NGC 1142}\label{sec:ngc1142}
\noindent \textbf{Data reduction and fitting:} No issues to report.
\textbf{Analysis of results:} The three models fit the data well and agree with each other. Although \mytorus\ and \borus\ predict the average density of the torus to be Compton-thin, in agreement with what was found by \citetalias{zhao_2021}, \uxclumpy\ requires the additional thick \textit{inner ring} to model the reflection component fully. Nevertheless, the three models fully agree with the measurements of the \nh. Interestingly, we measure a $\Delta$N$_{\rm H,los}\simeq 25\times 10^{22}$\,cm$^{-2}$ between the two subsequent \suzaku\ observations, over a time span of $\sim$ six months. This suggests the origin of the obscuration to be between $\sim$1-49 pc from the central engine. The \nh\ variable nature of this galaxy is also shown by the low tension and p-values derived by the statistical analysis, in agreement with \citetalias{zhao_2021}; we thus classify NGC 1142 as `\nh\ Variable'.

\subsection{IRAS 16288+3929}\label{sec:iras16288}
\noindent \textbf{Data reduction and fitting:} No issues to report.
\textbf{Analysis of results:} \mytorus\ and \borus\ fit the data well, while \uxclumpy\ shows poorer statistics. We note that the two former models favor a Compton-thin average torus density, in agreement with \citetalias{zhao_2021}, while \uxclumpy\ requires the addition of a geometrically thick ($\rm {CTKcover}>0.57$), Compton-thick reflector. Noticeably, IRAS 16288+3929 hosts a water megamaser, as reported by \citet{castangia2013}, suggesting the possible connection between the thick inner reflector predicted by \uxclumpy\ and the warped megamaser disk \citep[see, e.g.,][]{Masini2016,buchner2021}. The statistical analysis and the p-value derivation point towards a \nh-variable scenario; we then classify the source as such.

\subsection{ESO 263-13}\label{sec:eso263-13}
\noindent \textbf{Data reduction and fitting:} No issues to report.
\textbf{Analysis of results:} The three models are in agreement with each other, with \uxclumpy\ showing slightly better statistics than the other two models. Although \mytorus\ and \borus\ measure a Compton-thin \nhav ($\simeq 8\times 10^{22}$\,cm$^{-2}$), in agreement with the results obtained by \citetalias{zhao_2021}, \uxclumpy\ suggests the presence of a thick reflector to model the hard X-ray part of the spectrum. Both \borus\ and \uxclumpy\ favor a geometrically thin torus, in disagreement with the high covering factor (C$_F>0.30$) found by \citetalias{zhao_2021}. On the other hand, the models agree with \citetalias{zhao_2021} on the torus inclination angle, which is $\theta_i \gtrsim 40$. Despite no model yielding a $\rm T<3 $, introducing both \nh\ and flux variability is the scenario for which we derive the lowest tensions (T$_{\rm best-fit}=4.33-6.77  $, T$_{\rm No-var}=92.43-94.31  $). Thus, also considering the very low derived p-values, we classify the source as `\nh\ Variable'.

\subsection{Fairall 272}\label{sec:fairall272}
\noindent \textbf{Data reduction and fitting:} Due to the very low \chandra\ exposure time (2.9 ks), we bin the data to a minimum of 5 cts/bin, applying C-statistics; we then used a mix of $\chi^2$ and C-statistics to the total fit.
\textbf{Analysis of results:} The three models slightly overfit the data, being in good agreement with each other. \uxclumpy\ requires a geometrically thin (CTKcover $<0.16$) CT reflector to fit the reflection hump, while \mytorus\ and \borus\ find the overall torus density to be to the order of $\sim 7\times 10^{23}$\,cm$^{-2}$, in disagreement with the CT torus (\nhav$\sim 10^{23}$\,cm$^{-2}$) reported by \citetalias{zhao_2021}. We note that all three models yield a p-value below the threshold, while the analysis of the tensions cannot discern the source of variability needed to model the data. We classify the source as `Undetermined'.

\subsection{LEDA 2816387}\label{sec:leda2816387}
\noindent \textbf{Data reduction and fitting:} No issues to report.
\textbf{Analysis of results:} There is good agreement between the results of the three models, with \uxclumpy\ yielding slightly better statistics. The three models find a hard photon index ($\Gamma<1.4$), in agreement with what was found by \citetalias{zhao_2021}. \borus\ and \mytorus\ measure the \nhav\ to be Compton-thin, while \uxclumpy\ seems to favor the presence of a denser component (CTKcover $<0.27$). Both \uxclumpy\ and \borus\ favor the edge-on scenario. The p-value is above the threshold for all the models, and the tensions are below $3 $ both in the variable and non-variable case, leading us to classify the source as `Non-variable in \nh'.

\subsection{2MASX J06411806+3249313}\label{sec:2masxj}
\noindent \textbf{Data reduction and fitting:} For the \xmm\ observation, MOS1 was not used as it was corrupted.
\textbf{Analysis of results:} We opted not to include \texttt{apec} in the analysis due to the low signal below 2\,keV. The three models slightly overfit the data but are overall in good agreement. Within errors, \mytorus\ and \borus\ are in agreement with the \nhav measure obtained by \citetalias{zhao_2021}, although \uxclumpy\ requires a thicker reflector to model the hard X-ray part of the spectrum. \borus\ and \uxclumpy\ favor a face-on scenario, as also reported by \citetalias{zhao_2021}. We also notice how the best-fit \nh\ values are in the same range as \nhav, in a scenario similar to the one reported by \citetalias{nuria2023} for 3C 105, NGC 833 and NGC 612. Both the p-value and the tension analysis indicate a `Non-variable in \nh' scenario, and we classify the source as such.

\subsection{NGC 7479}\label{sec:ngc7479}
\noindent \textbf{Data reduction and fitting:} Refer to \citet{pizzetti2022} for details about the data reduction and fitting process.
\textbf{Analysis of results:} The three models agree in the classification of NGC 7479 as `Non-variable in \nh', yielding a p-value above the threshold and tensions above $3 $.

\subsection{NGC 6300}\label{sec:ngc6300}
\noindent \textbf{Data reduction and fitting:} Refer to Sengupta et al., in prep. for details about the data reduction and fitting process.
\textbf{Analysis of results:} The three models used to analyze this source are \borus, \uxclumpy\ and \texttt{XClumpy} \citep{Tanimoto2019}. All of them slightly over-fit the data but are in good agreement with each other. The p-value and tension analysis shows this is a variable source, but `Non-variable in \nh'. However, a p-value analysis of flux shows an indication of flux variability.
\end{document}